\documentclass[final,3p,times,authoryear]{elsarticle}
\usepackage{amssymb}
\usepackage[colorlinks,linkcolor=blue,urlcolor=black,citecolor=blue]{hyperref}
\usepackage{subcaption}
\usepackage{booktabs}
\usepackage{amsmath}
\usepackage{amsthm}
\usepackage{graphicx}
\usepackage{multirow}
\usepackage{floatrow}
\usepackage[title]{appendix}
\usepackage{animate}
\usepackage{diagbox}
\usepackage{float}
\usepackage{threeparttable}
\usepackage{xcolor}
\usepackage[displaymath,mathlines,edtable]{lineno}
\usepackage{url}
\usepackage{enumitem}

\usepackage[ruled,linesnumbered,lined,boxed,commentsnumbered]{algorithm2e}

\newfloatcommand{figurebox}{figure}[\nocapbeside][\dimexpr(\textwidth-\columnsep)/2\relax]
\newfloatcommand{tablebox}{table}[\nocapbeside][\dimexpr(\textwidth-\columnsep)/2\relax]
\def\textBF#1{{\fontseries{b}\selectfont{#1}}}

\makeatletter
\newcommand{\rmnum}[1]{\romannumeral #1}
\newcommand{\Rmnum}[1]{\expandafter\@slowromancap\romannumeral #1@}
\makeatother

\journal{arXiv}

\begin{document}

\begin{frontmatter}
\title{
Real-time Traffic Simulation and Management for Large-scale Urban Air Mobility: Integrating Route Guidance and Collision Avoidance
}
\fntext[co-author]{
The authors contributed equally as the first author.
}
\cortext[cor]{
Corresponding author.
}
\author[inst1]{Canqiang Weng\fnref{co-author}}
\ead{wengcq@mail2.sysu.edu.cn}
\author[inst1,inst3]{Can Chen\fnref{co-author}}
\ead{can-caesar.chen@connect.polyu.hk}
\author[inst1]{Jingjun Tan}
\ead{tanjj7@mail2.sysu.edu.cn}
\author[inst4]{Tianlu Pan}
\ead{pantl@pcl.ac.cn}
\author[inst1]{Renxin Zhong\corref{cor}}
\ead{zhrenxin@mail.sysu.edu.cn}

\affiliation[inst1]{
organization={Guangdong Provincial Key Laboratory of Intelligent Transportation System, School of Intelligent Systems Engineering, Shenzhen Campus of  Sun Yat-sen University},
city={Shenzhen},
country={China}
}
% \affiliation[inst2]{
% organization={Guangdong Provincial Key Laboratory of Intelligent Transportation Systems},
% city={Shenzhen},
% country={China}
% }
\affiliation[inst3]{
organization={Department of Civil and Environmental Engineering, The Hong Kong Polytechnic University}, city={Hong Kong}, country={China}
}
\affiliation[inst4]{
organization={Department of Network Intelligence, Peng Cheng Laboratory},
city={Shenzhen},
country={China}
}

% \pagewiselinenumbers

\begin{abstract}
With a vision to expand transportation supply using low-altitude airspace, urban air mobility (UAM) has emerged as a promising alternative to provide point-to-point travel in congested areas.
The rapid development of electric vertical take-off and landing vehicles is expected to make UAM a viable and sustainable transportation mode. 
Given the spatial heterogeneity of land use patterns in most cities, large-scale UAM deployments will likely focus on specific areas, such as intertransfer traffic between suburbs and city centers. 
However, large-scale UAM operations connecting multiple origin-destination pairs raise concerns about air traffic safety and efficiency due to potential conflict movements, particularly at major conflict points analogous to roadway junctions.
To meet the safety and efficiency requirements of future UAM operations, this work proposes an air traffic management framework that integrates route guidance and collision avoidance.
The route guidance mechanism optimizes aircraft distribution across both spatial and temporal dimensions by regulating their paths (composed of waypoints). 
Given the optimized paths, the collision avoidance algorithm generates collision-free aircraft trajectories between waypoints in the 3D space. 
To enable large-scale applications, we develop fast approximation methods for centralized path planning and adopt the velocity obstacle model for distributed collision avoidance.
To our knowledge, this work is one of the first to integrate route guidance and collision avoidance for UAM. 
Simulation results demonstrate that the proposed framework enables efficient and flexible UAM operations, including air traffic assignment, local congestion mitigation, and dynamic no-fly zone management.
Compared with a collision-free baseline strategy, the proposed framework achieves considerable improvements in traffic safety and efficiency, with increases in the average minimum separation (+98.2\%), the average travel speed (+70.2\%), and the trip completion rate (+130\%), along with a reduction in the energy consumption (-23.0\%).
The proposed framework demonstrates its potential for real-time traffic simulation and management in large-scale UAM systems.
\end{abstract}


% Proposes a novel air traffic management framework for large-scale UAM operations.

% Integrates centralized route guidance with distributed collision avoidance for multiple aircraft

% Enables flexible UAM operations, including local congestion mitigation and dynamic no-fly zone management.

% Ensures traffic homogeneity assumption for applying the airspace Macroscopic Fundamental Diagram model

\end{frontmatter}

% \pagewiselinenumbers

\section{Introduction}\label{sec:Introduction}
With network capacity approaching saturation, traffic congestion in metropolises becomes a persistent challenge.
Simply expanding roadway infrastructure is no longer a sustainable solution to meet the increasing demand.
Instead, exploring innovative transportation modes is a more viable way to expand the supply of urban mobility.
Recently, low-altitude airspace (an underused resource for urban mobility) has gained increased attention due to its potential to support point-to-point air travel in congested areas.
Advances in machine intelligence, autonomy, and battery technologies are expected to enable electric vertical take-off and landing vehicles (eVTOLs) to become a safe, affordable, and environmentally friendly mode \citep{holden2018uber,kasliwal2019role}. As a result, urban air mobility (UAM) has emerged with the vision of unlocking altitude dimensions for urban transport to mitigate ground traffic congestion.
The feasibility of large-scale UAM has been further evidenced by the successful test flights of full-size eVTOLs, supported by the operational experience of ride-hailing platforms such as Lyft and Uber \citep{dietrich2020urban,garrow2021urban}. According to a recent report by \cite{whitepaper2023}, the UAM market in China alone is projected to contribute up to \$700 billion (RMB 5 trillion) to the national economy in the coming decade.

As the large-scale deployment of UAM systems approaches, concerns about air traffic safety and efficiency are growing.
Safety concerns primarily focus on collision avoidance for individual aircraft. Recent studies suggest that new approaches, distinct from the conventional scheme of pre-planned routes and fixed schedules, are compelling to address potential conflicts among multiple aircraft \citep{haddad2021traffic,safadi2023macroscopic}.
Efficiency concerns, on the other hand, highlight the challenge of alleviating traffic congestion in low-altitude airspace.
Emerging research indicates that UAM systems face congestion issues similar to ground traffic \citep{cummings2024airspace,cummings2024comparing}.
To address these critical concerns, we need to investigate: \textit{(\rmnum{1}) the individual traffic behavior of UAM aircraft}, and \textit{(\rmnum{2}) the collective traffic behavior of UAM systems}.

The microscopic behavior of individual aircraft is typically modeled as point-to-point motion in the 2D or 3D space, subject to the requirement of avoiding collisions with obstacles and other aircraft.
Collision avoidance is a critical issue in UAM systems and, more broadly, in multi-agent systems. 
Various algorithms have been reported over the past decade, as reviewed in \cite{huang2019collision,yasin2020unmanned} and the references therein.
Some representative works are summarized in \autoref{tab:review}.
In general, collision avoidance strategies can be categorized into two main types: centralized and distributed approaches.
In centralized approaches, a control center collects relevant information (e.g., agents' positions and velocities) and uses it to optimize decision making for all agents \citep{bahabry2019low,tang2021automated}.
These approaches often assume that all agents perfectly comply with the control commands, without accounting for uncertainties such as pilots' compliance.
In contrast, distributed approaches rely on each agent independently collecting limited information within its local detection range to make decisions \citep{van2011reciprocal,long2017deep}. 
An agent may observe the behavior of neighboring agents, but it cannot directly influence their actions.
It is widely accepted that centralized strategies can achieve globally optimal solutions at the expense of high computational cost, whereas distributed strategies offer locally optimal solutions with enhanced computational efficiency.

\begin{table}[!ht]\scriptsize
    \centering
    \begin{threeparttable}[b]
    \captionsetup{font=footnotesize}
    \caption{A brief survey of collision avoidance algorithms.}
    \vspace{-2mm}
    \label{tab:review}
    \renewcommand\arraystretch{1.3}
    \begin{tabular}{p{4cm}p{1.5cm}p{3cm}p{1.5cm}p{4cm}}
    \toprule
    & Strategy & Scenario & Methodology & Objective (\textit{minimize})\\
    \hline
    \cite{jose2016task} & Centralized & 2D grid network & A$^*$ and GA\tnote{a} & Total time spent / fuel consumption\\
    \cite{xue2019scenario} & Centralized & 2D space without obstacle & MILP\tnote{b} & Deviation from the reference trajectory\\
    \cite{alrifaee2014centralized} & Centralized & 2D space with obstacle & MPC\tnote{c} & Deviation from the reference trajectory\\
    \cite{bahabry2019low,tang2021automated} & Centralized & 3D airline network & MILP\tnote{b} & Total time spent / total flying cost\\
    \cite{yang2020scalable} & Distributed & 2D space without obstacle & Game Theory\tnote{d} & Total time spent\\
    \cite{quan2021practical} & Distributed & 3D space with obstacle & APF\tnote{e} & Deviation from the reference trajectory\\
    \cite{van2011reciprocal} & Distributed & 2D/3D space with obstacle & VO\tnote{f} & Deviation from the reference velocity\\
    \cite{long2017deep} & Distributed & 2D/3D space without obstacle & DNN\tnote{g} & ---\\
    \bottomrule
    \end{tabular}
    \begin{tablenotes}
    \item[a] Genetic Algorithm
    \item[b] Mixed Integer Linear Program
    \item[c] Model Predictive Control
    \item[d] Logit level-k model 
    \item[e] Artificial Potential Field
    \item[f] Velocity Obstacles
    \item[g] Deep Neural Network
   \end{tablenotes}
   \end{threeparttable}
\vspace{-2mm}
\end{table}

\begin{figure}[!ht]
\centering
\begin{subfigure}[t]{0.245\linewidth}
    \centering
    \includegraphics[width=\linewidth]{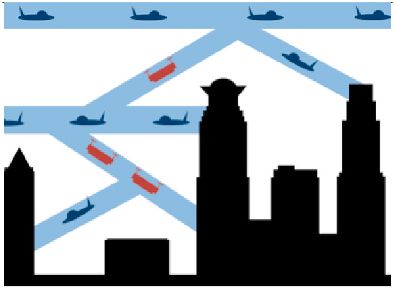}
    \captionsetup{font=footnotesize}
    \subcaption{Tube}
    \label{fig:tubes}
\end{subfigure}
\begin{subfigure}[t]{0.245\linewidth}
    \centering
    \includegraphics[width=\linewidth]{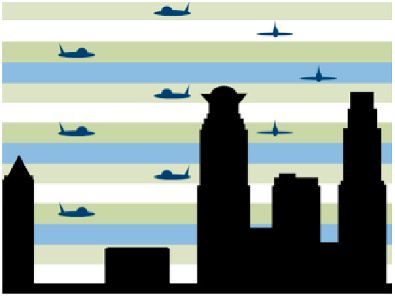}
    \captionsetup{font=footnotesize}
    \subcaption{Layer}
    \label{fig:layers}
\end{subfigure}
\begin{subfigure}[t]{0.245\linewidth}
    \centering
    \includegraphics[width=\linewidth]{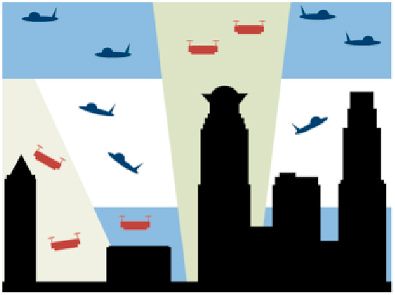}
    \captionsetup{font=footnotesize}
    \subcaption{Zone}
    \label{fig:zones}
\end{subfigure}
\begin{subfigure}[t]{0.245\linewidth}
    \centering
    \includegraphics[width=\linewidth]{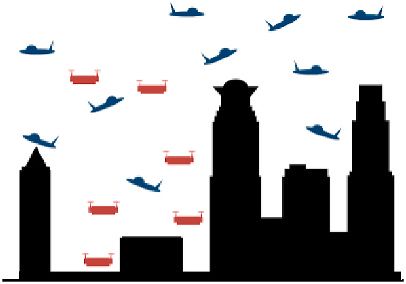}
    \captionsetup{font=footnotesize}
    \subcaption{Full Mix}
    \label{fig:full_mix}
\end{subfigure}
\captionsetup{font=footnotesize}
\caption{Different airspace structure designs for UAM, ordered by degrees of freedom \citep{sunil2015metropolis}.}
\label{fig:airspace}
\end{figure}

In the early stage of UAM deployment, structured airspace is suggested to reduce the risk of aircraft collisions.
Currently, airspace structures proposed by both academia and industry can be roughly categorized into four basic concepts
\footnote{Here, we treat the terms tube, corridor, sky lane, and airway as referring to the same structure concept, despite minor distinctions across existing studies \citep{jang2017concepts,bradford2020urban,quan2021sky}.}, 
i.e., tube, layer, zone, and full mix, as illustrated in \autoref{fig:airspace}.
The concept of airspace structure is to separate aircraft with different properties (e.g., speed, direction, and level of autonomy) by implementing virtual barriers.
For example, \cite{quan2021sky} proposed a tube-based structure to separate aircraft flying in opposing directions and to establish ``sky highways", which are conceptually analogous to roadway networks.
\cite{Liu2025Integrated} investigated aircraft take-off management and trajectory optimization for merging control in air-corridor networks.
Furthermore, tube-based structures are extended into multi-layer configurations to accommodate additional complexity and traffic density \citep{jang2017concepts,bradford2020urban}.
Alternatively, a zone-based air traffic network is explored to enhance aircraft safety and control through effective collision avoidance \citep{fu2022practical}.
For a comprehensive review of airspace structure designs and their performance trade-offs, readers are referred to \citet{bauranov2021designing}.
Typically, less structured airspace allows aircraft to fly with more degrees of freedom, leading to higher traffic density. 
However, high-density aircraft flying freely along self-preferred (often direct) routes would increase the chance of traffic congestion.

Although airspace structures provide a rule-based traffic scheme for UAM in its early stages, further investigation into air traffic management is needed to address future demands. Relevant research in this field is being led by efforts in the United States and Europe.
The original concept of operations (ConOps) for unmanned aerial vehicle traffic management (UTM) was proposed by NASA \citep{kopardekar2016unmanned}.
The European version of UTM, known as U-space, was introduced in \cite{uspace2017}. A detailed discussion and comparison of the UTM and U-space ConOps can be found in \cite{shrestha2021survey}.
The proposed ConOps is similar to a set of guidelines that describe the functions of system components and how they interact with each other \citep{shrestha2021survey}. Under the current UTM architecture, point-to-point operations relying on human operators have limitations on the deployment scale.
Nevertheless, both UTM and U-space envision a future of large-scale UAM services with high levels of autonomy.
In this context, the challenge of air traffic congestion resulting from high-density UAM operations has been receiving increasing attention. \cite{cummings2024airspace} simulated point-to-point aircraft movement in the 3D space using a decentralized conflict resolution approach.
Their simulations indicate that the UAM system is subject to congestion phenomena similar to those observed in roadway traffic.
\cite{cummings2024comparing} evaluated air traffic congestion under different airspace structures and suggested that advanced routing algorithms can help mitigate air traffic congestion.
\cite{safadi2023macroscopic} expanded the simulation scale and investigated fundamental relationships between air traffic flow, density, and speed. The air traffic flow is observed to follow a concave curve with respect to traffic density, where the critical density serves as a benchmark for identifying congestion \citep{cummings2024airspace}.
Based on the flow-density relationship, which is also known as the Macroscopic Fundamental Diagram (MFD), several approaches have been proposed to manage air traffic congestion \citep{haddad2021traffic,safadi2023aircraft,weng2025urban}.

\begin{figure}[!ht]
\centering
\begin{subfigure}[t]{0.492\linewidth}
    \centering
    \includegraphics[width=\linewidth]{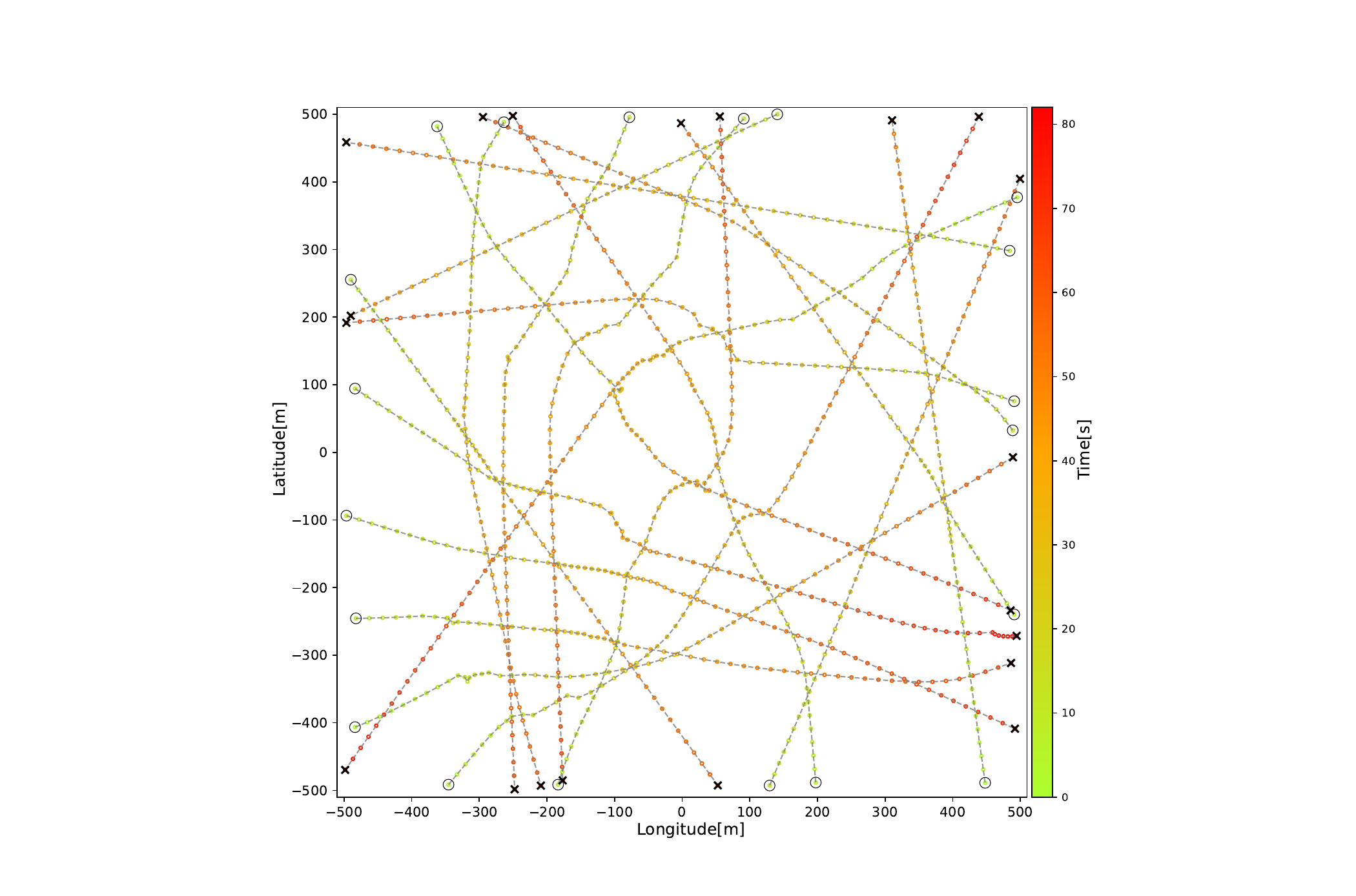}
    \captionsetup{font=footnotesize}
    \subcaption{Uniformly distributed OD pattern}
    \label{fig:homogeneous_OD}
\end{subfigure}
\hspace{1mm}
\begin{subfigure}[t]{0.492\linewidth}
    \centering
    \includegraphics[width=\linewidth]{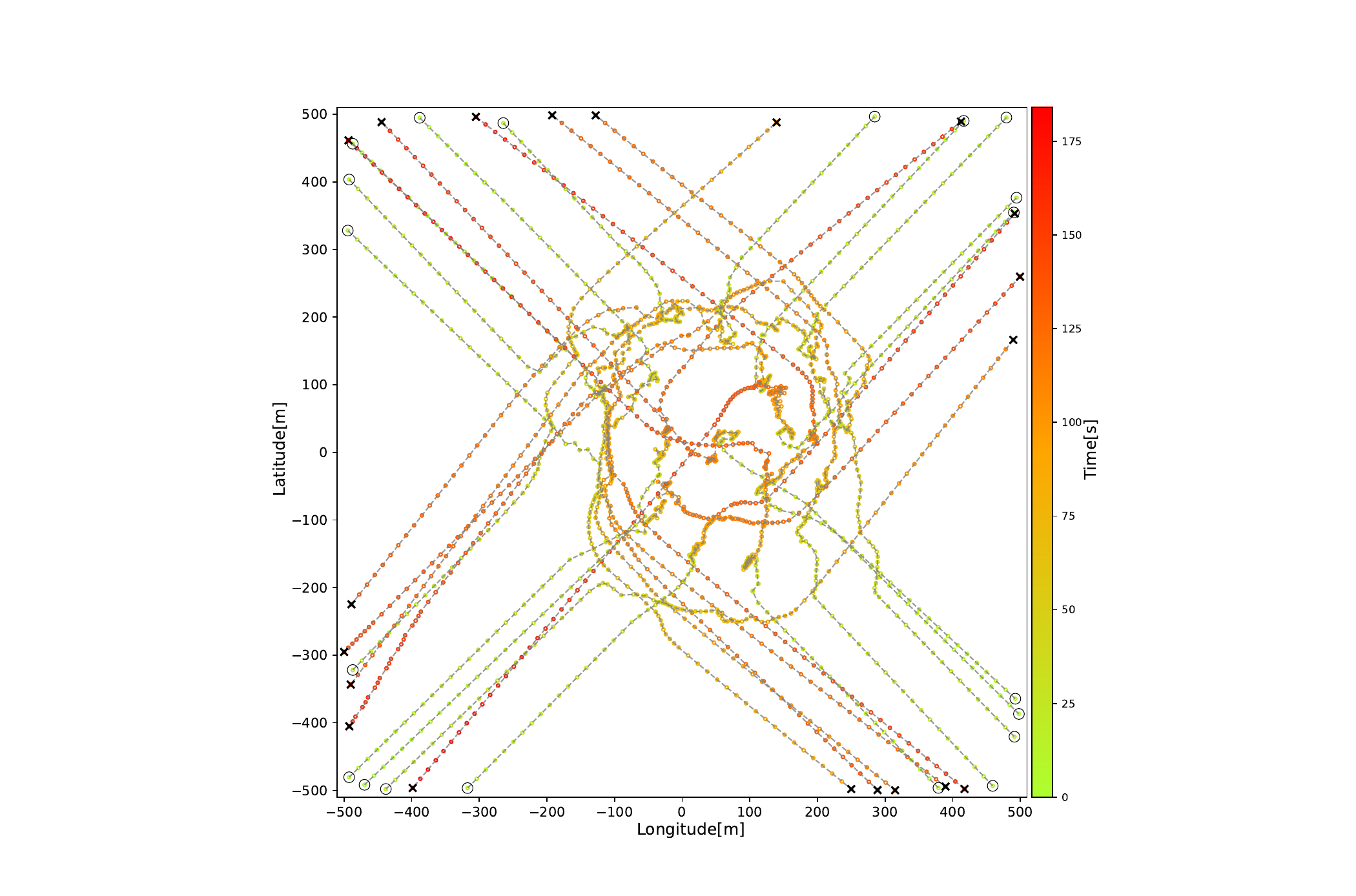}
    \captionsetup{font=footnotesize}
    \subcaption{Directionally distributed OD pattern}
    \label{fig:heterogeneous_OD}
\end{subfigure}
\captionsetup{font=footnotesize}
\caption{A motivating example:
Simulation of collision-free aircraft trajectories in the 2D space with a deployment scale of 20 aircraft, where ``O" marks the origin and ``X" marks the destination. 
Compared with the homogeneous OD pattern in (\hyperref[fig:homogeneous_OD]{a}), the heterogeneous OD pattern in (\hyperref[fig:heterogeneous_OD]{b}) poses higher risks of delay and congestion.
}
\label{fig:demand_comparison}
\end{figure}

For roadway traffic, the MFD framework implicitly assumes that traffic congestion is evenly distributed across space, i.e., the areas with an MFD should be (roughly) homogeneously loaded \citep{geroliminis2011properties}.
To fulfill this assumption, current studies often simplify UAM demand by adopting a uniformly distributed origin-destination (OD) pattern \citep{haddad2021traffic,safadi2024integrated}.
However, this simplification encounters practical limitations, as UAM infrastructure in most cities is unevenly distributed due to the spatial heterogeneity of land use patterns \citep{wu2021integrated}.
In reality, large-scale UAM services are more likely to be deployed along specific corridors with concentrated demand, such as between suburbs and city centers. 
To visualize the impact of OD distribution on air traffic operations, we depict simulated trajectories in \autoref{fig:demand_comparison}.
Potential aircraft conflicts are evenly distributed across the airspace under a uniformly distributed OD pattern, whereas they become concentrated in specific areas (similar to roadway junctions) under a directionally distributed OD pattern.
In this example, heterogeneous demand, compared with homogeneous demand, raises greater concerns about traffic safety and efficiency, as evidenced by reductions in average minimum separation (-74.1\%) and average travel speed (-53.1\%).
Similar phenomena of local traffic deadlock have also been reported in the literature \citep{dergachev2021distributed,cummings2024comparing}.
Investigating air traffic management schemes for spatially heterogeneous demand is of practical significance but has not yet received sufficient attention.

To address the above research gaps, we propose an air traffic management framework that integrates route guidance and collision avoidance for large-scale UAM systems.
The main contributions of this paper are summarized as follows:
\begin{itemize}[leftmargin=5mm]
    \item[\textBF{1.}] 
    To the best of our knowledge, this work is one of the first to integrate route guidance and collision avoidance for large-scale UAM operations.
    We develop a centralized path planning model and fast approximation methods to optimize the spatial-temporal distribution of aircraft within a multi-layer regional airspace network.
    Furthermore, we employ the velocity obstacle model to generate collision-free aircraft trajectories between airspace regions in a distributed manner.
    The integration of route guidance and collision avoidance effectively reduces the occurrence of aircraft conflicts and local traffic deadlock, thereby enhancing both traffic safety and efficiency.
    \item[\textBF{2.}] 
    The proposed framework enables efficient and flexible UAM operations, such as air traffic assignment, local congestion mitigation, and dynamic no-fly zone management.
    In addition, the framework can ensure air traffic homogeneity even under spatially heterogeneous demand.
    Compared with a collision-free baseline strategy, the proposed framework achieves considerable improvements in traffic safety and efficiency, with increases in the average minimum separation (+98.2\%), the average travel speed (+70.2\%), and the trip completion rate (+130\%), along with a reduction in the energy consumption (-23.0\%).
\end{itemize}

The remainder of this paper is organized as follows. 
\autoref{sec:Methodology} presents the methodology of the proposed framework, including the approaches for route guidance and collision avoidance. 
\autoref{sec:result} evaluates the performance of the proposed framework through a series of experiments. 
Finally, \autoref{sec:Conclusion} concludes the paper and discusses potential directions for future work.

\section{Large-scale UAM operations with route guidance and collision avoidance}\label{sec:Methodology}

In this section, we first propose a multi-layer regional airspace network design, which defines the operational scope for aircraft motion control. 
Then, we introduce the proposed air traffic management framework, including its overall workflow and key components. 
Furthermore, we formulate the centralized path planning model and the distributed collision avoidance approach in detail.

\subsection{The proposed multi-layer regional airspace network}\label{sec:network_design}

\begin{figure}[!ht]
    \centering
    \includegraphics[width=0.58\linewidth]{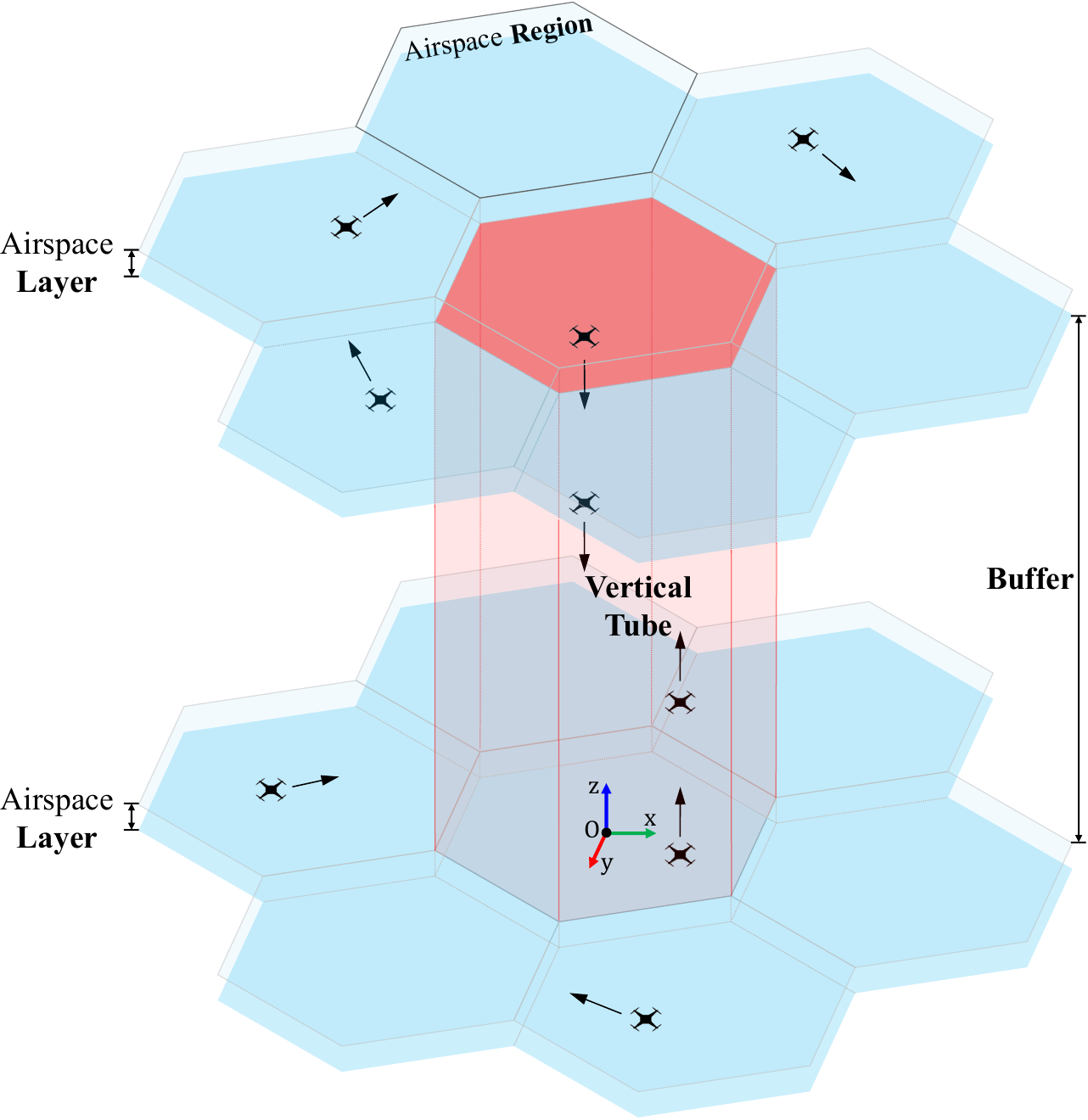}
    \captionsetup{font=footnotesize}
    \caption{Multi-layer regional airspace network design for large-scale UAM operations.}
    \label{fig:multi-layer-region}
\end{figure}

As discussed, the individual traffic behavior of UAM aircraft is directly influenced by the airspace structure.
Inspired by \cite{jang2017concepts,haddad2021traffic}, we adopt a multi-layer regional airspace network design in this paper.
As illustrated in \autoref{fig:multi-layer-region}, the low-altitude airspace is partitioned into multiple layers along the altitude ($z$) axis. 
Each layer is further subdivided into regions based on longitude ($x$) and latitude ($y$).
Within each region, the full-mix structure is employed, allowing aircraft to freely select their flight directions.
To ensure vertical separation and reduce the risk of conflicts, altitude buffers are introduced between adjacent layers.
Additionally, specific regions are connected by vertical tubes that function as dedicated corridors for cross-layer transitions \citep{samir2019internet}.
The proposed airspace network integrates the concepts of layer, zone, and full mix, aiming to achieve a reasonable trade-off between aircraft safety and airspace capacity.

\begin{figure}[!ht]
    \centering
    \includegraphics[width=0.7\linewidth]{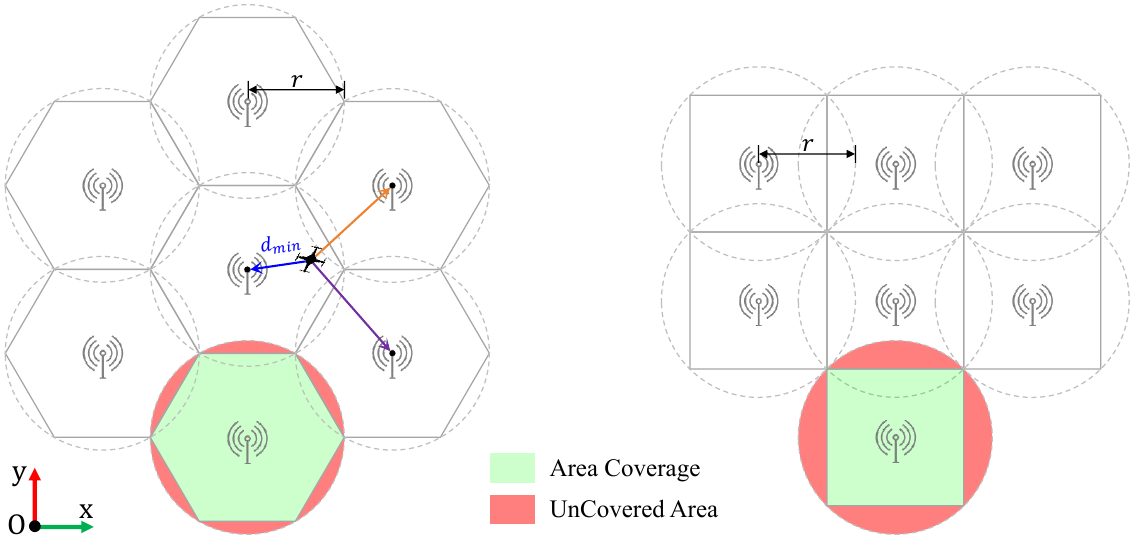}
    \captionsetup{font=footnotesize}
    \caption{
    Hexagonal and square grid network partitions. In a hexagonal grid network, the cell containing a point corresponds to the one whose center is closest to that point. 
    For the same cell size and number, a hexagonal grid network covers a larger area than a square grid network.
    }
\label{fig:hexagonal-division}
\end{figure}

In this paper, we adopt a uniform hexagonal shape and size for each airspace region.
This hexagonal grid has been widely employed across various scientific domains (e.g., biological applications, mobile communications, and digital imaging), due to its advantages \citep{prvan2019review}.
First, hexagonal grids can cover the plane completely without overlaps or gaps, and provide higher area coverage compared with other polygonal shapes, such as triangular or rectangular ones \citep{hepsibha2013comparative}.
\autoref{fig:hexagonal-division} presents a visual comparison between hexagonal and square grid networks in terms of area coverage.
Another important advantage is that each hexagonal cell maintains the same distance from all its neighboring cells, enabling efficient aircraft localization.
Specifically, identifying the region where an aircraft is located can be simplified to finding the closest cell center, as illustrated in \autoref{fig:hexagonal-division}.

\begin{figure}[!ht]
\centering
\begin{subfigure}[t]{0.4\linewidth}
    \centering
    \includegraphics[height=5cm]{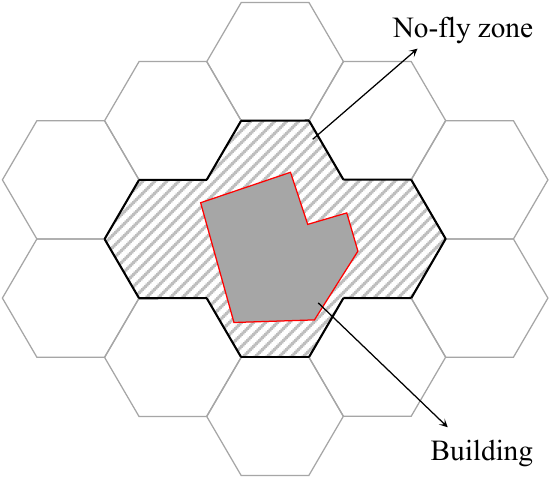}
    \captionsetup{font=footnotesize}
    \subcaption{}
    \label{fig:}
\end{subfigure}
\begin{subfigure}[t]{0.4\linewidth}
    \centering
    \includegraphics[height=5cm]{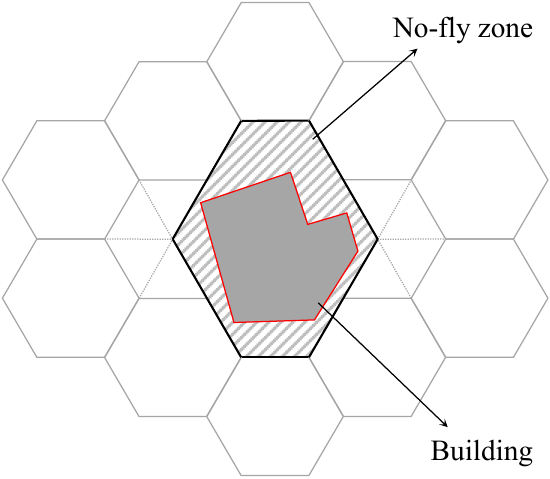}
    \captionsetup{font=footnotesize}
    \subcaption{}
    \label{fig:}
\end{subfigure}
\captionsetup{font=footnotesize}
\caption{
Two approaches for handling no-fly zones induced by buildings in a hexagonal airspace layer:  
(a) closing entire hexagonal cells that overlap with buildings;  
(b) subdividing hexagonal cells into triangular sub-cells and closing only those that overlap with buildings.
}
\label{fig:buliding}
\end{figure}

The layer-region division presented above is based on the defined scope of usable airspace, which generally refers to low-altitude airspace outside of no-fly zones.
Identifying no-fly zones requires consideration of multiple factors, including buildings, weather conditions, noise pollution, and privacy concerns \citep{bauranov2021designing}.
In structured airspace configurations, no-fly zones typically exhibit regular shapes. 
For instance, in this study, strategies for designing no-fly zones are illustrated in \autoref{fig:buliding}. 
One approach is to close all hexagonal cells that overlap with buildings, whereas an alternative method subdivides each hexagonal cell into smaller cells (e.g., triangles) and closes only those sub-cells overlapping with buildings.
Compared with the second method, the first approach is easier to implement but may result in a waste of usable airspace.
By contrast, no-fly zones in unstructured airspace may assume more complex, irregular shapes. 
\cite{stevens2020generating} explored irregular polygonal keep-out geofences by accounting for vehicle performance constraints and environmental factors, such as minimum turning radius and persistent wind.
\cite{vagal2021new} developed an algorithm based on alpha shapes and Voronoi diagrams to generate dynamic geofences in urban environments. 
Within a hexagonal airspace network, we validate the proposed route guidance mechanism for accommodating no-fly zone management using experimental examples in \autoref{sec:flexible UAM operations}.
Nevertheless, extending the proposed methods to different airspace configurations remains an important direction for future work.

\subsection{Framework for integrating route guidance and collision avoidance}

\begin{figure}[!ht]
    \centering
    \includegraphics[width=\linewidth]{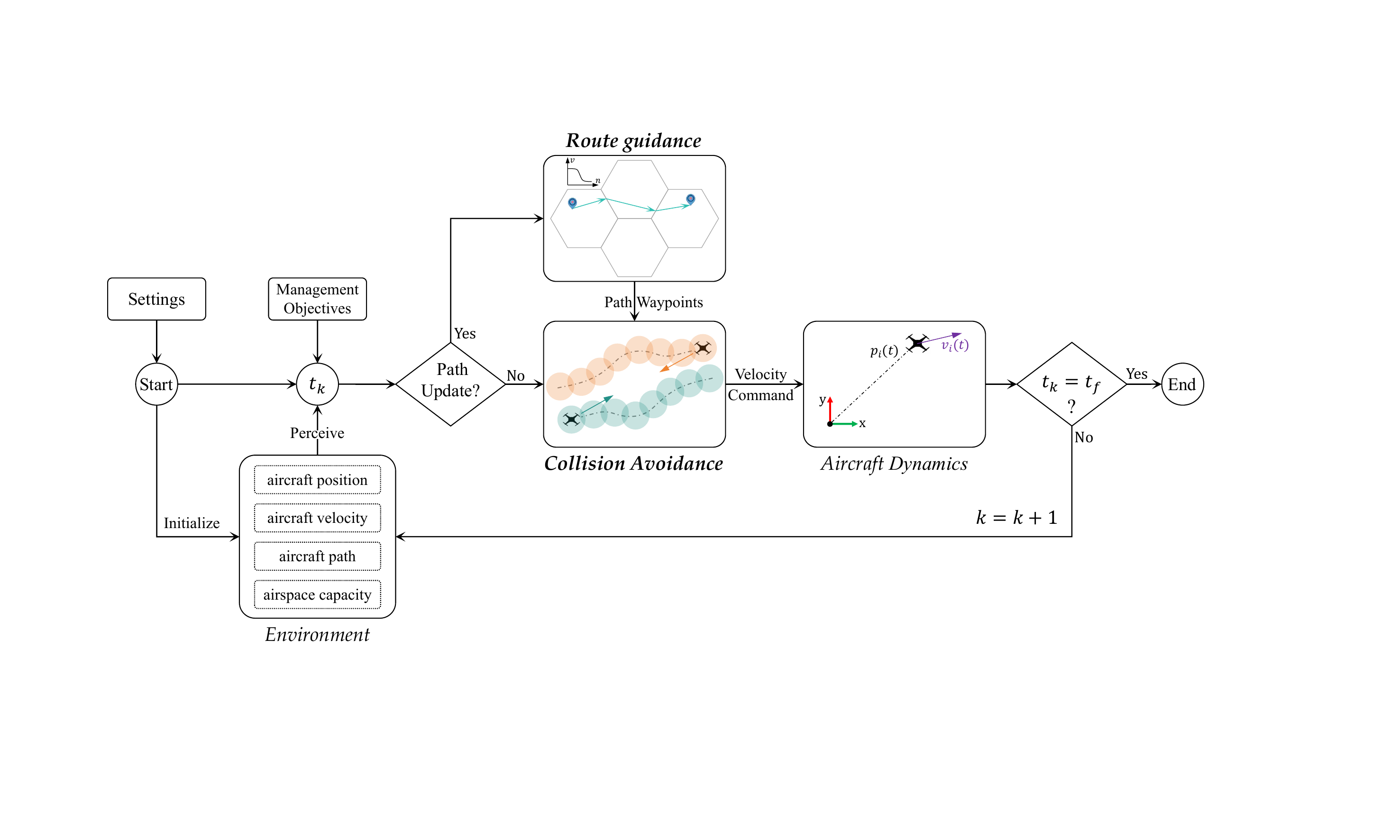}
    \captionsetup{font=footnotesize}
    \caption{Proposed air traffic management framework integrating route guidance and collision avoidance for large-scale UAM operations.}
    \label{fig:framework}
\end{figure}

The proposed air traffic management framework is illustrated in \autoref{fig:framework}, and its workflow is outlined as follows.
First, the UAM system is initialized by configuring relevant parameters, including airspace settings (e.g., layer altitude and region radius), aircraft parameters (e.g., safety radius and maximum speed), and demand patterns (e.g., OD distribution and departure flow).
At each discrete time step $t_k$, environmental information (e.g., aircraft positions, velocities, and paths) is collected to support the decision-making process: %, which consists of two core components: 
route guidance and collision avoidance.
The system manager determines whether aircraft paths require updating based on the current system state and management objectives. If an update is triggered, the route guidance algorithm generates all aircraft paths in a centralized manner. 
Otherwise, aircraft paths (composed of waypoints) are retained\,\footnote{Each aircraft's initial path is a direct route from its origin to its destination.}.
Given the current paths of all aircraft, the collision avoidance algorithm computes velocity commands and safety envelopes for each aircraft in a distributed manner.
These velocity commands are then passed to aircraft dynamics to steer the aircraft between waypoints while ensuring safe separation.
This decision-making loop continues until the mission is completed.

%The will be presented in later sections.
Before formulating the route guidance and collision avoidance problems, we introduce the following assumptions to simplify the modeling and ensure the tractability of the proposed framework.
\begin{itemize}[leftmargin=5.5mm]
\setlength{\itemsep}{0mm}
    \item [(1)] Aircraft are equipped with sufficient battery power or fuel to complete their designated flight missions.
    \item [(2)] Aircraft are capable of executing control commands without unacceptable deviations.
    \item [(3)] Time delays and uncertainties in aircraft dynamics, such as those induced by communication disturbances or wake turbulence, are not considered.
\end{itemize}
Since this paper focuses on operational management of air traffic flow in terms of route guidance and collision avoidance strategies, we adopt the above assumptions to avoid introducing additional complexities arising from energy management, actuator limitations, or dynamic uncertainties.
Nevertheless, extending the proposed framework to accommodate more realistic operational constraints is an important direction for future research.

\subsection{Region-based route guidance for collective aircraft management}\label{sec:Route_guidance}

As discussed in \autoref{sec:Introduction}, relying solely on local optimal collision avoidance strategies for aircraft motion control presents significant challenges, particularly under conditions of over-saturation or spatially heterogeneous travel demand.
On the one hand, aircraft behavior driven by local (or user) optimal collision avoidance is often egoistic.
Intuitively, each aircraft seeks the shortest path as illustrated in \autoref{fig:demand_comparison}.
As a result, certain airspace regions may experience excessive aircraft density, thereby increasing the risk of traffic congestion or even deadlock.
Similar findings have been reported in the literature, accompanied by recommendations for advanced routing algorithms \citep{safadi2023macroscopic, cummings2024comparing}.
On the other hand, previous studies on urban ground traffic management show that an appropriate regional route guidance strategy can not only alleviate traffic congestion in specific regions (e.g., the CBD area) but also improve overall network performance \citep{li2022quasi,chen2024iterative}.
Motivated by these insights, we propose a region-based optimal route guidance strategy, which aims to optimize the 4D spatial-temporal trajectories of aircraft by regulating their paths.
As demonstrated in \autoref{sec:single-layer}, the proposed route guidance strategy can effectively mitigate local traffic congestion and retain air traffic homogeneity, even in cases of over-saturation or spatially heterogeneous demand.

\subsubsection{Optimal path planning based on 4D trajectory prediction}\label{sec:OptPP}

\begin{figure}[!h]
    \centering
    \includegraphics[width=0.5\linewidth]{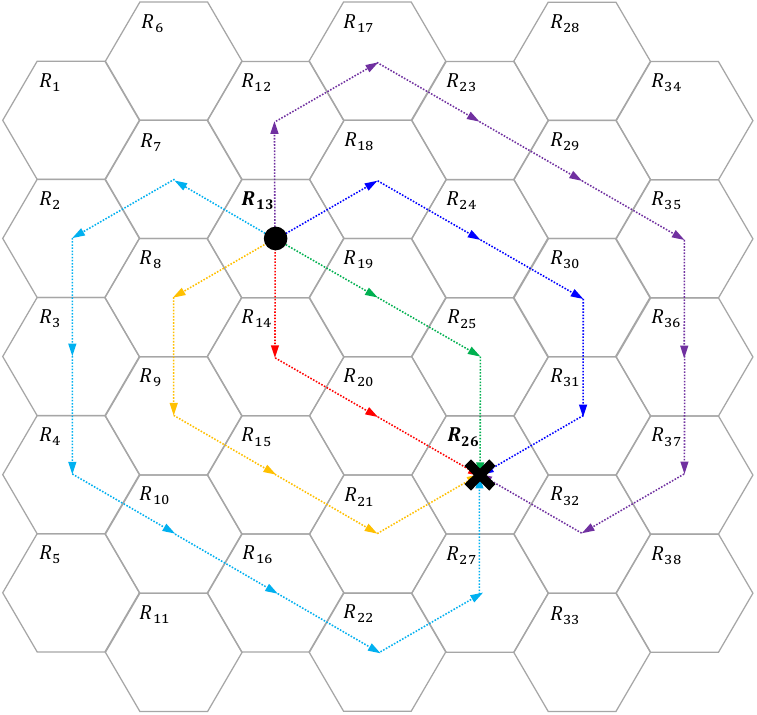}
\caption{Schematic of candidate paths from region $R_{13}$ to $R_{26}$.}
\label{fig:candidate_path}
\end{figure}

We consider multiple aircraft $A_i, i=1,2,\dots,N$ with given ODs.
For convenience, suppose that given ODs are located within the airspace network illustrated in \autoref{fig:multi-layer-region}, which allows us to omit the control of aircraft during takeoff and landing phases.
Based on the topology of the airspace network, multiple candidate paths can be constructed between each OD pair.
Let $\Omega_i$ denote the set of candidate paths available to aircraft $A_i$, and let $P_i \in \Omega_i$ represent the selected path for aircraft $A_i$.
To overcome the limitations arising from individual egoistic behaviors, the path choices of multiple aircraft must exhibit some form of coordination.
For instance, when the number of aircraft assigned to a certain path becomes excessive, other aircraft should avoid selecting that path.
In the literature, path planning is typically driven by costs associated with candidate paths, implying that the cost for an aircraft choosing a path needs to account for the influence of other aircraft's path decisions.
We introduce $\boldsymbol{P} \triangleq (P_1,P_2,\dots,P_N) \in \boldsymbol{\Omega}$ to denote the joint path decision of all aircraft, where $\boldsymbol{\Omega}=\Omega_1 \times \Omega_2 \times \dots \times \Omega_N$ represents the Cartesian product of all candidate path sets.
Accordingly, the cost for aircraft $A_i$ selecting path $P_i$ is expressed as a function of the joint path decision, denoted by $\mathcal{C}_i(\boldsymbol{P})$.
In this paper, collaborative path planning aims to find the optimal joint path decision $\boldsymbol{P}^*$ that minimizes the total cost across all aircraft, i.e., 
\begin{align}\label{eq:system_path_planning}
    \boldsymbol{P}^* = \mathop{\arg\min}_{\boldsymbol{P} \in \boldsymbol{\Omega}} \ \mathcal{J}(\boldsymbol{P}) = \sum_{i=1}^{N} \mathcal{C}_i\big(\boldsymbol{P}\big)
\end{align}
The optimization problem \eqref{eq:system_path_planning} represents a generalized form of collaborative path planning for multiple agents.
In the remainder of this section, we introduce detailed definitions of candidate paths and path costs to further refine the problem. First, we define a path $P_i$ for aircraft $A_i$ as a sequence of waypoints $\boldsymbol{p}^w \in \mathbb{R}^3$ in the 3D space, i.e.,
\begin{align}\label{eq:waypint_sequence}
P_i \triangleq
\langle\,\boldsymbol{p}^w_{i,o}\,,\,\boldsymbol{p}^w_{i,1}\,,\,\boldsymbol{p}^w_{i,2}\,,\,\dots\,,\,\boldsymbol{p}^w_{i,m_i}\,,\,\boldsymbol{p}^w_{i,d} \,\rangle
\end{align}
where the first waypoint $\boldsymbol{p}^w_{i,o}$ and the last waypoint $\boldsymbol{p}^w_{i,d}$ correspond to the given origin and destination, respectively. 
The intermediate waypoints $\boldsymbol{p}^w_{i,1}\,,\,\boldsymbol{p}^w_{i,2}\,,\,\dots\,,\,\boldsymbol{p}^w_{i,m_i}$ correspond to the pass-by airspace regions along path $P_i$. Based on the above definitions, the candidate path set $\Omega_i$ for aircraft $A_i$ is constructed following \autoref{alg:candidate_paths}.
For better illustration, an example of candidate path construction is provided in \autoref{fig:candidate_path}. Intuitively, candidate paths for each aircraft are designed such that they do not overlap the same airspace regions, thereby promoting a balanced utilization of the airspace. Moreover, restricting each aircraft to up to six candidate paths helps to limit the search space for path planning.

\begin{algorithm}[!h]
    \caption{Candidate Paths Construction}
    \label{alg:candidate_paths}
\vspace{1mm}
\SetKwInOut{Input}{input}\SetKwInOut{Output}{output}
\Input{\ 
    Airspace network $\mathcal{G}$, represented as a directed graph, where\\
    \, -- Nodes represent airspace regions;\\
    \, -- Edges indicate connectivity between regions;\\
    \, -- Weights of edges correspond to the distances between the centroids of connected regions.
}\vspace{1mm}

\Input{\ 
    Origin $\boldsymbol{p}_{i,o}$ and destination $\boldsymbol{p}_{i,d}$.
}\vspace{1mm}

\Output{\ 
    Candidate paths set $\Omega_i$.
}\vspace{1mm}

Initialize candidate paths set $\Omega_i = \{\}$.
\vspace{1mm}

Calculate the airspace regions in which origin $\boldsymbol{p}_{i,o}$ and destination $\boldsymbol{p}_{i,d}$ are located, denoted as $R_{i,o}$ and $R_{i,d}$, respectively.
\vspace{1mm}

\While{ a path exists from $R_{i,o}$ to $R_{i,d}$ in graph $\mathcal{G}$ }
{\vspace{1mm}

    Search the shortest path from node $R_{i,o}$ to node $R_{i,d}$ in graph $\mathcal{G}$, denoted as $\langle\,R_{i,o}, R_{i,1}, R_{i,2},\dots, R_{i, m_i}, R_{i,d}\,\rangle$.
    \vspace{1mm}
    
    Construct path $P_i$ using equation \eqref{eq:waypint_sequence}, where $\boldsymbol{p}^w_{i,o} = \boldsymbol{p}_{i,o}$, $\boldsymbol{p}^w_{i,d} = \boldsymbol{p}_{i,d}$, and $\boldsymbol{p}^w_{i,j}$ is the centroid of region $R_{i,j}$, $\forall j =1,2,\dots,m_i$.
    \vspace{1mm}
    
    Add $P_i$ to the candidate path set $\Omega_i = \Omega_i \bigcup \{P_i\}$.
    \vspace{1mm}
    
    Remove intermediate nodes $R_{i,1}, R_{i,2},\dots, R_{i, m_i}$ from graph $\mathcal{G}$.
    
}\vspace{1mm}

Export data $\Omega_i$.
\vspace{1mm}

\end{algorithm}

\begin{figure}[!h]
    \centering
    \includegraphics[width=0.38\linewidth]{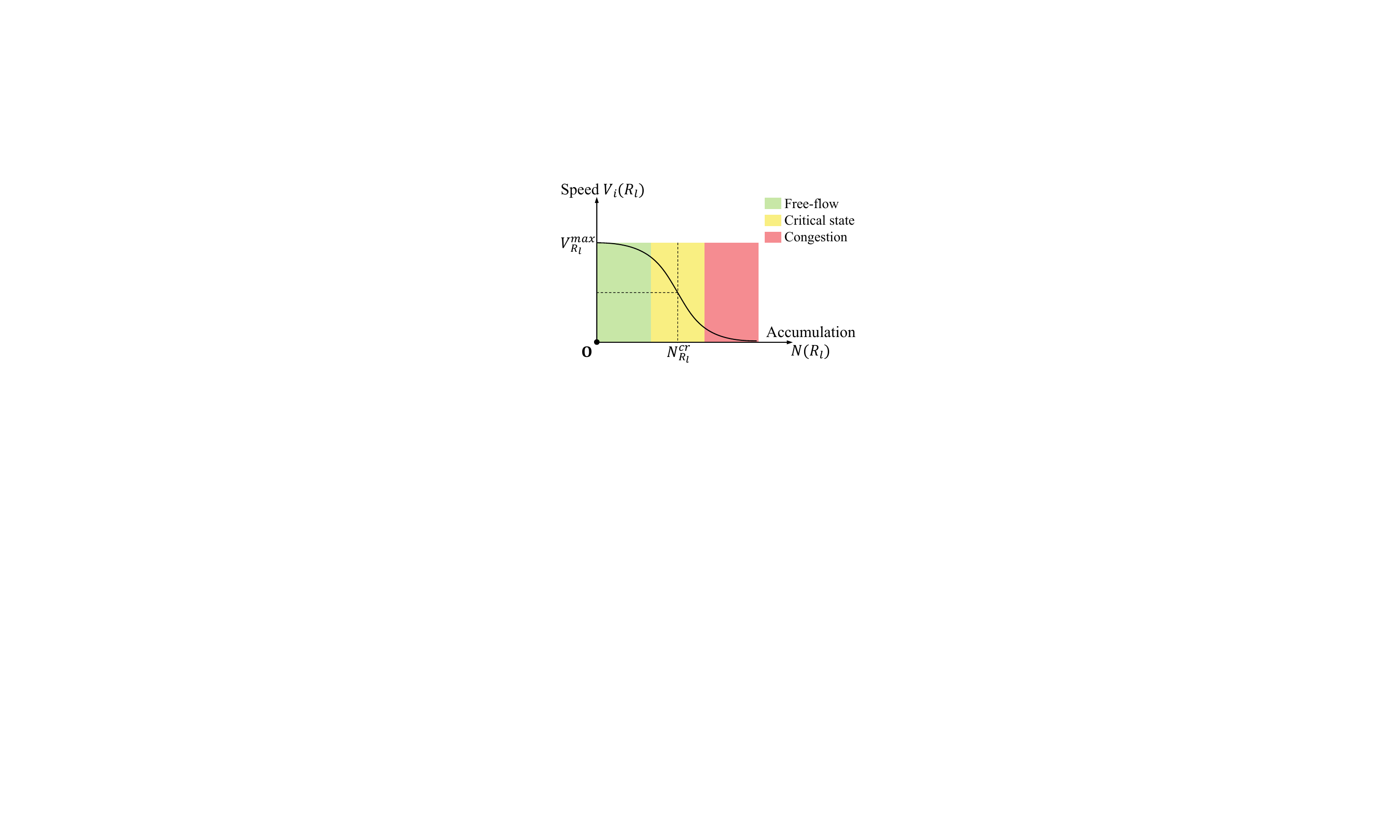}
\caption{When calculating path costs, the speed of aircraft $A_i$ needs to respond to the traffic states of region $R_l$ where aircraft $A_i$ is located.}
\label{fig:sigmoid}
\end{figure}

We further define $\mathcal{C}_i(\boldsymbol{P})$ as the travel time of aircraft $A_i$ along path $P_i$.
When calculating path costs, we treat aircraft as point particles with no collision volume for simplicity, implying that aircraft travel along straight-line segments between waypoints.
Under this simplification, if aircraft $A_i$ flies at a constant speed, the path cost $\mathcal{C}_i(\boldsymbol{P})$ depends solely on the length of path $P_i$.
To account for the impact of traffic congestion on path costs, we regulate aircraft speed according to the following rule:
\begin{align}\label{eq:sigmoid}
    V_i({R_l}) = \frac{ \exp\Big( -N({R_l}) + N^{cr}_{R_l}\, \Big)}{1+ \exp\Big( -N({R_l}) + N^{cr}_{R_l}\, \Big)}V^{max}_{R_l}
\end{align}
where $V_i({R_l})$ denotes the speed of aircraft $A_i$ within airspace region $R_l$; 
$N({R_l})$ denotes the number of aircraft currently present in region $R_l$;
$N^{cr}_{R_l}$ and $V^{max}_{R_l}$ are given parameters, representing the critical accumulation and maximum speed of aircraft within region $R_l$, respectively.
A schematic illustration of this speed regulation rule is provided in \autoref{fig:sigmoid}, where traffic congestion increases the cost of traversing region $R_l$ by reducing aircraft speed.
Note that regional traffic congestion results from path decisions of all aircraft.
In other words, the regulation rule \eqref{eq:sigmoid} captures the interactions among aircraft path choices by coupling their path costs.
This coupling is essential to achieving collaborative path planning, but it also complicates cost computation, as path travel times can no longer be calculated analytically.
One approach to addressing this challenge is to predict the 4D trajectories of all aircraft.
Once the predicted trajectories, denoted by $\hat{\boldsymbol{p}}_i(t), i = 1,2,\dots,N$, are given, the cost $\mathcal{C}_i(\boldsymbol{P})$ can be calculated by definition
\begin{align}
    \mathcal{C}_i(\boldsymbol{P}) = 
    \big(\mathop{\arg\min}_{t} \, \Vert \,\hat{\boldsymbol{p}}_i(t) - \boldsymbol{p}_{i, d}\,\Vert\,\big) - 
    \big(\mathop{\arg\min}_{t} \, \Vert \,\hat{\boldsymbol{p}}_i(t) - \boldsymbol{p}_{i, o}\,\Vert\,\big)
\end{align}
where the first and second terms on the right-hand side represent the times at which aircraft $A_i$ reaches its destination and origin, respectively.
\autoref{alg:4D_trajectories} outlines the procedure used to predict 4D aircraft trajectories.

As formulated above, regional traffic congestion would significantly increase path costs.
In turn, the solution to the path planning problem \eqref{eq:system_path_planning}, which minimizes the total cost, is promising to avoid local traffic congestion.
More generally, collaborative path planning promotes a homogeneous distribution of aircraft trajectories across both temporal and spatial dimensions.

\begin{algorithm}[!h]
    \caption{Aircraft 4D Trajectories Prediction}
    \label{alg:4D_trajectories}
\vspace{1mm}
\SetKwInOut{Input}{input}\SetKwInOut{Output}{output}
\Input{\ 
    Joint path $\boldsymbol{P} = (P_1, P_2,\dots, P_N)$.
}\vspace{1mm}

\Input{\ 
    Prediction step $\Delta t$.
}\vspace{1mm}

\Output{\ 
    Predicted aircraft trajectories $\hat{\boldsymbol{p}}_i(t), i=1,2,\dots,N$.
}\BlankLine

Initialize $k=0$, $\Psi = \{1,2,\dots,N\}$.
\vspace{1mm}

\For{$i \in \Psi$}
{\vspace{1mm}

    Initialize aircraft's position $\hat{\boldsymbol{p}}_i(t_k) = \boldsymbol{p}^w_{i,o}$.
    \vspace{1mm}
    
    Initialize aircraft's target waypoint $\boldsymbol{p}^{tar}_i = \boldsymbol{p}^w_{i,1}$.
    
}\vspace{1mm}

\While{$ \Psi \neq \varnothing$}
{\vspace{1mm}

    Calculate the traffic accumulation in each airspace region based on the current positions of all aircraft.
    \vspace{1mm}
    
    \For{$i \in \Psi$}
    {\vspace{1mm}
    
        \If{$\Vert\,\hat{\boldsymbol{p}}_i(t_k) - \boldsymbol{p}^w_{i,d}\,\Vert \leq \epsilon$} 
        {\vspace{1mm}
        
            $\Psi = \Psi \setminus \{i\}$.
            \vspace{1mm}
            
            Export data \ $\hat{\boldsymbol{p}}_i(t), \ t=t_0,t_1,\dots,t_k$.
            
        }\vspace{1mm}
        
        Update the target waypoint $\boldsymbol{p}^{tar}_i$ for aircraft $A_i$ based on its current position $\hat{\boldsymbol{p}}_i(t_k)$ and path $P_i$.
        \vspace{1mm}
        
        Calculate aircraft's moving direction 
        $\boldsymbol{U}_i = \displaystyle{\frac{\boldsymbol{p}^{tar}_i - \hat{\boldsymbol{p}}_i(t_k)}
        {\Vert\,\boldsymbol{p}^{tar}_i - \hat{\boldsymbol{p}}_i(t_k)\,\Vert}}$.
        \vspace{1mm}

        Calculate aircraft's moving speed $V_i$ using equation \eqref{eq:sigmoid}.
        \vspace{1mm}
        
        $t_{k+1} = t_k + \Delta t$.
        \vspace{1mm}
        
        Update aircraft's position  $\hat{\boldsymbol{p}}_i(t_{k+1}) = \hat{\boldsymbol{p}}_i(t_{k}) + V_i \cdot \boldsymbol{U}_i \cdot \Delta t$.
        
    }\vspace{1mm}
    
    $k = k+1$.
    
}\vspace{1mm}

\end{algorithm}

\subsubsection{Fast approximation methods for approaching the optimal route guidance strategy}\label{sec:FAM}

Solving the collaborative path planning problem \eqref{eq:system_path_planning} is a challenging task.
On the one hand, the search space of the joint path decision grows exponentially, i.e., $|\boldsymbol{\Omega}| = \prod_{i=1}^N |\Omega_i| = |\Omega_1|\times|\Omega_2|\times\cdots\times|\Omega_N|$.
As a result, traditional approaches such as exhaustive search become computationally infeasible for large-scale problems.
On the other hand, calculating path costs based on precise aircraft trajectory prediction is computationally expensive.
To address these challenges, we propose fast approximation methods (FAMs) to approximate the optimal route guidance strategy.

First, we introduce a simplified method for estimating path costs, as outlined in \autoref{alg:path cost}.
This method approximates precise aircraft trajectories using the sequence of pass-by regions along each path.
In addition, it enforces temporal alignment of the pass-by region sequences across all aircraft.
Specifically, each column of the matrix $\mathcal{R}$ in \autoref{alg:path cost} represents a state of aircraft traveling in the corresponding regions, and state transitions (from one column to the next) occur synchronously for all aircraft.
In this way, the calculation of regional traffic accumulation, regional travel speed, and regional travel time is reduced to a series of matrix operations.
Compared with the model predictive approach, this method significantly improves computational efficiency.

\begin{algorithm}[!h]
    \caption{Fast Path Cost Estimation}
    \label{alg:path cost}
\vspace{1mm}
\SetKwInOut{Input}{input}\SetKwInOut{Output}{output}
\Input{\ 
    Joint path $\boldsymbol{P} = (P_1, P_2,\dots, P_N)$.
}\vspace{1mm}

\Output{\ 
    Estimated total path cost $\hat{\mathcal{J}}(\boldsymbol{P})$.
}\vspace{1mm}

\For{$i = 1,2,\dots,N$}
{\vspace{1mm}

    Calculate the pass-by regions along path $P_i$, denoted as $\langle\, R_{i,1}\,,\,R_{i,2}\,,\,\dots\,,\,R_{i,m_i}\,\rangle$.
    
}\vspace{1mm}

Construct the pass-by region matrix
\begin{align*}
\mathcal{R} = 
\begin{bmatrix}
    R_{1,1}, &R_{1,2}, &\ldots, &R_{1,M}\\
    R_{2,1}, &R_{2,2}, &\ldots, &R_{2,M}\\
    \vdots &\vdots &\ddots &\vdots \\
    R_{N,1}, &R_{N,2}, &\ldots, &R_{N,M}\\
\end{bmatrix}
\end{align*}
where $M = \max\{m_1,m_2,\dots,m_N\}$ represents the maximum number of pass-by regions, and the undefined elements $R_{i,m_{i+1}},R_{i,m_{i+2}},\dots,R_{i,M}$ are padded with null values.
\vspace{1mm}

\For{$j = 1,2,\dots,M$}
{\vspace{1mm}

    Count the number of non-empty elements $R_{i,j}$ in column $j$ of the matrix $\mathcal{R}$, denoted as $N_{R_{i,j}}$.
    \vspace{1mm}
    
    Calculate the aircraft speed corresponding to $N_{R_{i,j}}$ using equation \eqref{eq:sigmoid}, denoted as $V_{i}(R_{i,j})$.
    
}\vspace{1mm}

Construct the regional travel time matrix
\begin{align*}
\mathcal{T} = 
\begin{bmatrix}
    \frac{L}{V_{1}(R_{1,1})}, &\frac{L}{V_{1}(R_{1,2})}, &\ldots, &\frac{L}{V_{1}(R_{1,M})}\\
    \frac{L}{V_{2}(R_{2,1})}, &\frac{L}{V_{2}(R_{2,2})}, &\ldots, &\frac{L}{V_{2}(R_{2,M})}\\
    \vdots &\vdots &\ddots &\vdots \\
    \frac{L}{V_{N}(R_{N,1})}, &\frac{L}{V_{N}(R_{N,2})}, &\ldots, &\frac{L}{V_{N}(R_{N,M})}\\
\end{bmatrix}
\end{align*}
where $L$ represents the average travel distance within each region, and the undefined elements $\frac{L}{V_{i}(R_{i,m_{i+1}})},\frac{L}{V_{i}(R_{i,m_{i+2}})},\dots,\frac{L}{V_{i}(R_{i,M})}$ are padded with zeros.
\vspace{1mm}

Calculate the sum of all elements in the matrix $\mathcal{T}$, i.e., $\text{sum}(\mathcal{T})=\sum_{i=1}^{N}\sum_{j=1}^{M} \frac{L}{V_{i}(R_{i,j})}$.
\vspace{1mm}

Export data $\hat{\mathcal{J}}(\boldsymbol{P})=\text{sum}(\mathcal{T})$.
\vspace{1mm}

\end{algorithm}

The simplification adopted in \autoref{alg:path cost} (i.e., temporal alignment) requires that the travel speeds of all aircraft be (approximately) uniform.
Notably, such uniformity in travel speeds emerges from a balanced distribution of regional traffic.
As discussed, collaborative path planning promotes regional traffic homogeneity, which, to some extent, justifies the reasonableness of \autoref{alg:path cost}.
Based on the simulation results presented in \autoref{sec:Evaluation_FAM}, we find this simplified method acceptable in practice, while a detailed theoretical analysis of its accuracy is beyond the scope of this paper.

We further propose an approximate optimal path search method, as detailed in \autoref{alg:path search}.
This method employs incremental path planning for individual aircraft to approximate the optimal joint path for the entire fleet.
Specifically, aircraft are prioritized based on their origin-to-destination distances, with shorter distances assigned higher priority.
The first aircraft (with the highest priority) is allowed to search for its optimal path independently, without accounting for the presence of other aircraft (with lower priorities).
Once the first path is determined, the algorithm performs an exhaustive search over the candidate paths of the second aircraft, aiming to optimize the joint path of these two aircraft.
A similar process continues iteratively for each subsequent aircraft, where the algorithm searches for the optimal path of the current aircraft while considering only the influence of higher-priority aircraft.

The rationale behind the distance-based priority assignment is twofold. 
First, prioritizing aircraft with shorter OD distances facilitates earlier trip completion, which helps reduce potential disruptions to lower-priority aircraft. 
Second, aircraft with longer origin-to-destination distances typically have more routing flexibility. Consequently, adjusting the paths of these long-distance aircraft to avoid congestion in critical regions is preferable, as it can be achieved without significantly increasing travel distances.
\autoref{alg:path search} efficiently produces an approximate optimal solution by decomposing the original combinatorial optimization problem into a sequence of simpler subproblems.
This approach significantly reduces the search space, lowering the computational complexity from exponential to linear, i.e., $\prod_{i=1}^N |\Omega_i| \rightarrow \sum_{i=1}^N |\Omega_i|$.
The simulation results presented in \autoref{sec:Evaluation_FAM} demonstrate that the approximate solution closely matches the exact solution.

\begin{algorithm}[!h]
    \caption{Approximate Optimal Path Search}
    \label{alg:path search}
\vspace{1mm}
\SetKwInOut{Input}{input}\SetKwInOut{Output}{output}
\Input{\ 
    Candidate path sets $\Omega_i, i=1,2,\dots,N$.
}\vspace{1mm}

\Output{\ 
    Approximate optimal joint path $\hat{\boldsymbol{P}^*}$.
}\vspace{1mm}

Sort the aircraft by the distance between their origin $\boldsymbol{p}^w_{i,o}$ and destination $\boldsymbol{p}^w_{i,d}$ in ascending order, denoted as $\langle A_1,A_2,\dots,A_N\rangle$.
\vspace{1mm}

Search for the shortest path within the path set $\Omega_1$, denoted as $\hat{P}^*_1$.
\vspace{1mm}

\For{$i = 2,3,\dots,N$}
{\vspace{1mm}

    \For{$P_i \in \Omega_i$}
    {\vspace{1mm}
    
        Define $\boldsymbol{P} \triangleq (\hat{P}^*_1, \hat{P}^*_2,\dots, \hat{P}^*_{i-1}, P_i)$, and calculate the corresponding cost $\hat{\mathcal{J}}(\boldsymbol{P})$ using \autoref{alg:path cost}.
        
    }\vspace{1mm}
    
    Search for the optimal path $\hat{P}^*_i = \mathop{\arg\min}\limits_{P_i \in \Omega_i} \ \hat{\mathcal{J}}(\boldsymbol{P})$.
    
}\vspace{1mm}

Construct the approximate optimal solution to problem \eqref{eq:system_path_planning} as $\hat{\boldsymbol{P}^*} = (\hat{P}^*_1, \hat{P}^*_2,\dots, \hat{P}^*_N)$.
\vspace{1mm}

\end{algorithm}
\subsection{Velocity obstacle model for individual aircraft collision avoidance}\label{sec:Collision_avoidance}

The route guidance strategy specifies the airspace regions that aircraft should pass through.
However, the actual trajectories between regional waypoints are determined by aircraft motion control algorithms.
These algorithms are responsible for steering each aircraft toward its target waypoint while avoiding collisions with other aircraft.
As summarized in \autoref{tab:review}, a variety of collision avoidance algorithms have been proposed for multi-agent systems.
Within the context of UAM, \cite{safadi2023macroscopic} recently evaluated the effectiveness of the Artificial Potential Field (APF) method for aircraft collision avoidance.
To provide an alternative solution, this paper adopts the Velocity Obstacles (VO) model for aircraft collision avoidance in high-density scenarios.
Among its advantages, the VO model offers an intuitive geometric interpretation and demonstrates high computational efficiency \citep{van2011reciprocal, long2017deep,han2022reinforcement}.

We consider an aircraft $A_i$ tasked with reaching its target waypoint $\boldsymbol{p}^{tar}_i$.
The motion dynamics of the aircraft are described by the following discrete-time kinematic equations
\begin{subequations}\label{eq:dynamics}
\begin{align}
    \boldsymbol{p}_i(t_{k+1}) & = \boldsymbol{p}_i(t_k) + \boldsymbol{v}_i(t_k) \Delta t 
    \label{eq:dynamics_a}
    \\
    \boldsymbol{v}_i(t_{k+1}) & = \boldsymbol{v}_i(t_k) + \boldsymbol{a}_i(t_k) \Delta t
    \label{eq:dynamics_b}
\end{align}
\end{subequations}
where $\boldsymbol{p}_i(\cdot)\in \mathbb{R}^3$, $\boldsymbol{v}_i(\cdot)\in \mathbb{R}^3$ and $\boldsymbol{a}_i(\cdot)\in \mathbb{R}^3$ denote the position, velocity, and acceleration of aircraft $A_i$, respectively.
The time steps $t_k$ and $t_{k+1}$ correspond to the current and next discrete moments, and $\Delta t$ denotes the time interval between them.
Many organizations and companies have developed open-source semi-autonomous autopilot systems, which are capable of automatically regulating acceleration based on given velocity commands.
Following \cite{quan2021practical}, we assume the acceleration obeys the following control model
\begin{align}\label{eq:acceleration}
    \boldsymbol{a}_i(t_k) = - l_i \,\big(\,\boldsymbol{v}_i(t_k)- \boldsymbol{v}^{c}_i(t_k)\,\big)
\end{align}
where $\boldsymbol{v}^{c}_i(\cdot) \in \mathbb{R}^3$ and $l_i > 0$ are the velocity command and control gain of aircraft $A_i$, respectively.
Substituting \eqref{eq:acceleration} into \eqref{eq:dynamics_b}, we derive the following expression for the velocity update
\begin{align}
    \boldsymbol{v}_i(t_{k+1}) = (1-\omega_i) \,\boldsymbol{v}_i(t_k) + \omega_i \,\boldsymbol{v}^{c}_i(t_k)
\end{align}
where $\omega_i = l_i \,\Delta t$.
This expression reveals that the updated velocity is a weighted combination of the current velocity and the commanded velocity.
Consequently, the aircraft velocity asymptotically tracks the commanded velocity and can reach it exactly within a finite time horizon $\tau = 1/l_i$, assuming constant command.
The control gain $l_i$ thus serves as a measure of the aircraft's acceleration capability, which can be obtained through flight experiments.
Intuitively, a larger $l_i$ implies faster convergence to the commanded velocity, enabling more responsive maneuvering.

The velocity command serves two purposes: (\rmnum{1}) guiding the aircraft toward its target waypoint, and (\rmnum{2}) ensuring collision avoidance with other aircraft. 
To fulfill these objectives, we introduce two fundamental concepts: preferred velocity and safe velocity.
Firstly, the preferred velocity for aircraft $A_i$ is defined as the maximum velocity vector directed from its current position toward its target waypoint. 
Formally, it is given by
\begin{align}
    \boldsymbol{v}^p_i = \frac{\boldsymbol{p}^{tar}_i - \boldsymbol{p}_i}{\Vert\, \boldsymbol{p}^{tar}_i - \boldsymbol{p}_i \,\Vert} v^{max}_i
\end{align}
where $v^{max}_i$ denotes the maximum speed of aircraft $A_i$.
The preferred velocity represents the aircraft's most efficient effort to reach its target waypoint.
Secondly, the safe velocity for aircraft $A_i$ refers to any velocity that enables aircraft $A_i$ to avoid collisions with other aircraft over a time horizon of at least $\tau$. 
We denote the set of all such collision-free velocities by $\Psi_i$.
Based on above definitions, the velocity command can be given by solving the following quadratic programming problem
\begin{align}\label{eq:orca_objective}
    \boldsymbol{v}^{c}_i = \mathop{\arg\min}_{\boldsymbol{v}\in \Psi_i} \Vert\, \boldsymbol{v} -  \boldsymbol{v}^{p}_i \,\Vert
\end{align}
which seeks a velocity within the safe set to be as close as possible to the preferred velocity.
The complexity of solving problem \eqref{eq:orca_objective} largely depends on the structure of the constraint set $\boldsymbol{v} \in \Psi_i$. 
Fortunately, the velocity obstacle model offers a tractable approach for constructing this constraint.
In particular, \cite{van2011reciprocal} derives sufficient conditions for safe velocities and reformulates them as a set of linear constraints.
To acknowledge their contribution, we encourage readers to consult the original work for further details.
Nonetheless, for completeness, we provide a brief derivation of the safe velocity constraints at the end of this section.

\begin{figure}[!h]
\centering
\begin{subfigure}[t]{0.3\linewidth}
    \centering
    \includegraphics[height=6.5cm]{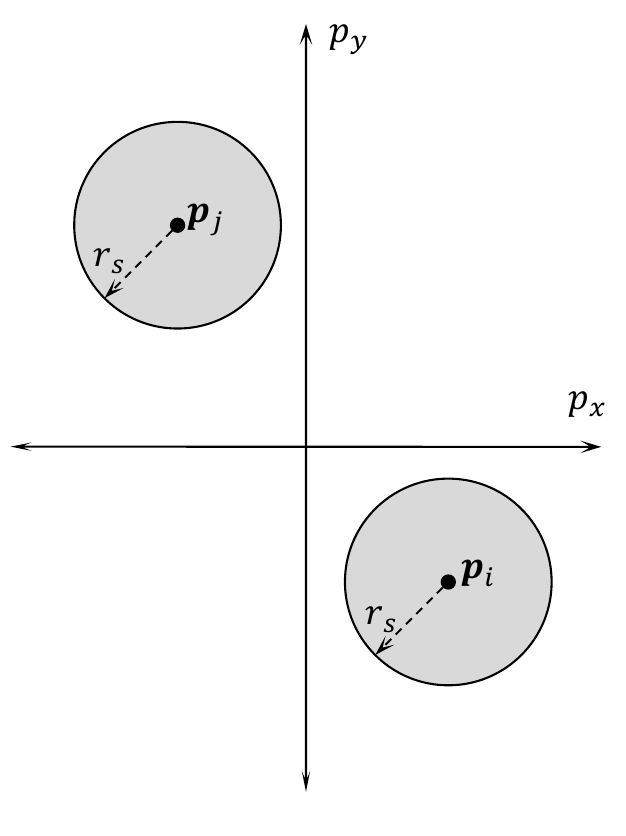}
    \captionsetup{font=footnotesize}
    \subcaption{}
    \label{fig:orca-a}
\end{subfigure}
\hspace{20mm}
\begin{subfigure}[t]{0.55\linewidth}
    \centering
    \includegraphics[height=6.5cm]{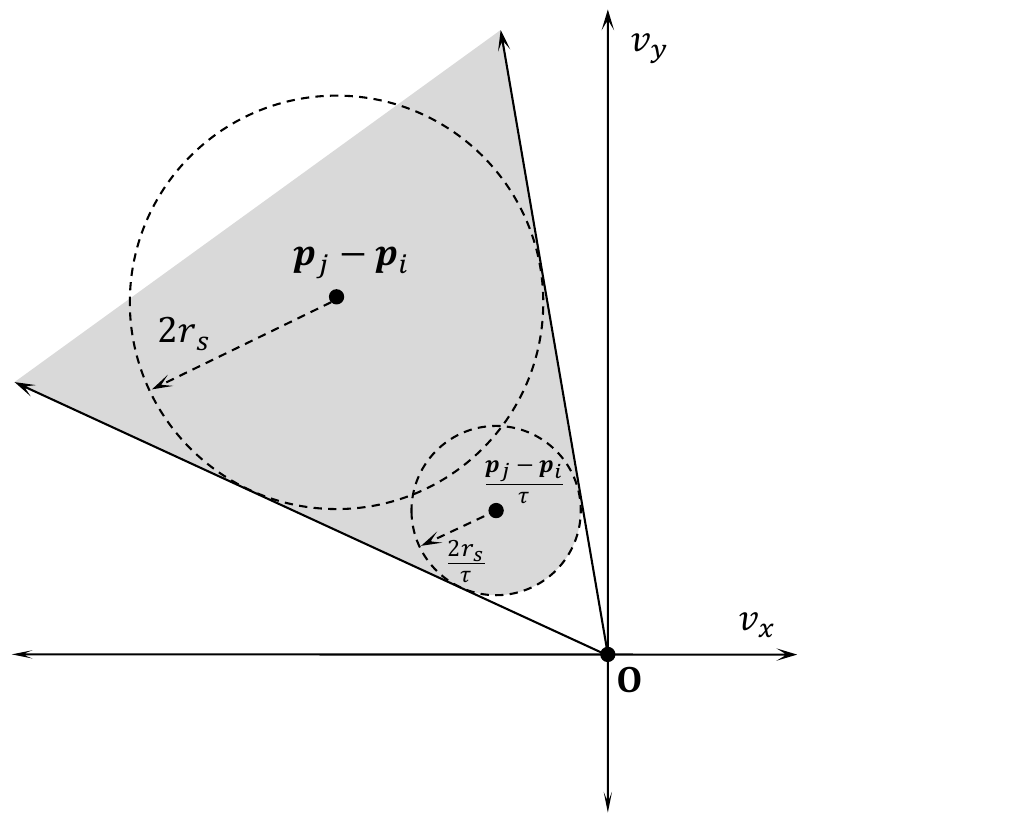}
    \captionsetup{font=footnotesize}
    \subcaption{}
    \label{fig:orca-b}
\end{subfigure}
\\
\begin{subfigure}[t]{0.49\linewidth}
    \centering
    \includegraphics[height=6.5cm]{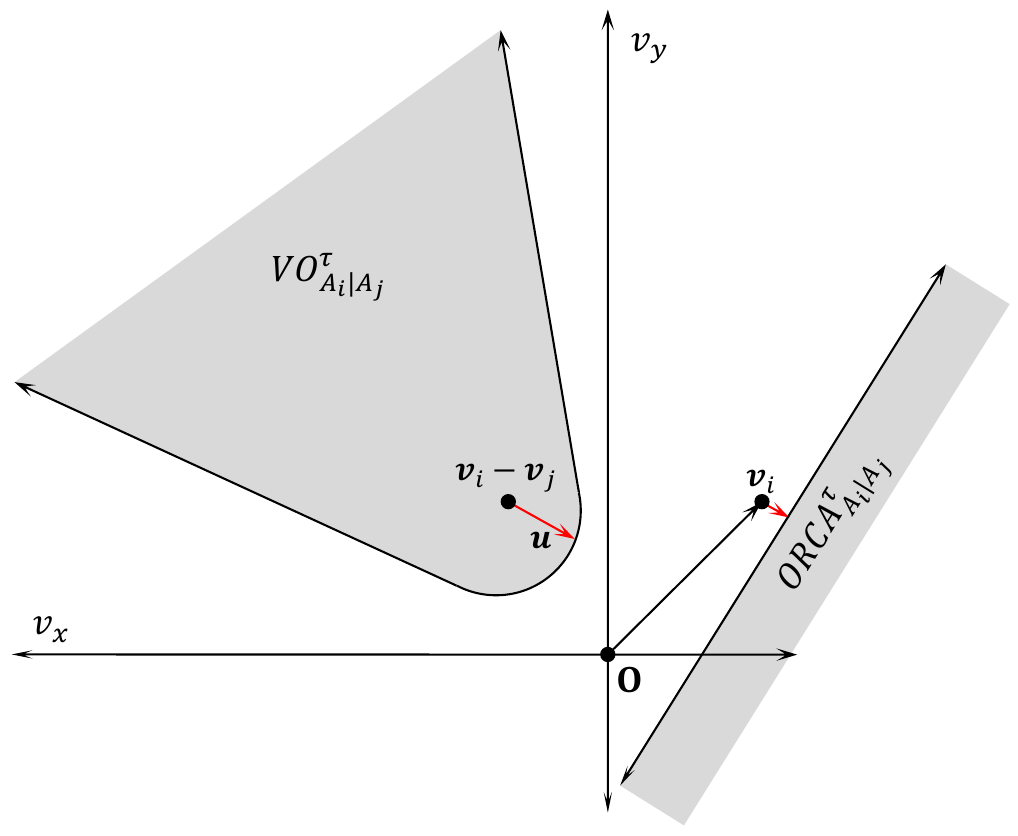}
    \captionsetup{font=footnotesize}
    \subcaption{}
    \label{fig:orca-c}
\end{subfigure}
\captionsetup{font={footnotesize,stretch=1.25}}
\caption{
(a) The collision-free configuration of aircraft $A_i$ and $A_j$ with non-overlapping safety spaces.
(b) The velocity obstacle $VO^\tau_{A_i|A_j}$ is a truncated cone (in velocity space) with its apex at the origin and its legs tangent to the disc of radius $2r_s$ centered at $\boldsymbol{p}_j-\boldsymbol{p}_i$.
(c) The set $ORCA^\tau_{A_i|A_j}$ of safe velocities for aircraft $A_i$ induced by aircraft $A_j$ is a half-plane pointing in the direction of $\boldsymbol{u}$ starting at the point $\boldsymbol{v}_i + \frac{1}{2}\boldsymbol{u}$, where $\boldsymbol{u}$ is the vector from $\boldsymbol{v}_i - \boldsymbol{v}_j$ to the closest point on the boundary of  $VO^\tau_{A_i|A_j}$.}
\end{figure}

We start by deriving the collision-free conditions for two aircraft, $A_i$ and $A_j$, each equipped with an identical safety radius $r_s$.
Let $D(\boldsymbol{p}, r)$ denote an open disc of radius $r$ centered at position $\boldsymbol{p}$, defined as
\begin{align}
    D(\boldsymbol{p}, r) = \Big\{ \boldsymbol{q} \, \big\vert\  \Vert\, \boldsymbol{q} - \boldsymbol{p} \,\Vert < r \Big\}
\end{align}
To ensure a collision-free configuration, the safety spaces of aircraft $A_i$ and $A_j$ must not overlap, i.e., $D(\boldsymbol{p}_i, r_s) \cap D(\boldsymbol{p}_j, r_s) = \varnothing$, as illustrated in \autoref{fig:orca-a}.
Next, we introduce the concept of the velocity obstacle, which defines a set of relative velocities that would lead to a collision between the two aircraft within a given time horizon.  
In particular, if the relative velocity of aircraft $A_i$ with respect to $A_j$ lies inside the velocity obstacle $VO^\tau_{A_i|A_j}$, then the two aircraft will collide at some moment before time $\tau$.
Formally, the velocity obstacle is defined as
\begin{align}
    VO^\tau_{A_i|A_j} = \Big\{ \boldsymbol{v} \, \big\vert\  \exists\, t \in [0, \tau] : \boldsymbol{v}t \in D(\boldsymbol{p}_j-\boldsymbol{p}_i, 2r_s)\Big\}
\end{align}
Geometrically, $VO^\tau_{A_i|A_j}$ corresponds to a truncated cone in the velocity space, as shown in \autoref{fig:orca-b}.
Given aircraft velocities $\boldsymbol{v}_i$ and $\boldsymbol{v}_j$, the relative velocity of aircraft $A_i$ with respect to $A_j$ is $\boldsymbol{v}_i - \boldsymbol{v}_j$.
By definition, if $\boldsymbol{v}_i - \boldsymbol{v}_j \in VO^\tau_{A_i|A_j}$, a collision will occur at some moment before time $\tau$, assuming aircraft velocities remain unchanged.
Conversely, if $\boldsymbol{v}_i - \boldsymbol{v}_j \notin VO^\tau_{A_i|A_j}$, the two aircraft are guaranteed to remain collision-free for at least time horizon $\tau$.
When a collision is imminent, we compute the minimum change in relative velocity required to avoid collision.
Let $\boldsymbol{u}$ be the vector from $\boldsymbol{v}_i - \boldsymbol{v}_j$ to the closest point on the boundary of the velocity obstacle $VO^\tau_{A_i|A_j}$, defined as
\begin{align}
    \boldsymbol{u} = \big(\mathop{\arg\min}\limits_{\boldsymbol{v}\in\partial VO^\tau_{A_i|A_j}} \Vert \boldsymbol{v} - (\boldsymbol{v}_i - \boldsymbol{v}_j) \Vert \,\big) - (\boldsymbol{v}_i - \boldsymbol{v}_j)
\end{align}
where $\partial VO^\tau_{A_i|A_j}$ denotes the boundary of the (closed) set $VO^\tau_{A_i|A_j}$.
The geometric interpretation of vector $\boldsymbol{u}$ is the minimum variation in relative velocity $\boldsymbol{v}_i - \boldsymbol{v}_j$ to exit the velocity obstacle, as illustrated in \autoref{fig:orca-c}.
To fairly share the responsibility for collision avoidance, we assume that aircraft $A_i$ and $A_j$ each take half of the responsibility for the velocity variation $\boldsymbol{u}$.
Thus, aircraft $A_i$ should adapt its velocity by at least $\frac{1}{2}\boldsymbol{u}$, 
As a result, the safe velocities for aircraft $A_i$ should lie in the half-plane pointing in the direction of $\boldsymbol{u}$ starting at the point $\boldsymbol{v}_i + \frac{1}{2}\boldsymbol{u}$. 
More formally, the set of safe velocities for aircraft $A_i$ induced by $A_j$ is defined as
\begin{align}
    ORCA^\tau_{A_i|A_j} = \Big\{ \boldsymbol{v} \, \big\vert\  (\boldsymbol{v}- (\boldsymbol{v}_i + \frac{1}{2}\boldsymbol{u})) \cdot \boldsymbol{u} \geq 0 \Big\}
\end{align}
Geometrically, $ORCA^\tau_{A_i|A_j}$ corresponds to a half-plane in the velocity space, as illustrated in \autoref{fig:orca-c}.
To handle multiple aircraft, we compute the safe velocity set with respect to each neighboring aircraft and take the intersection of all such half-planes, i.e.
\begin{align}
    \mathit{ORCA}^\tau_{A_i} = \bigcap\limits_{A_j \in \mathbb{N}_i} \mathit{ORCA}^\tau_{A_i|A_j}
\end{align}
where $\mathbb{N}_i$ denotes the set of aircraft within the detection range of aircraft $A_i$.
Typically, the aircraft's maximum speed constraint $\boldsymbol{v} \in D(\boldsymbol{0}, v^{max}_i)$ is also required.
In summary, the (complete) set of safe velocities for aircraft $A_i$ is formally defined as
\begin{align}
    \Psi_i = D(\boldsymbol{0}, v^{max}_i) \cap \mathit{ORCA}^\tau_{A_i}
\end{align}
As mentioned before, the optimization problem \eqref{eq:orca_objective} is efficiently solvable, since the constraint set $\boldsymbol{v} \in \Psi_i$ involves only a norm bound and linear inequalities.

\section{Simulation of large-scale air traffic flow}\label{sec:result}
In this section, we present a series of case studies to evaluate the performance of the proposed framework.  
First, under a single-layer airspace network configuration, we compare the proposed framework with a collision-free baseline strategy commonly used in free-flight systems.  
The compared performance indices include the average minimum separation, average travel speed, trip completion rate, energy consumption, and computational efficiency.  
Then, we further evaluate the effectiveness of the proposed framework within a two-layer airspace network.
Additionally, we present examples of dynamic no-fly zones to illustrate the proposed framework's flexibility in UAM operations.  
Lastly, we evaluate the performance of the fast approximation methods proposed in \autoref{sec:FAM}, in terms of computational efficiency and accuracy.

\subsection{Simulation settings}
To initialize the UAM system, the configuration of airspace settings, aircraft parameters, and demand patterns is required.
We consider both single-layer and two-layer airspace network designs.
For the single-layer network, examples with different altitude ranges are provided, including $z = 500\,\mathrm{m}$, $z \in [450, 550]\,\mathrm{m}$, and $z \in [400, 600]\,\mathrm{m}$.
For the two-layer network, the altitudes of the layers are set to $z = 400\,\mathrm{m}$ and $z = 600\,\mathrm{m}$.
Both network designs are configured with a region radius of $250\,\mathrm{m}$.
In addition, all aircraft are configured with the following parameters: detection radius $r_d = 200\,\mathrm{m}$, safety radius $r_s = 50\,\mathrm{m}$, maximum speed $v^{\mathrm{max}} = 20\,\mathrm{m/s}$ and control gain $l = 5\,\mathrm{[-]}$.
The simulation time step is set to $\Delta t = 1\,\mathrm{s}$.
The proposed route guidance and collision avoidance algorithms are implemented in Python 3.7 and executed on a 64-bit Windows PC with a 2.9-GHz Intel Core i7 processor and 36 GB of RAM.

\subsection{Simulation results}
\subsubsection{Results of the single-layer airspace network} \label{sec:single-layer}

\begin{figure}[!h]
\centering
\begin{subfigure}[t]{0.329\linewidth}
    \centering
    \includegraphics[width=\linewidth]{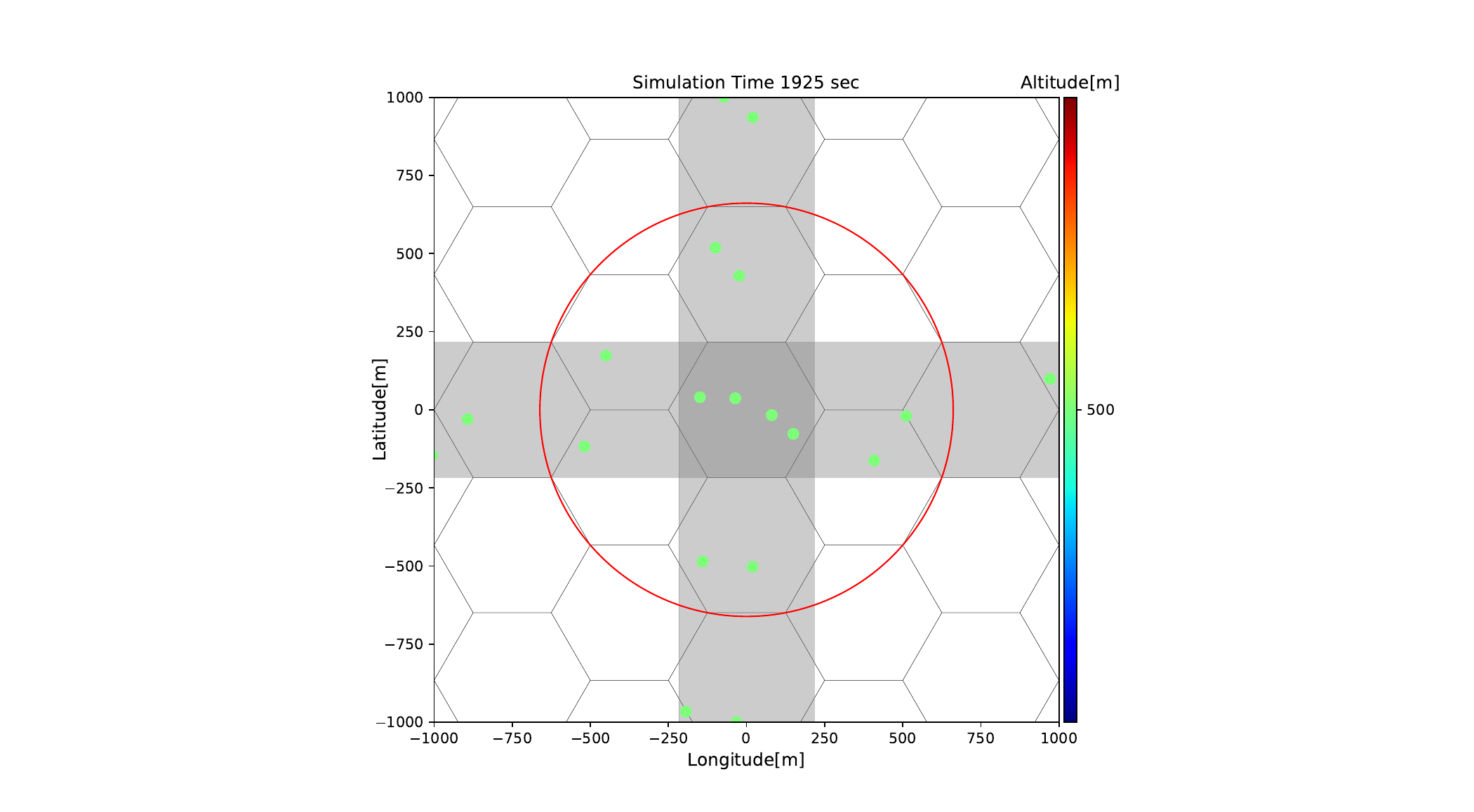}
    \captionsetup{font=footnotesize}
    \subcaption{The ``$+$'' type scenario ($z = 500\,\mathrm{m}$)}
    \label{}
\end{subfigure}
\begin{subfigure}[t]{0.329\linewidth}
    \centering
    \includegraphics[width=\linewidth]{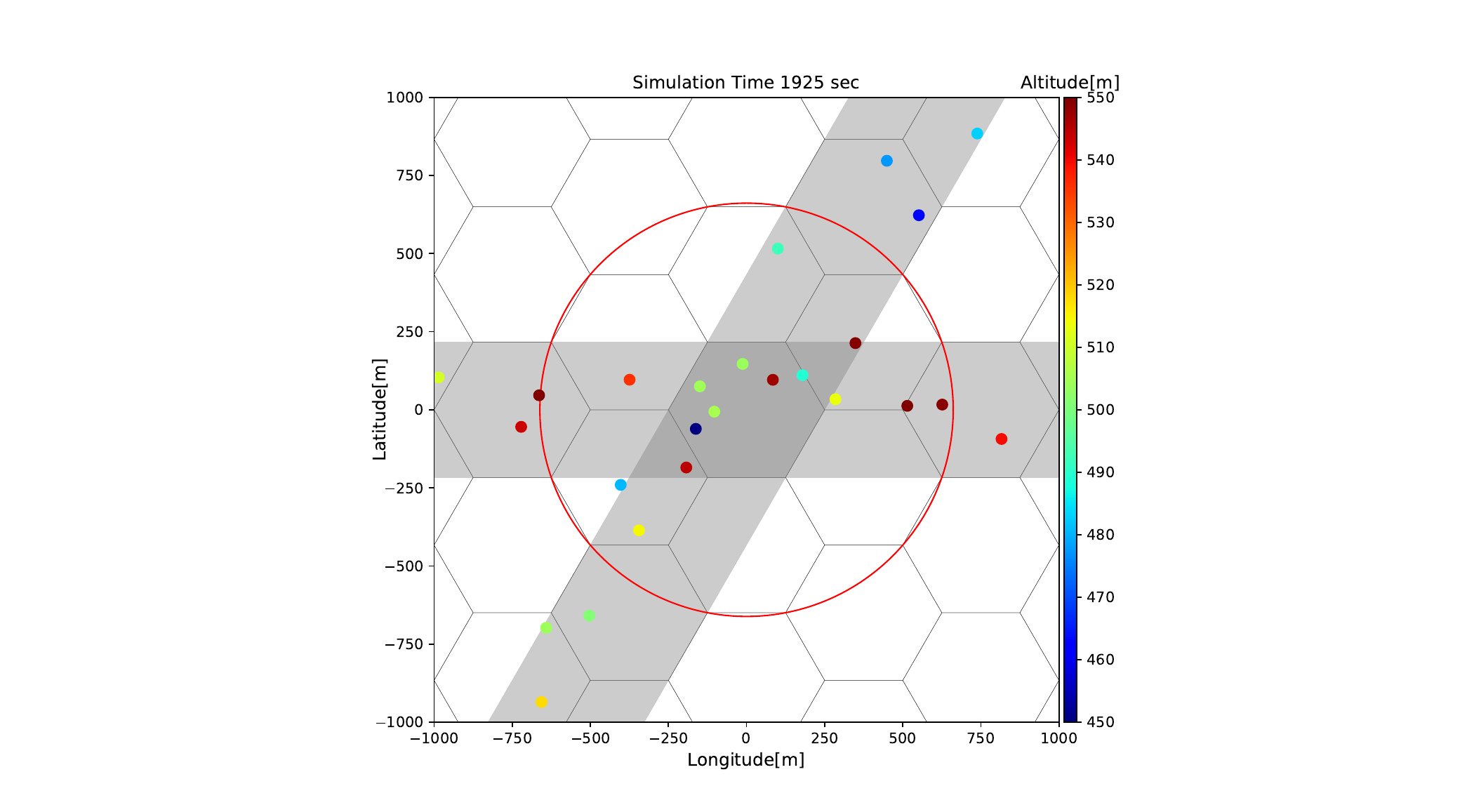}
    \captionsetup{font=footnotesize}
    \subcaption{The ``$\#$'' type scenario ($z \in [450, 550]\,\mathrm{m}$)}
    \label{}
\end{subfigure}
\begin{subfigure}[t]{0.329\linewidth}
    \centering
    \includegraphics[width=\linewidth]{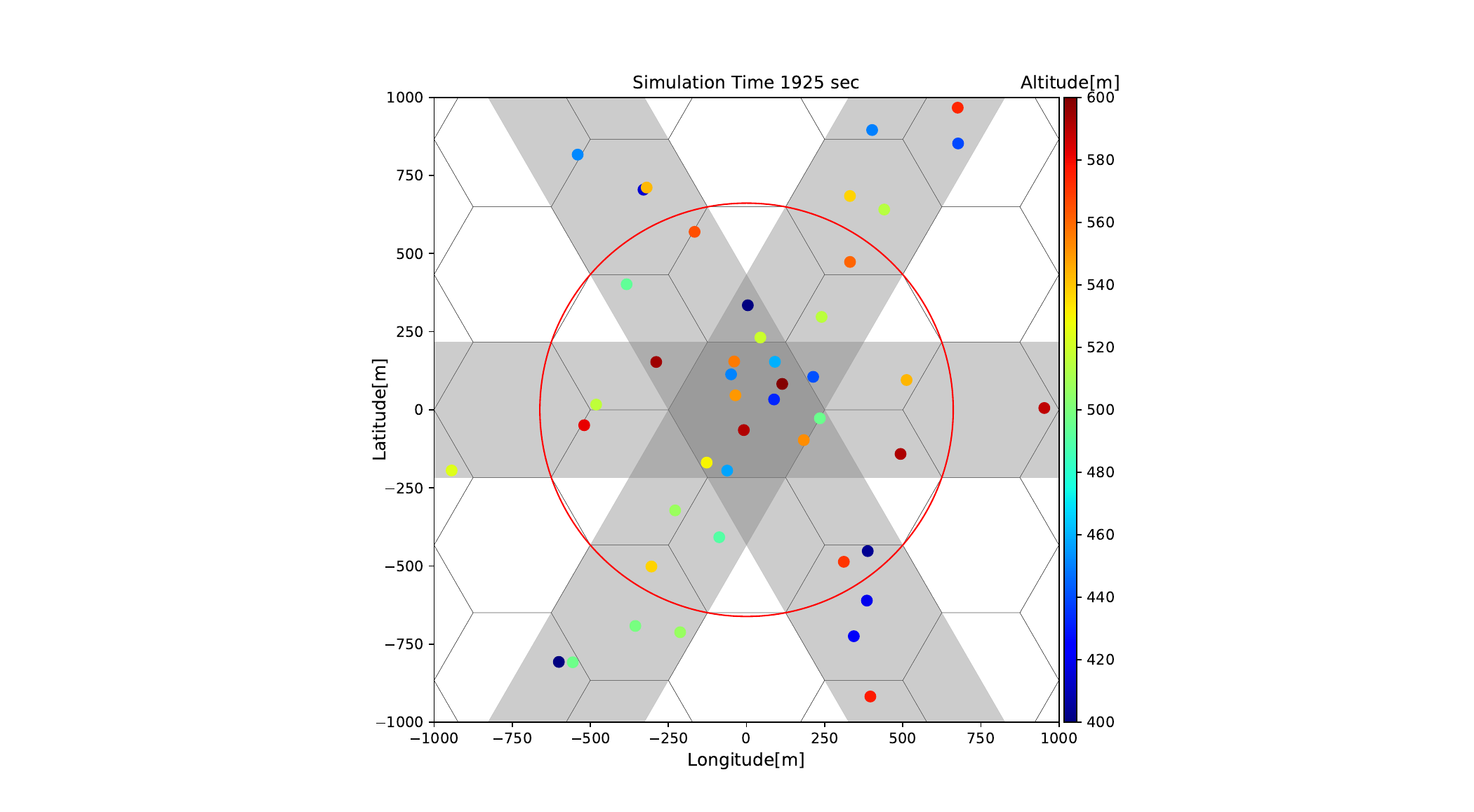}
    \captionsetup{font=footnotesize}
    \subcaption{``$*$'' type scenario ($z \in [400, 600]\,\mathrm{m}$)}
    \label{}
\end{subfigure}
\captionsetup{font=footnotesize}
\caption{
Testing scenarios: 
(a) intersection of two orthogonal corridors; 
(b) intersection of two oblique corridors; 
(c) intersection of three corridors. 
Virtual corridors are indicated by gray shaded areas, and aircraft are represented by scatter points with colors indicating altitude.
}
\label{fig:Test_scenarios}
\end{figure}

\begin{figure}[!h]
\centering
\begin{subfigure}[t]{0.329\linewidth}
    \centering
    \includegraphics[width=\linewidth]{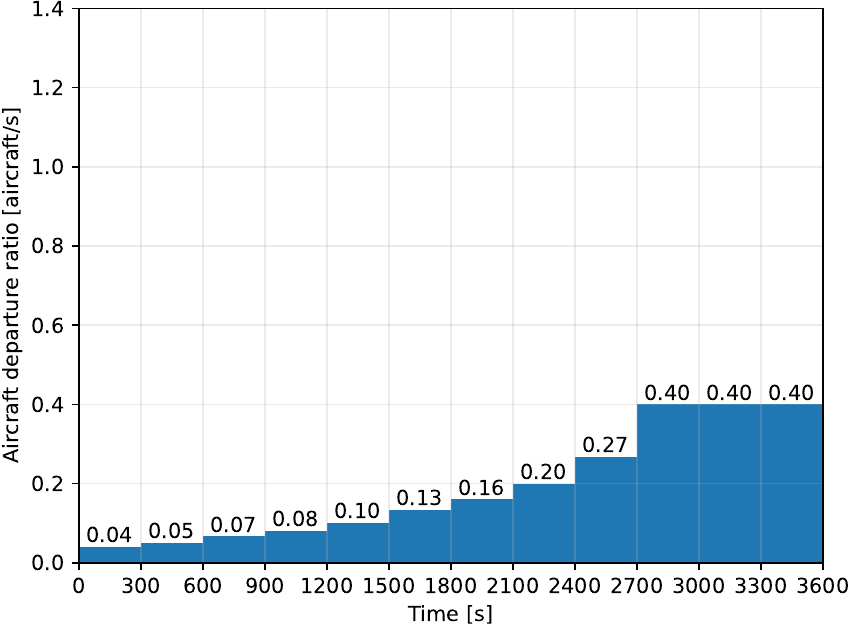}
    \captionsetup{font=footnotesize}
    \subcaption{}
    \label{}
\end{subfigure}
\begin{subfigure}[t]{0.329\linewidth}
    \centering
    \includegraphics[width=\linewidth]{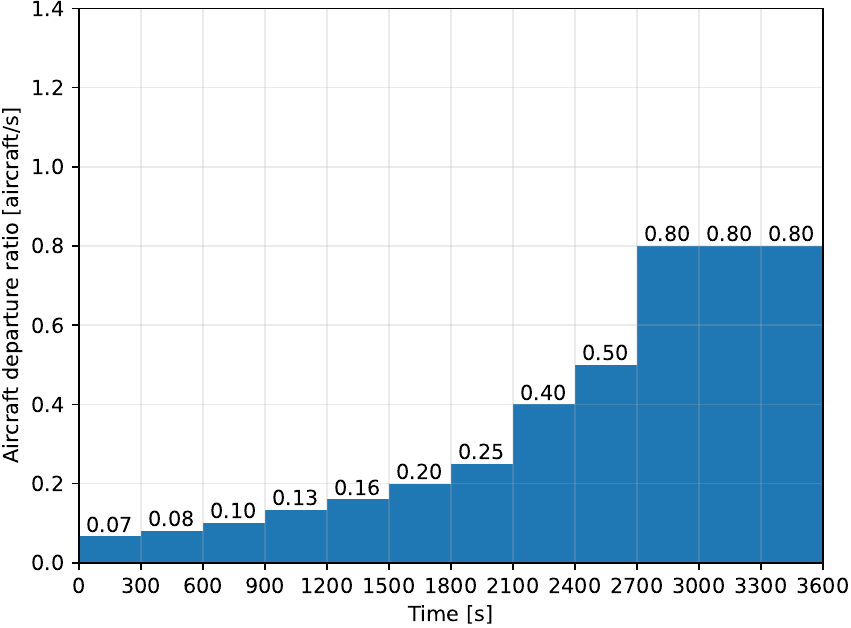}
    \captionsetup{font=footnotesize}
    \subcaption{}
    \label{}
\end{subfigure}
\begin{subfigure}[t]{0.329\linewidth}
    \centering
    \includegraphics[width=\linewidth]{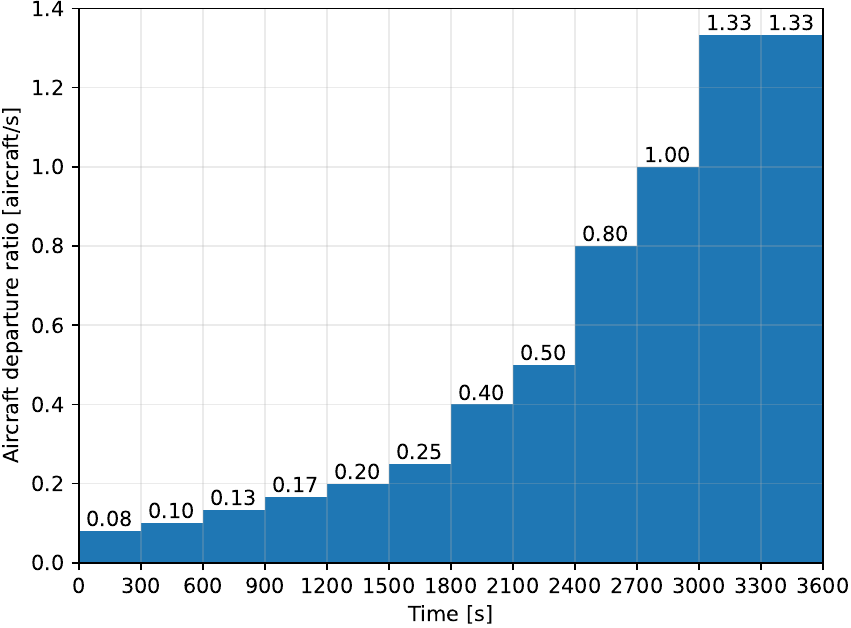}
    \captionsetup{font=footnotesize}
    \subcaption{}
    \label{}
\end{subfigure}
\caption{
Aircraft inflow patterns for: 
(a) 2D cases with $z = 500\,\mathrm{m}$; 
(b) 3D cases with $z \in [450, 550]\,\mathrm{m}$; 
(c) 3D cases with $z \in [400, 600]\,\mathrm{m}$.
}
\label{fig:Aircraft_inflow}
\end{figure}

We first evaluate the performance of the proposed framework within a single-layer airspace network.
As illustrated in \autoref{fig:Test_scenarios}, three types of scenarios, corresponding to different OD patterns, are considered.
For comparison, a collision-free baseline scheme is introduced.
The baseline scheme simulates point-to-point operations, wherein aircraft operate along virtual corridors, flying from one end to the other while relying on their onboard autonomy to avoid collisions.
To accommodate heavy demand conditions, the physical constraints of virtual corridors are relaxed by allowing aircraft to fly outside the corridors (in terms of longitude and latitude) when necessary.
For each scenario, both the 2D cases ($z = 500\,\mathrm{m}$) and the 3D cases ($z \in [450, 550]\,\mathrm{m}$ and $z \in [400, 600]\,\mathrm{m}$) are evaluated.
Since the 3D configurations are expected to support higher UAM demand than their 2D counterparts, three distinct aircraft inflow patterns are accordingly designed for both the 2D and 3D cases, as illustrated in \autoref{fig:Aircraft_inflow}.

We conduct simulations and compute the average aircraft accumulation within each airspace region throughout $300\,\mathrm{s}$, with heat maps illustrating the evolution of air traffic.
The results for the baseline scheme in 2D cases are presented in \autoref{fig:comparison_a}, \autoref{fig:comparison_c}, and \autoref{fig:comparison_e}.
Across different scenarios, the evolution of air traffic exhibits consistent patterns.
As demand increases, more aircraft enter the intersections of air corridors, thereby elevating the risk of traffic congestion.
Once a small number of aircraft become near-hovering in local deadlocks, congestion tends to propagate and intensify.
Regardless of the corridor intersection geometry, traffic congestion typically originates at the intersection and subsequently spreads throughout the entire network.
Similar traffic evolution patterns have also been documented in roadway traffic systems \citep{geroliminis2011properties}.

The results for the proposed framework in 2D cases are presented in \autoref{fig:comparison_b}, \autoref{fig:comparison_d}, and \autoref{fig:comparison_f}.
By design, the route guidance mechanism enables aircraft to choose diverse paths rather than strictly adhering to direct routes from their origins to destinations.
Moreover, the objective of minimizing total travel time promotes a better homogeneous distribution of aircraft across both spatial and temporal dimensions. The simulation results align well with these expectations.
Compared with the baseline, aircraft are distributed across a wider range of airspace regions, with accumulation levels in these regions remaining approximately uniform.
More importantly, aircraft accumulation at the intersections is significantly reduced, effectively mitigating the risk of air traffic congestion.
Across different scenarios, regardless of the spatial distribution of demand, the proposed framework consistently ensures air traffic homogeneity.

The qualitative evolution of air traffic in the 3D cases exhibits patterns consistent with the 2D results presented above and is omitted here for brevity.
Instead, quantitative results based on the concept of airspace MFD are presented.
Traffic outflow and accumulation at the intersections are measured, with the measurement area indicated by a red circle in \autoref{fig:Test_scenarios}.
The resulting outflow-accumulation relationships are illustrated in \autoref{fig:mfd+}-\autoref{fig:mfd*}, covering nine examples in both 2D and 3D cases across different scenarios.
In the baseline cases, the traffic outflow is observed to initially increase and subsequently decrease once aircraft accumulation exceeds a critical value.
By contrast, the outflow–accumulation curves for the proposed framework exhibit only the ascending phase.
Research on MFD indicates that the ascending phase corresponds to the free-flow state, while the descending phase corresponds to traffic congestion \citep{safadi2023macroscopic, cummings2024airspace}.
Thus, it can be concluded that the proposed framework consistently maintains air traffic in the free-flow state across all examples, whereas the baseline fails to do so.

It is also observed that the critical points of the MFDs (marked by ``$\blacktriangle$'') exhibit significant disparities between the 2D and 3D cases.
Following \cite{safadi2023macroscopic}, we estimate the MFDs using the following function
\begin{align*}
   \widetilde{G}(N)=\alpha N \exp\big( -\frac{1}{\beta} (\frac{N}{N^{cr}})^{\beta} \big)
\end{align*}
where $\alpha$, $\beta$, and $N^{cr}$ are parameters to be calibrated. 
The critical point of the fitted curve is given by $(N^{cr}, \alpha N^{cr} e^{-\frac{1}{\beta}})$.  
For each case with the same altitude range, the critical points across different scenarios exhibit no significant differences. 
This suggests that the critical point primarily depends on the altitude range of the airspace. More specifically, the critical point mainly depends on the traffic volume in the airspace.   
Therefore, we aggregate results from different scenarios and represent them as flow-density relationships to eliminate the impact of the measurement range.  
The aggregated results are illustrated in \autoref{fig:airsapce_capacity}.  
Owing to the increased sample size, the aggregated results exhibit reduced variance and are thus more suitable for parameter estimation.  
\autoref{tab:airspace_capacity} summarizes the estimated values of the jam density $\widetilde{K}_{jam}$, critical density $\widetilde{K}_{cr}$, and critical flow $\widetilde{Q}(\widetilde{K}_{cr})$.
In the literature, these metrics serve as benchmarks for evaluating airspace capacity \citep{cummings2024comparing}.   
The capacity metrics are observed to increase with the expansion of the altitude range. %, in an approximately linear manner.  
An interpretation is that the airspace corresponding to $z \in [400, 600]\,\mathrm{m}$ can be regarded as a stack of two airspace layers corresponding to $z \in [450, 550]\,\mathrm{m}$.

\begin{table}[!h]
    \centering
    \captionsetup{font=footnotesize}
    \caption{Airspace capacity estimation.}
    \label{tab:airspace_capacity}
    \renewcommand\arraystretch{1.25}
    \footnotesize
    \begin{tabular}{llll}
    % \hline
    \toprule
    Layer's altitude range & $\widetilde{K}_{jam}\,[\mathrm{aircraft/km}^2]$ & $\widetilde{K}_{cr}\,[\mathrm{aircraft/km}^2]$ & $\widetilde{Q}(\widetilde{K}_{cr})\,[\mathrm{aircraft/s/km}^2]$\\
    % \hline
    \midrule
    $z = 500\,\mathrm{m}$ & 75 & 13.67 & 0.10 \\
    % \hline
    $z \in [450, 550]\,\mathrm{m}$ & 200 & 27.66 & 0.18 \\
    % \hline
    $z \in [400, 600]\,\mathrm{m}$ & 300 & 40.75 & 0.28 \\
    % \hline
    \bottomrule
    \end{tabular}
    \vspace{4mm}
\end{table}

{
\newpage
\begin{figure}[!h]
\centering
\begin{subfigure}[t]{\linewidth}
    \centering
    \includegraphics[height=3.05cm]{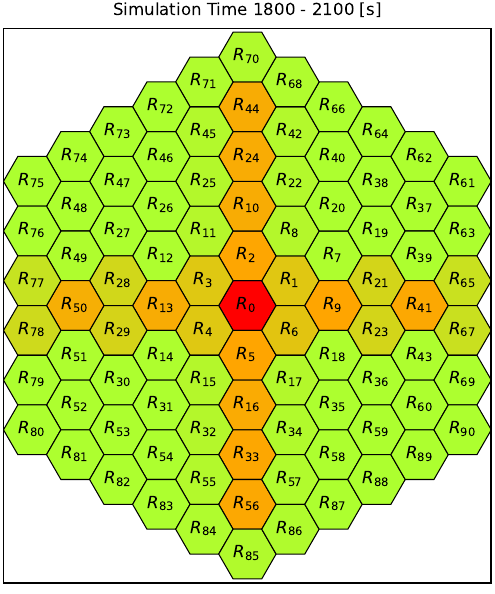}
    \hspace{-1.8mm}
    \includegraphics[height=3.05cm]{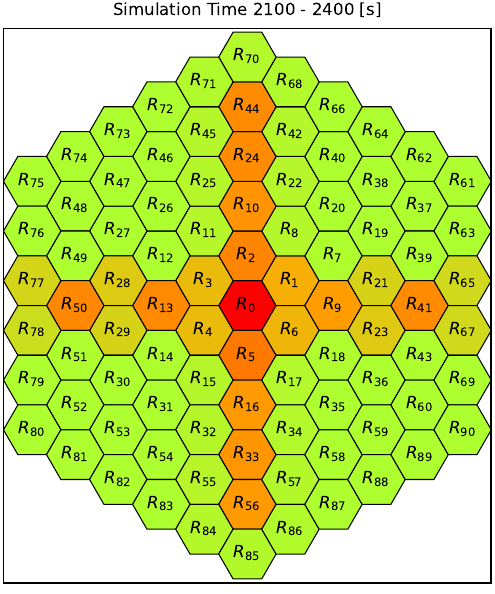}
    \hspace{-1.8mm}
    \includegraphics[height=3.05cm]{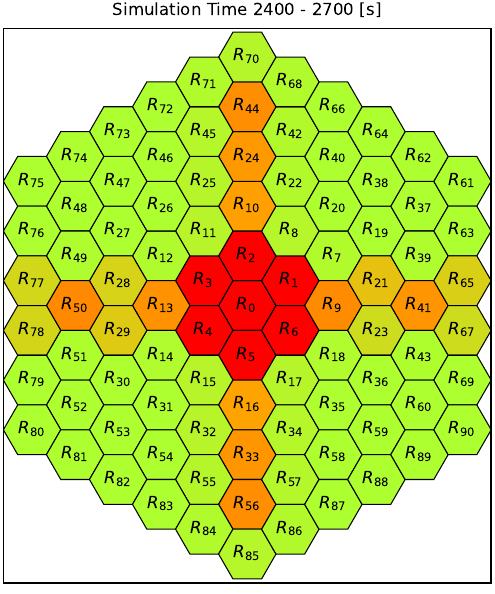}
    \hspace{-1.8mm}
    \includegraphics[height=3.05cm]{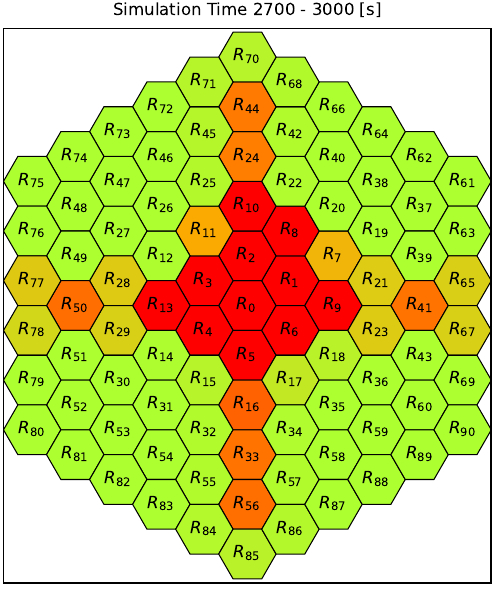}
    \hspace{-1.8mm}
    \includegraphics[height=3.05cm]{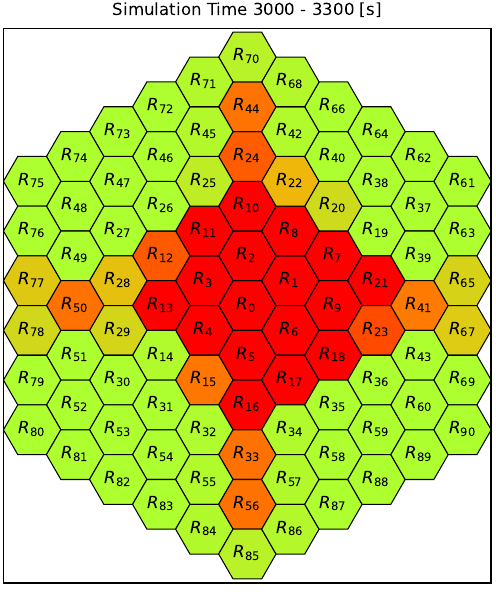}
    \hspace{-1.8mm}
    \includegraphics[height=3.05cm]{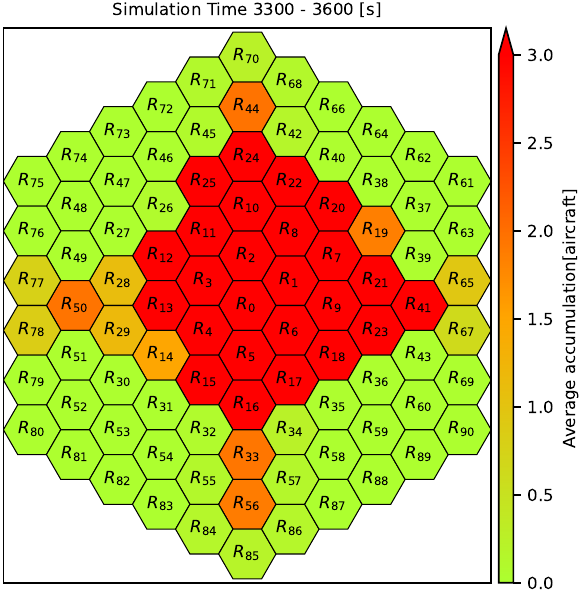}
    \captionsetup{font=scriptsize}
    \vspace{-1.8mm}
    \subcaption{Traffic evolution of the baseline method in the ``$+$'' type scenario}
    \vspace{0.8mm}
    \label{fig:comparison_a}
\end{subfigure}
\begin{subfigure}[t]{\linewidth}
    \centering
    \includegraphics[height=3.05cm]{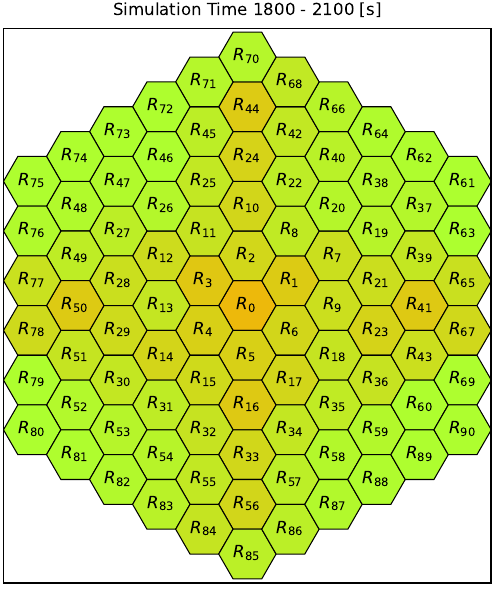}
    \hspace{-1.8mm}
    \includegraphics[height=3.05cm]{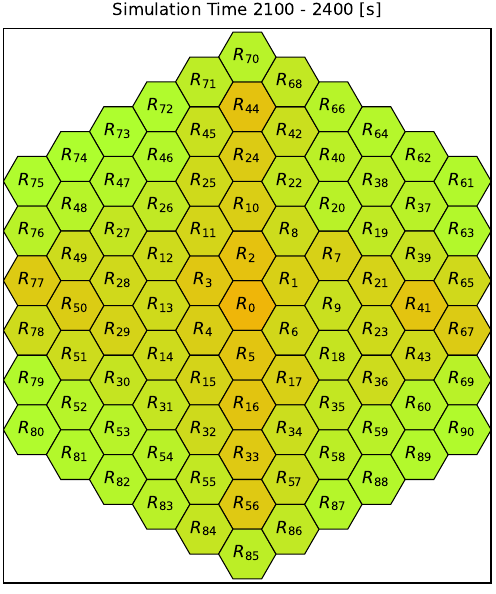}
    \hspace{-1.8mm}
    \includegraphics[height=3.05cm]{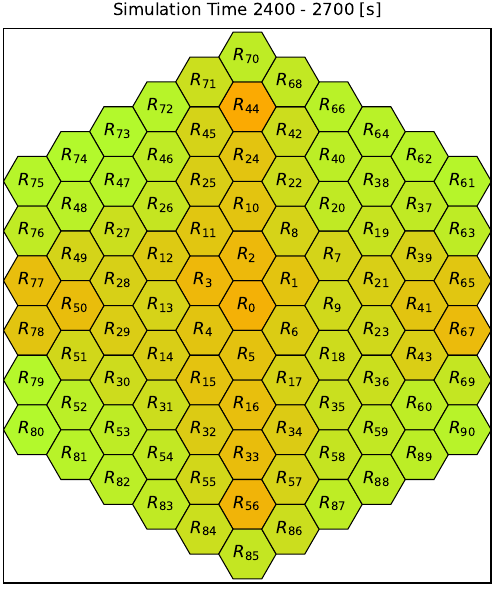}
    \hspace{-1.8mm}
    \includegraphics[height=3.05cm]{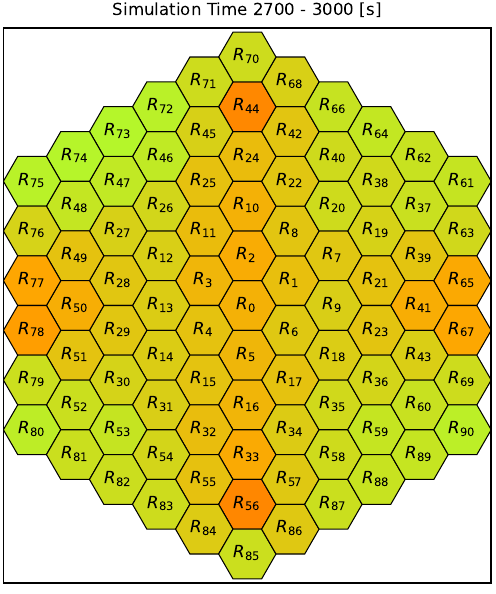}
    \hspace{-1.8mm}
    \includegraphics[height=3.05cm]{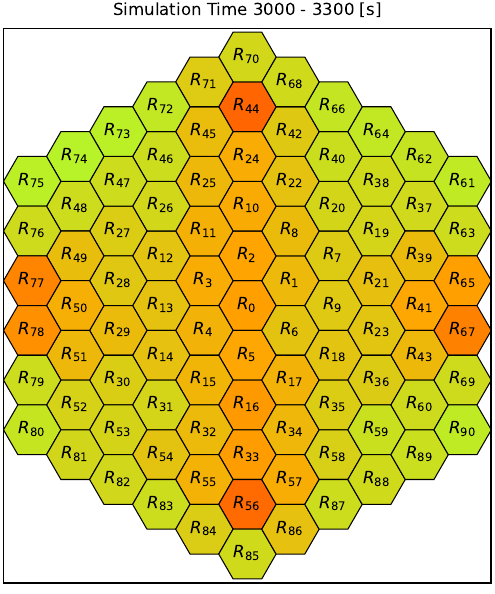}
    \hspace{-1.8mm}
    \includegraphics[height=3.05cm]{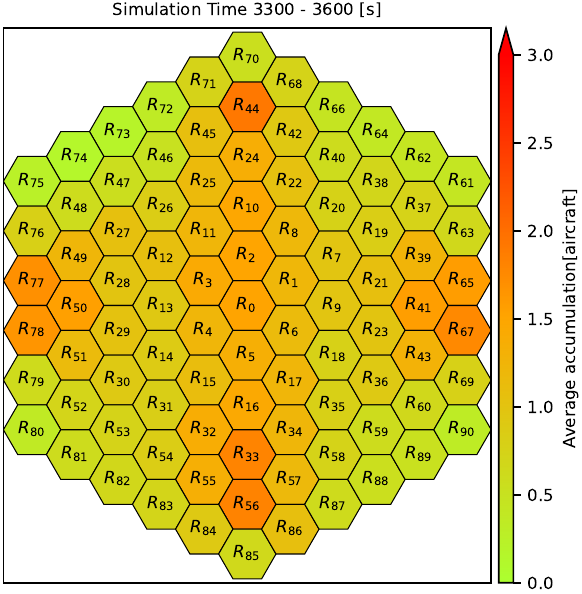}
    \captionsetup{font=scriptsize}
    \vspace{-1.8mm}
    \subcaption{Traffic evolution of the proposed method in the ``$+$'' type scenario}
    \vspace{0.8mm}
    \label{fig:comparison_b}
\end{subfigure}
\begin{subfigure}[t]{\linewidth}
    \centering
    \includegraphics[height=3.05cm]{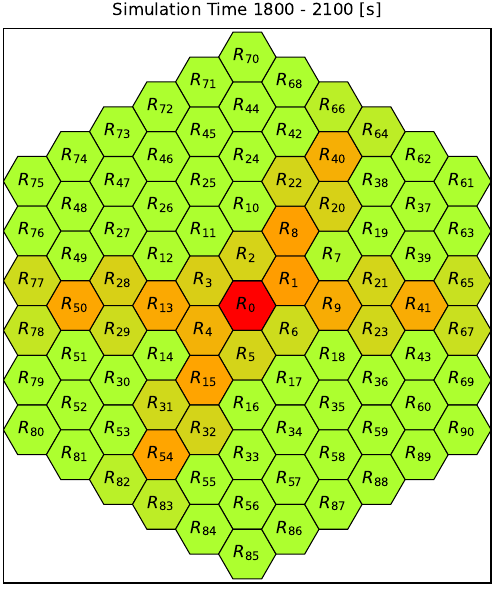}
    \hspace{-1.8mm}
    \includegraphics[height=3.05cm]{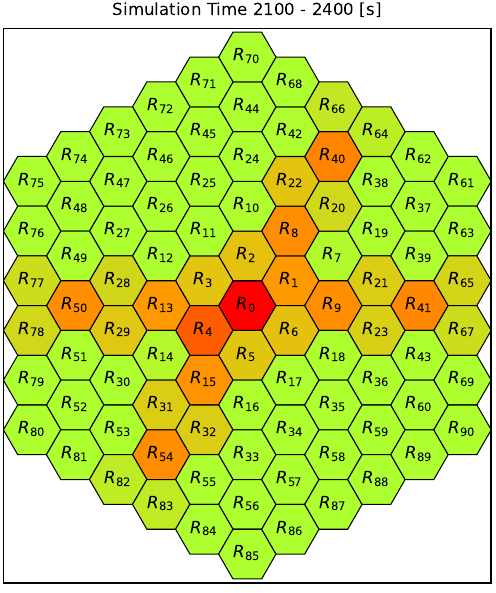}
    \hspace{-1.8mm}
    \includegraphics[height=3.05cm]{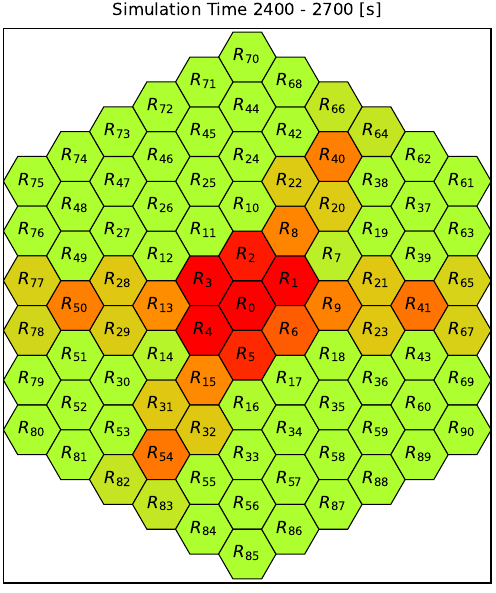}
    \hspace{-1.8mm}
    \includegraphics[height=3.05cm]{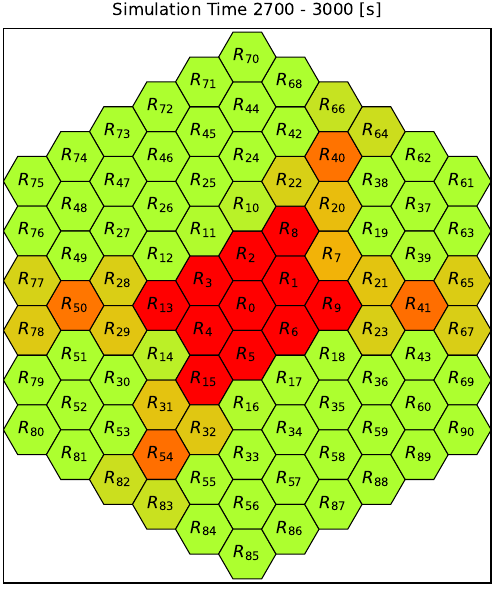}
    \hspace{-1.8mm}
    \includegraphics[height=3.05cm]{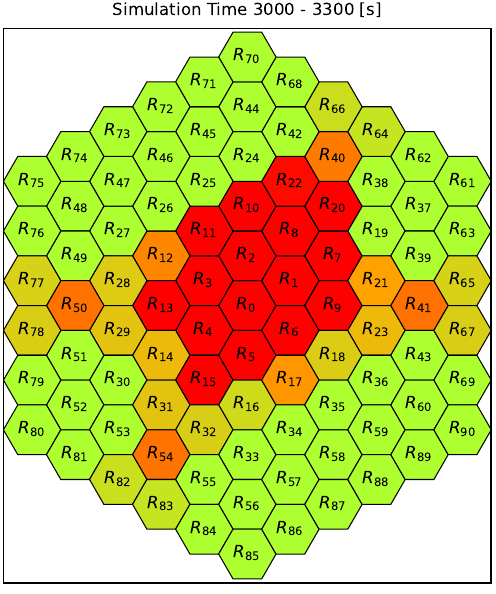}
    \hspace{-1.8mm}
    \includegraphics[height=3.05cm]{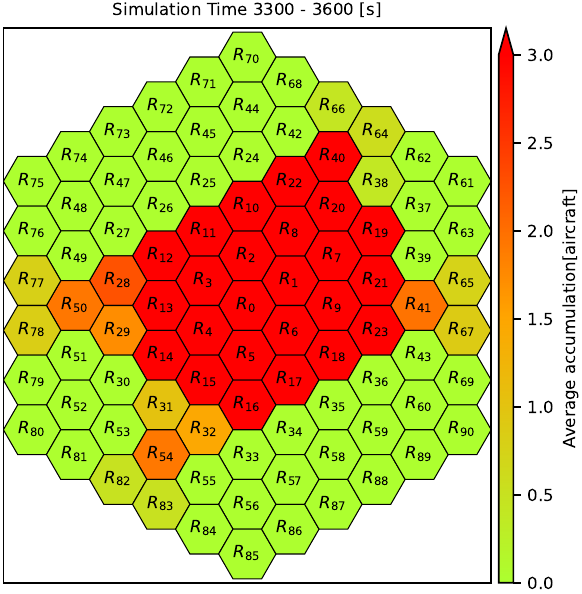}
    \captionsetup{font=scriptsize}
    \vspace{-1.8mm}
    \subcaption{Traffic evolution of the baseline method in the ``$\#$'' type scenario}
    \vspace{0.8mm}
    \label{fig:comparison_c}
\end{subfigure}
\begin{subfigure}[t]{\linewidth}
    \centering
    \includegraphics[height=3.05cm]{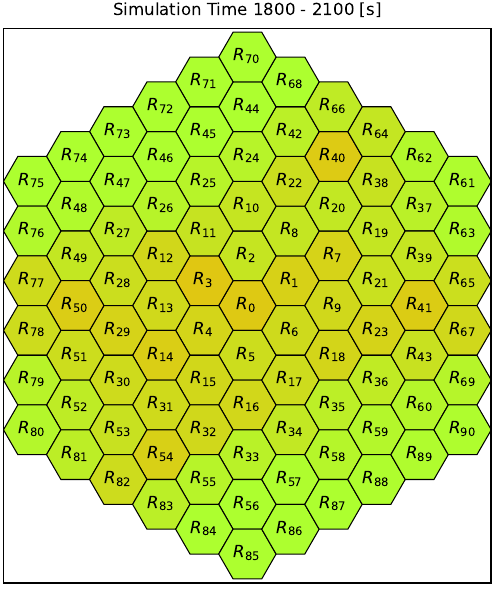}
    \hspace{-1.8mm}
    \includegraphics[height=3.05cm]{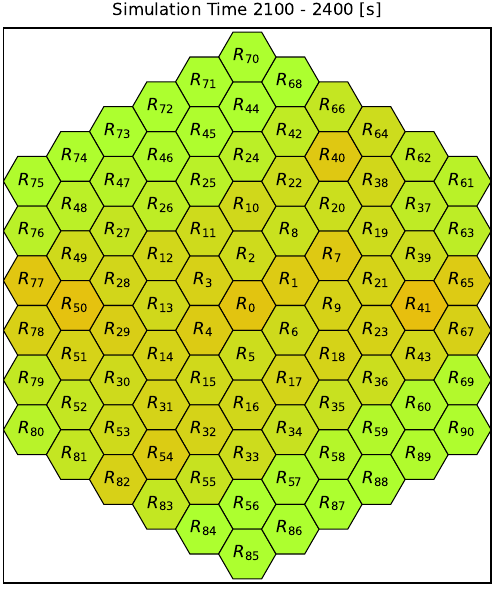}
    \hspace{-1.8mm}
    \includegraphics[height=3.05cm]{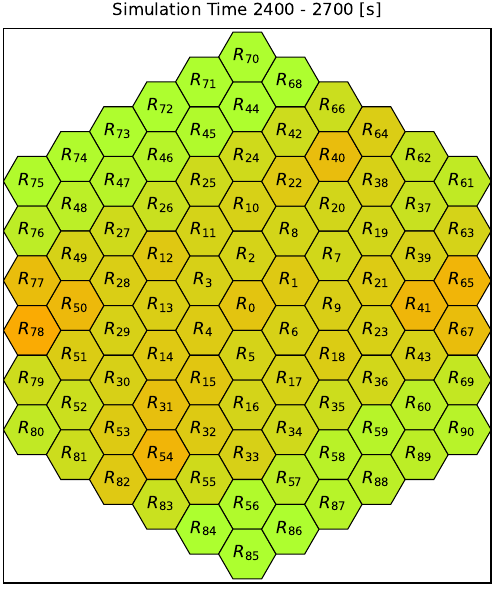}
    \hspace{-1.8mm}
    \includegraphics[height=3.05cm]{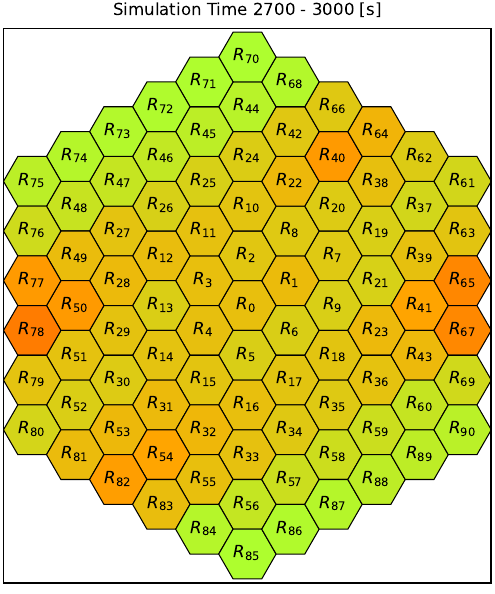}
    \hspace{-1.8mm}
    \includegraphics[height=3.05cm]{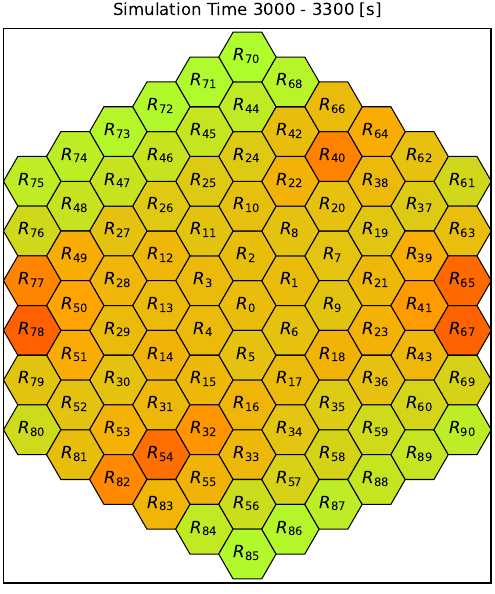}
    \hspace{-1.8mm}
    \includegraphics[height=3.05cm]{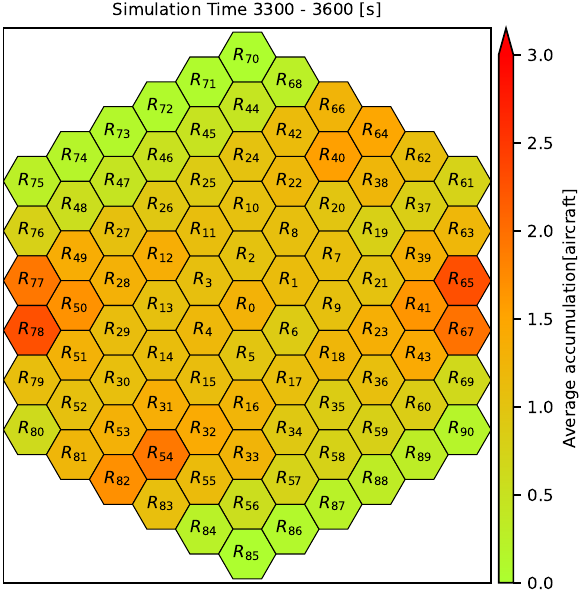}
    \captionsetup{font=scriptsize}
    \vspace{-1.8mm}
    \subcaption{Traffic evolution of the proposed method in the ``$\#$'' type scenario}
    \vspace{0.8mm}
    \label{fig:comparison_d}
\end{subfigure}
\begin{subfigure}[t]{\linewidth}
    \centering
    \includegraphics[height=3.05cm]{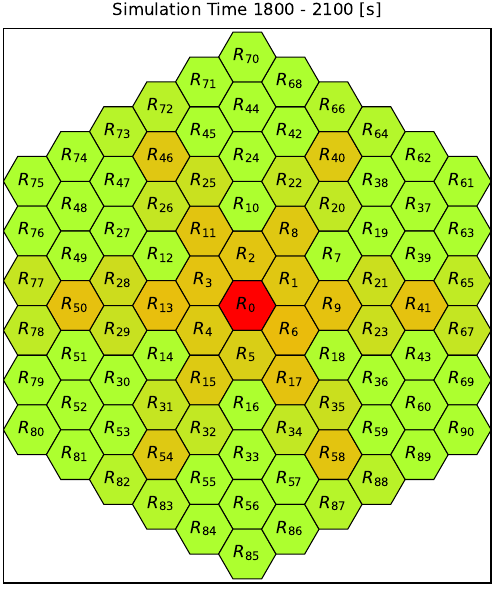}
    \hspace{-1.8mm}
    \includegraphics[height=3.05cm]{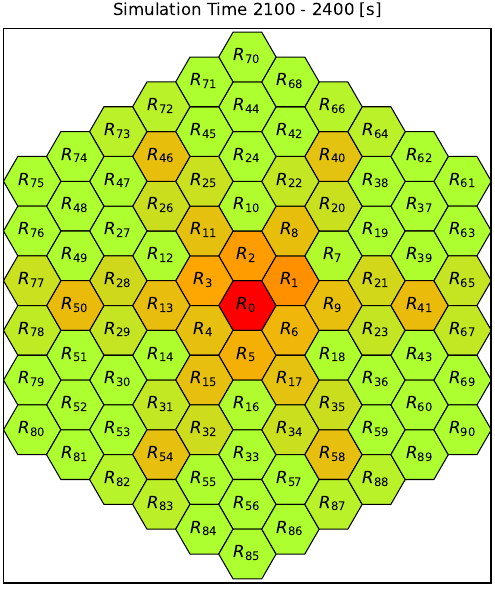}
    \hspace{-1.8mm}
    \includegraphics[height=3.05cm]{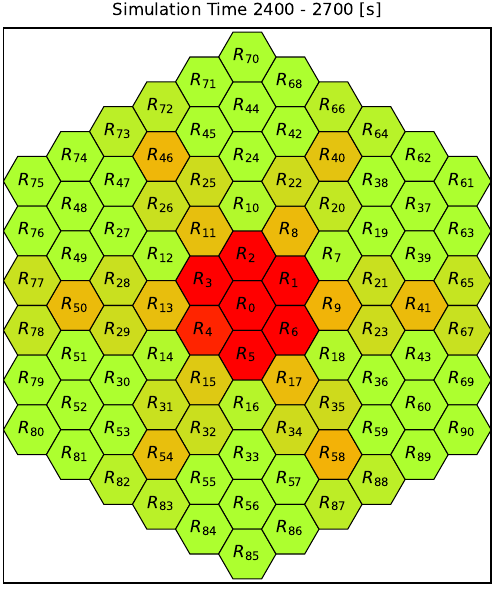}
    \hspace{-1.8mm}
    \includegraphics[height=3.05cm]{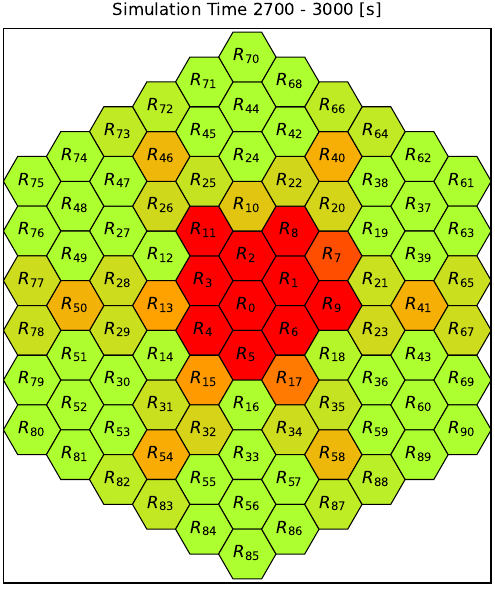}
    \hspace{-1.8mm}
    \includegraphics[height=3.05cm]{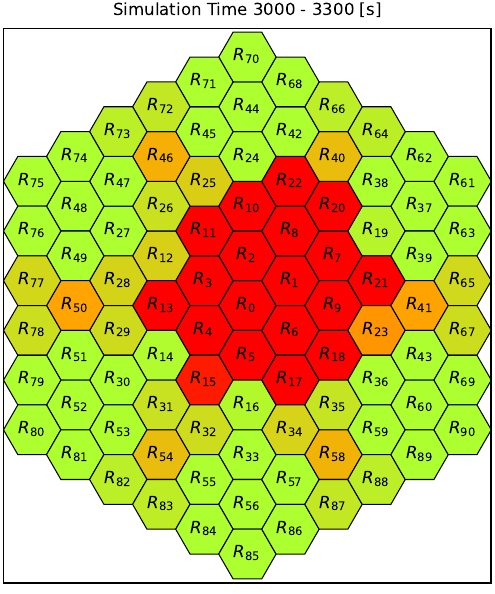}
    \hspace{-1.8mm}
    \includegraphics[height=3.05cm]{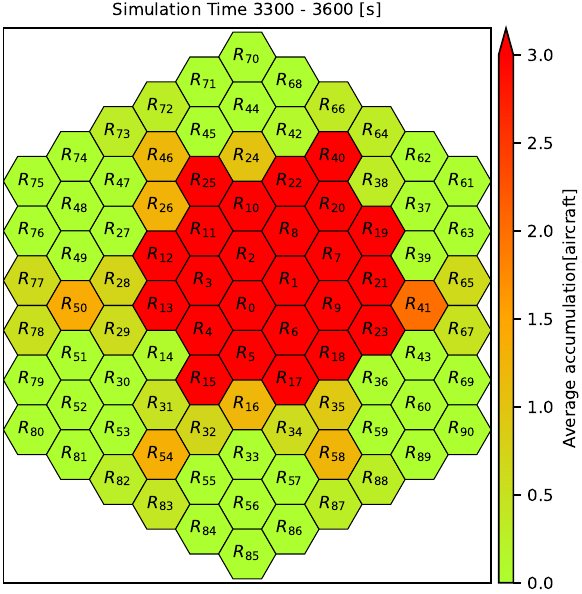}
    \captionsetup{font=scriptsize}
    \vspace{-1.8mm}
    \subcaption{Traffic evolution of the baseline method in the ``$*$'' type scenario}
    \vspace{0.8mm}
    \label{fig:comparison_e}
\end{subfigure}
\begin{subfigure}[t]{\linewidth}
    \centering
    \includegraphics[height=3.05cm]{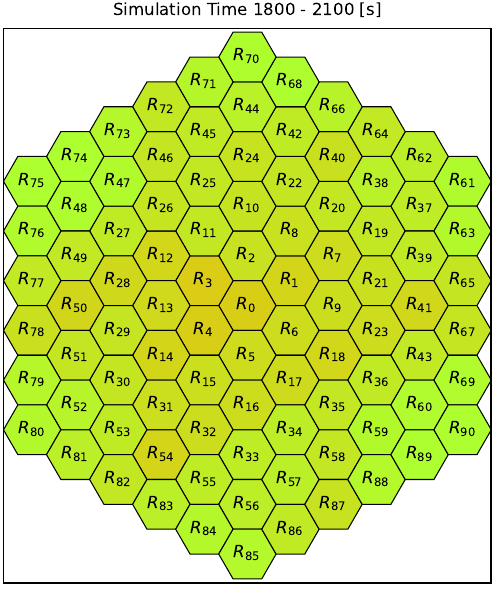}
    \hspace{-1.8mm}
    \includegraphics[height=3.05cm]{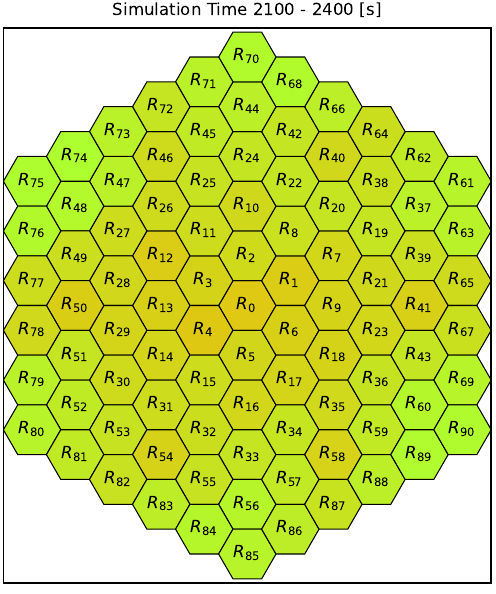}
    \hspace{-1.8mm}
    \includegraphics[height=3.05cm]{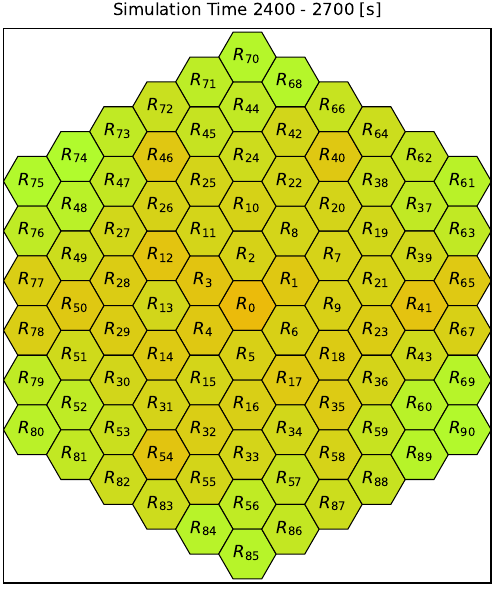}
    \hspace{-1.8mm}
    \includegraphics[height=3.05cm]{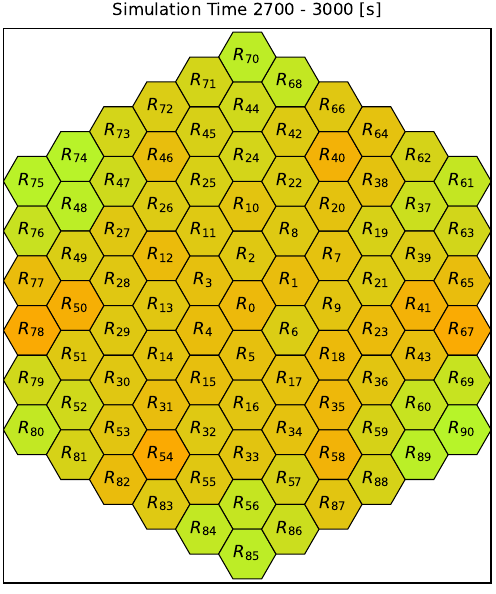}
    \hspace{-1.8mm}
    \includegraphics[height=3.05cm]{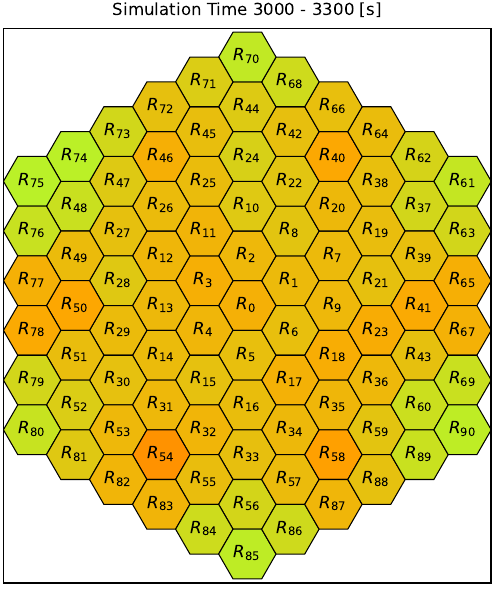}
    \hspace{-1.8mm}
    \includegraphics[height=3.05cm]{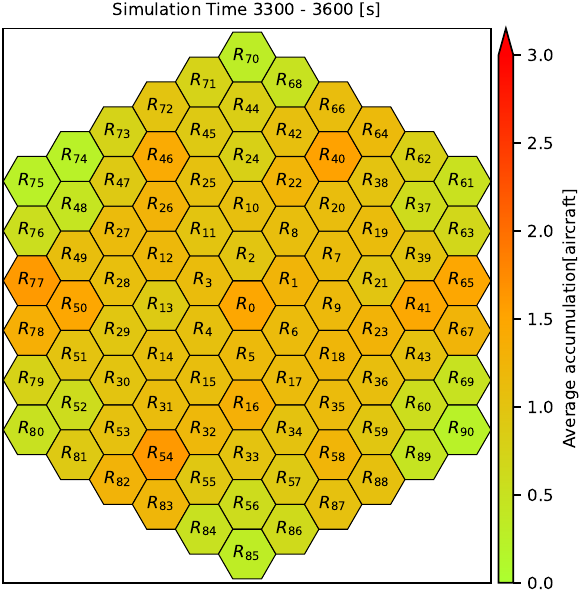}
    \captionsetup{font=scriptsize}
    \vspace{-1.8mm}
    \subcaption{Traffic evolution of the proposed method in the ``$*$'' type scenario}
    \label{fig:comparison_f}
\end{subfigure}
\vspace{-2mm}
\captionsetup{font=footnotesize}
\caption{Performance comparison of air traffic evolution in 2D cases.}
\label{fig:comparison}
\end{figure}
}

{
\newpage
\begin{figure}[!h]
\centering
\begin{subfigure}[t]{0.329\linewidth}
    \centering
    \includegraphics[width=\linewidth]{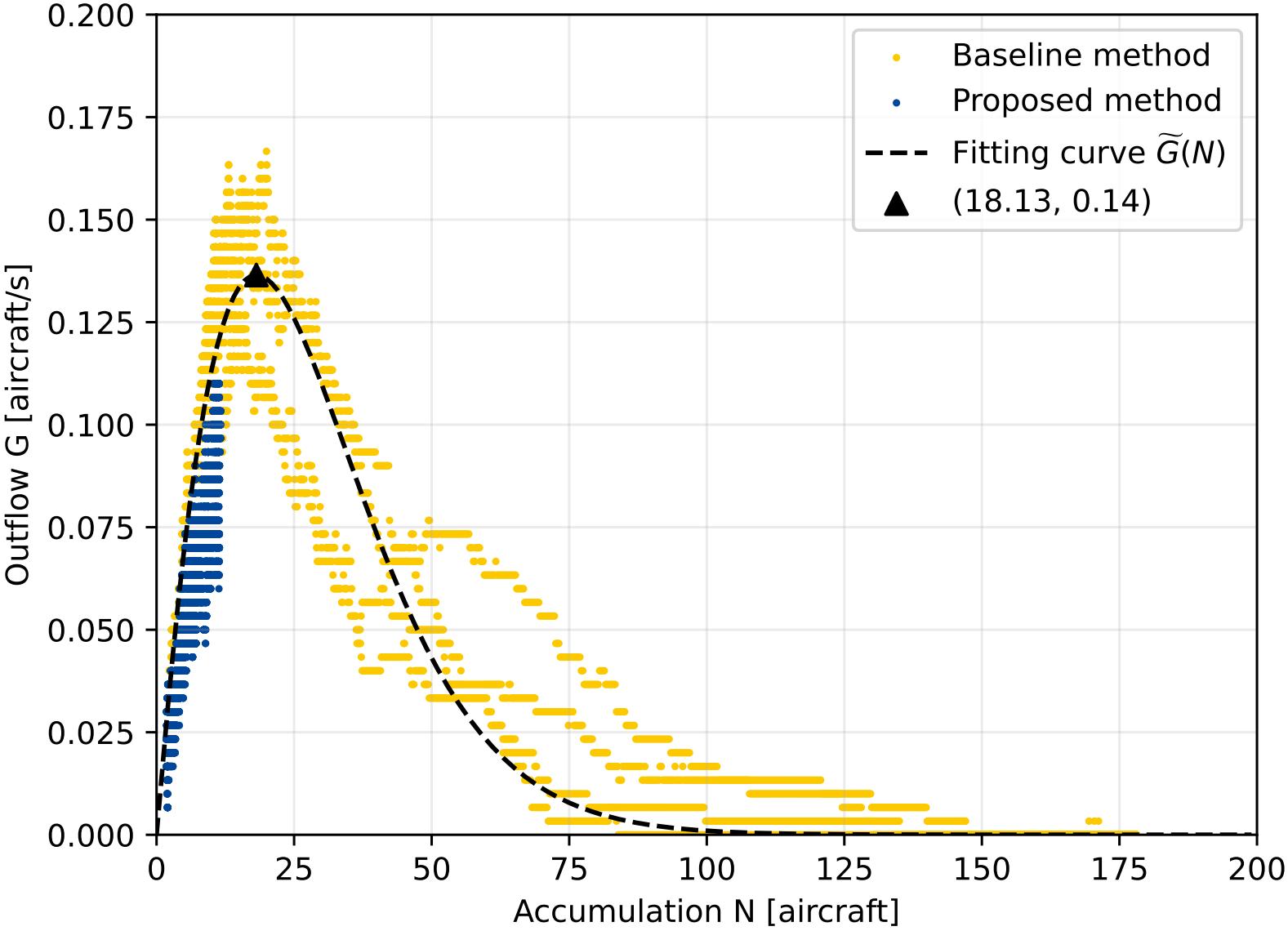}
    \captionsetup{font=scriptsize}
    \subcaption{2D examples with $z = 500\,\mathrm{m}$}
    \label{}
\end{subfigure}
\begin{subfigure}[t]{0.329\linewidth}
    \centering
    \includegraphics[width=\linewidth]{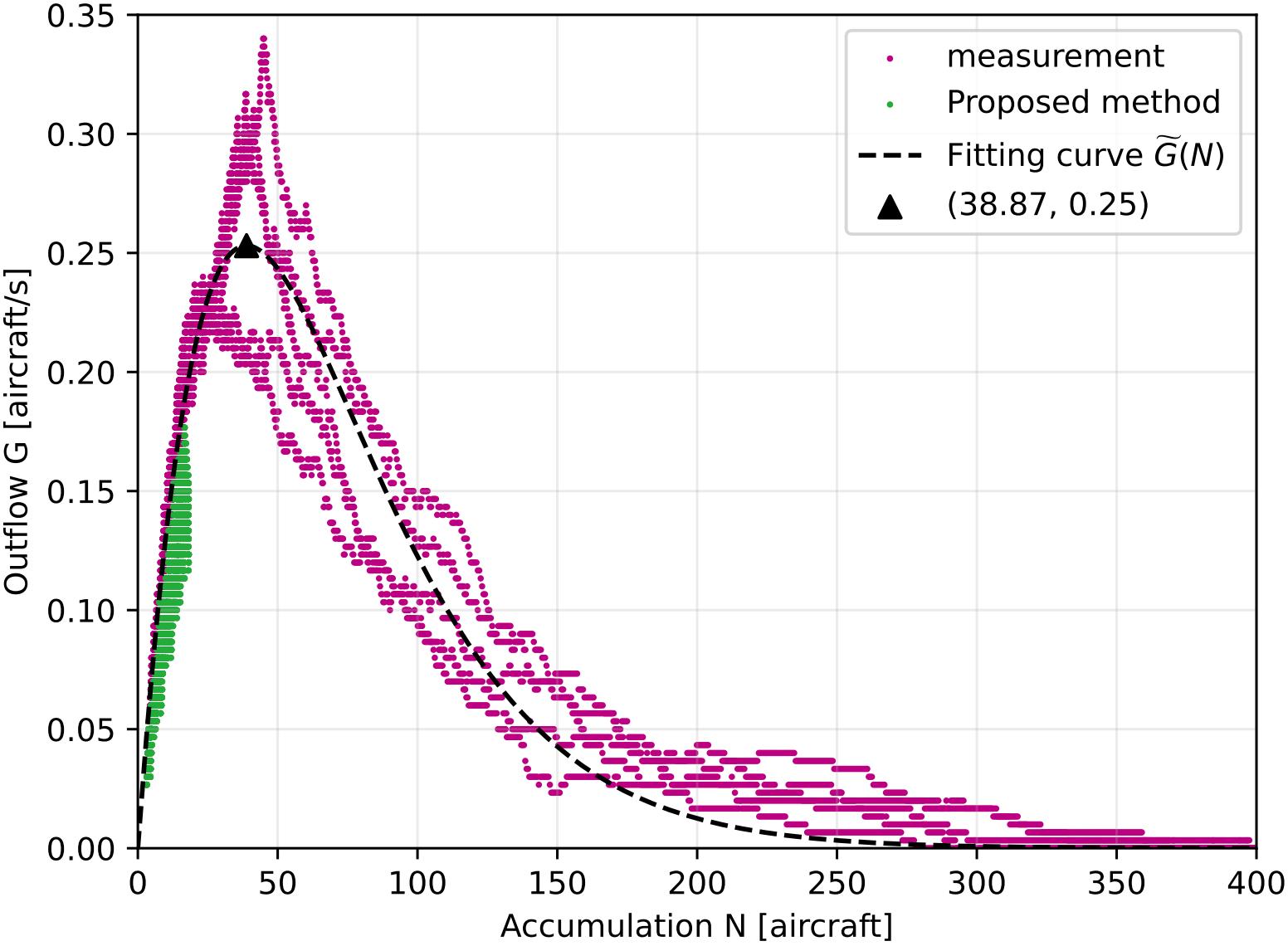}
    \captionsetup{font=scriptsize}
    \subcaption{3D examples with $z \in [450, 550]\,\mathrm{m}$}
    \label{}
\end{subfigure}
\begin{subfigure}[t]{0.329\linewidth}
    \centering
    \includegraphics[width=\linewidth]{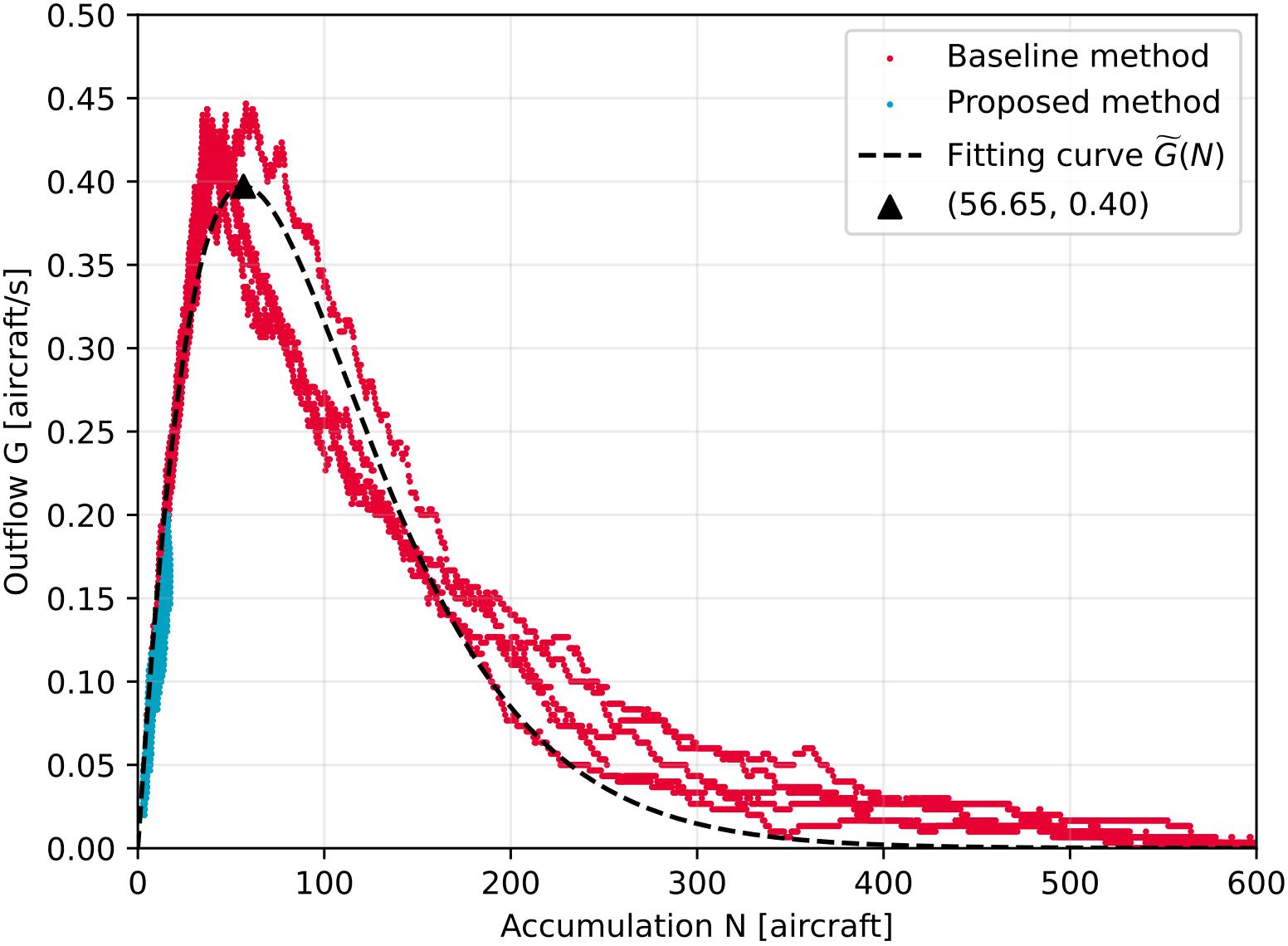}
    \captionsetup{font=scriptsize}
    \subcaption{3D examples with $z \in [400, 600]\,\mathrm{m}$}
    \label{}
\end{subfigure}
\vspace{-2mm}
\caption{The results of outflow-accumulation relationships in the ``$+$'' type scenario.}
\label{fig:mfd+}
\end{figure}
\begin{figure}[!h]
\vspace{-4mm}
\centering
\begin{subfigure}[t]{0.329\linewidth}
    \centering
    \includegraphics[width=\linewidth]{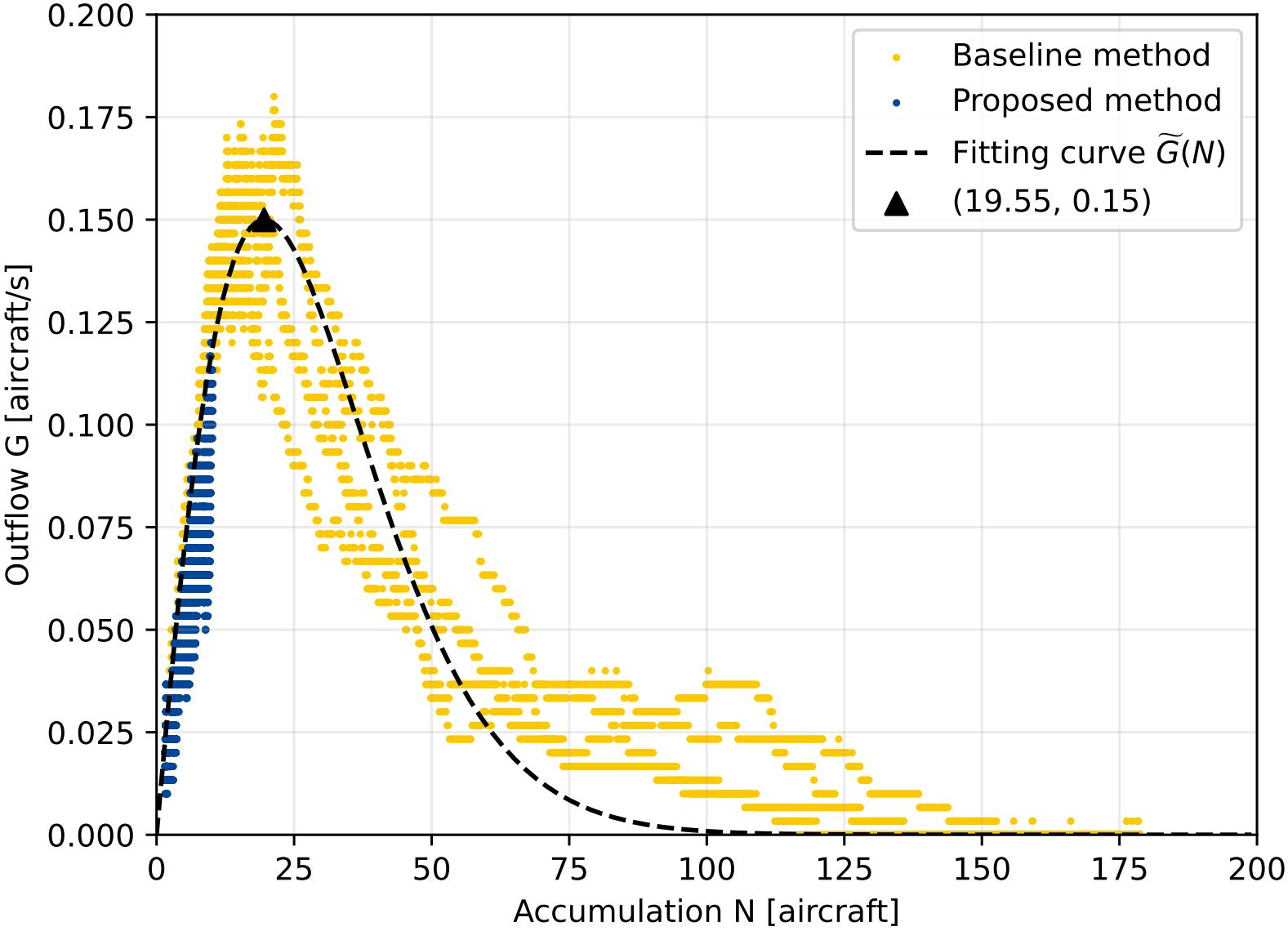}
    \captionsetup{font=scriptsize}
    \subcaption{2D examples with $z = 500\,\mathrm{m}$}
    \label{}
\end{subfigure}
\begin{subfigure}[t]{0.329\linewidth}
    \centering
    \includegraphics[width=\linewidth]{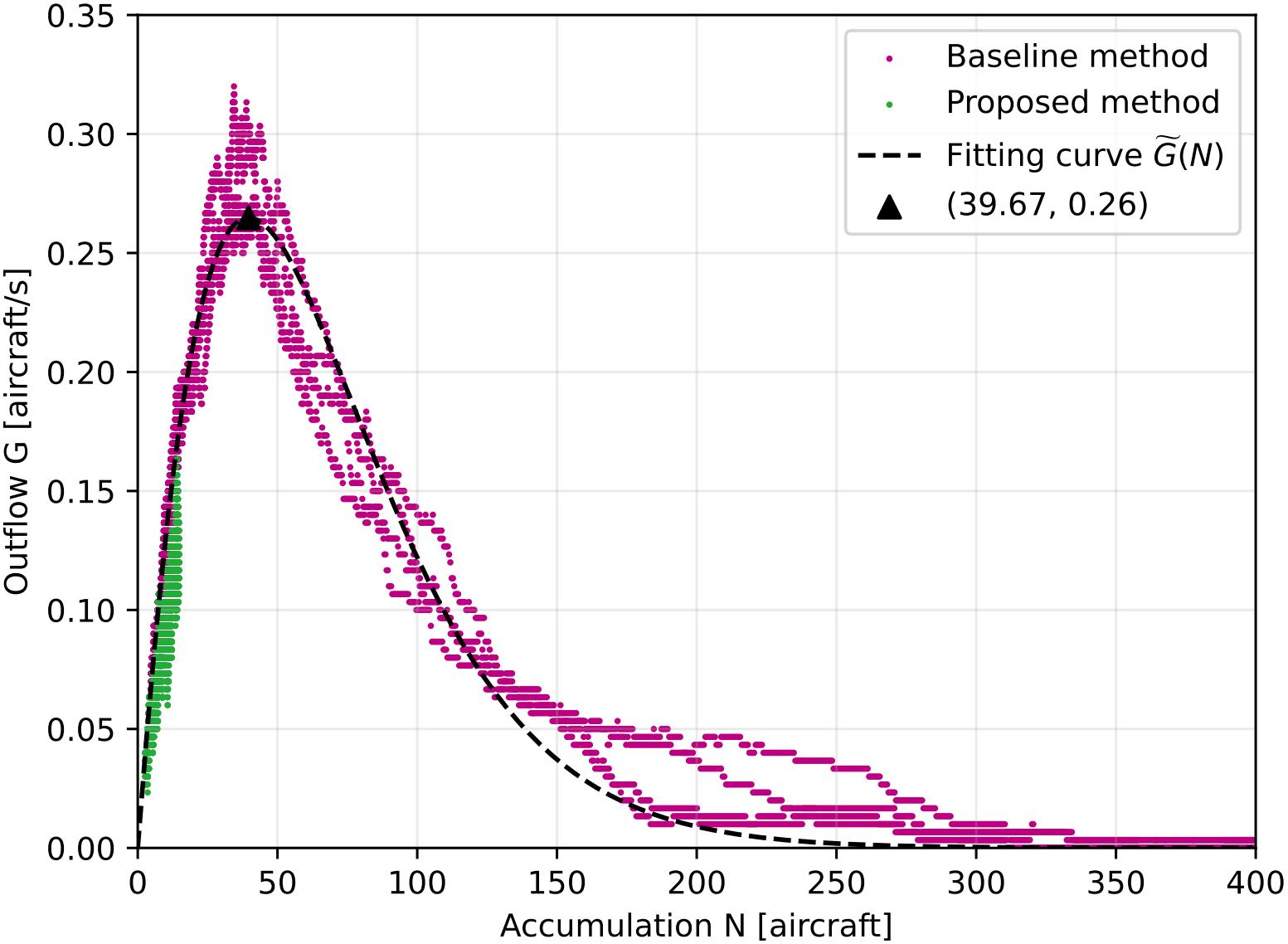}
    \captionsetup{font=scriptsize}
    \subcaption{3D examples with $z \in [450, 550]\,\mathrm{m}$}
    \label{}
\end{subfigure}
\begin{subfigure}[t]{0.329\linewidth}
    \centering
    \includegraphics[width=\linewidth]{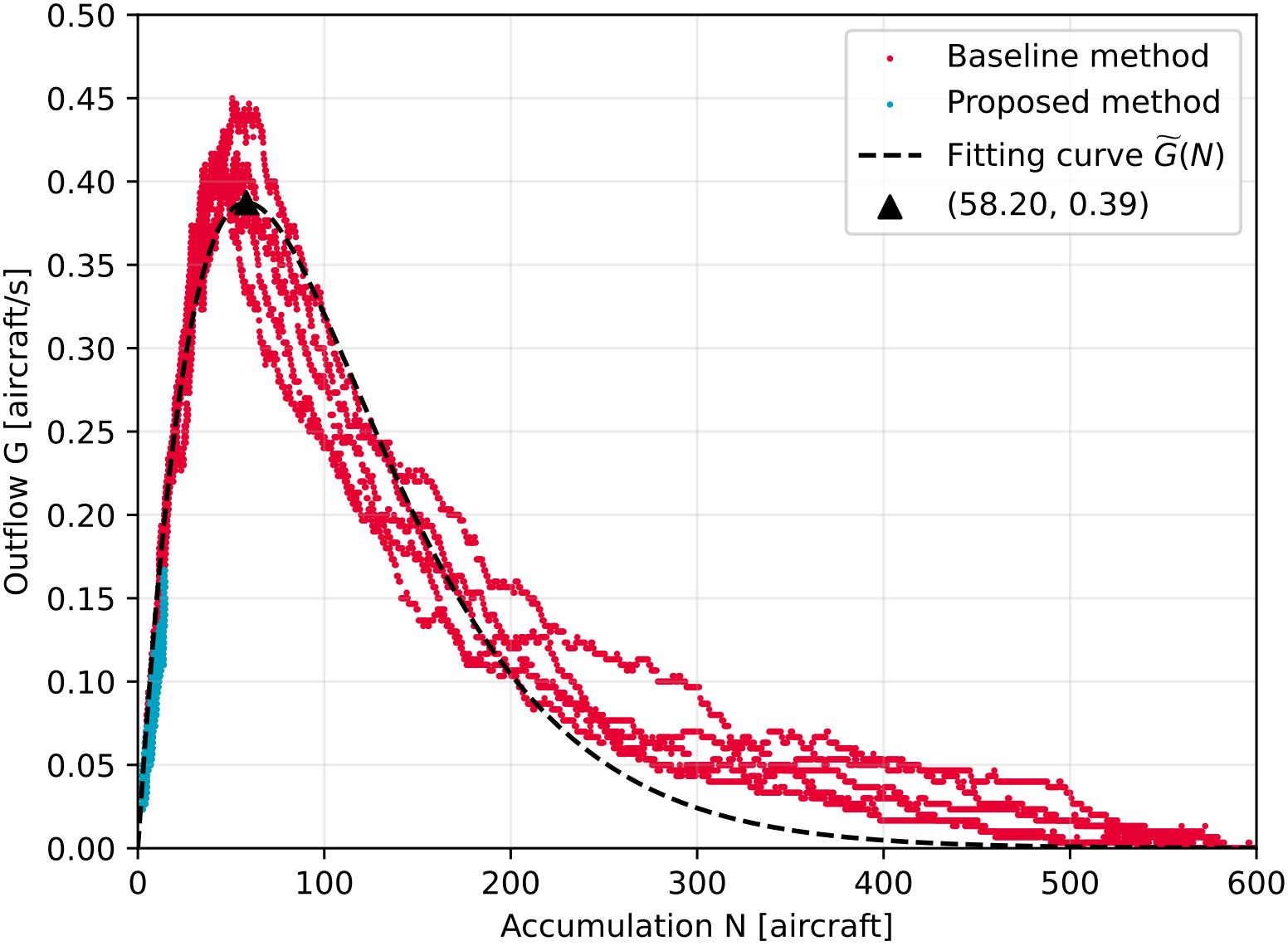}
    \captionsetup{font=scriptsize}
    \subcaption{3D examples with $z \in [400, 600]\,\mathrm{m}$}
    \label{}
\end{subfigure}
\vspace{-2mm}
\caption{The results of outflow-accumulation relationships in the ``$\#$'' type scenario.}
\label{fig:mfd井}
\end{figure}
\begin{figure}[!h]
\vspace{-4mm}
\centering
\begin{subfigure}[t]{0.329\linewidth}
    \centering
    \includegraphics[width=\linewidth]{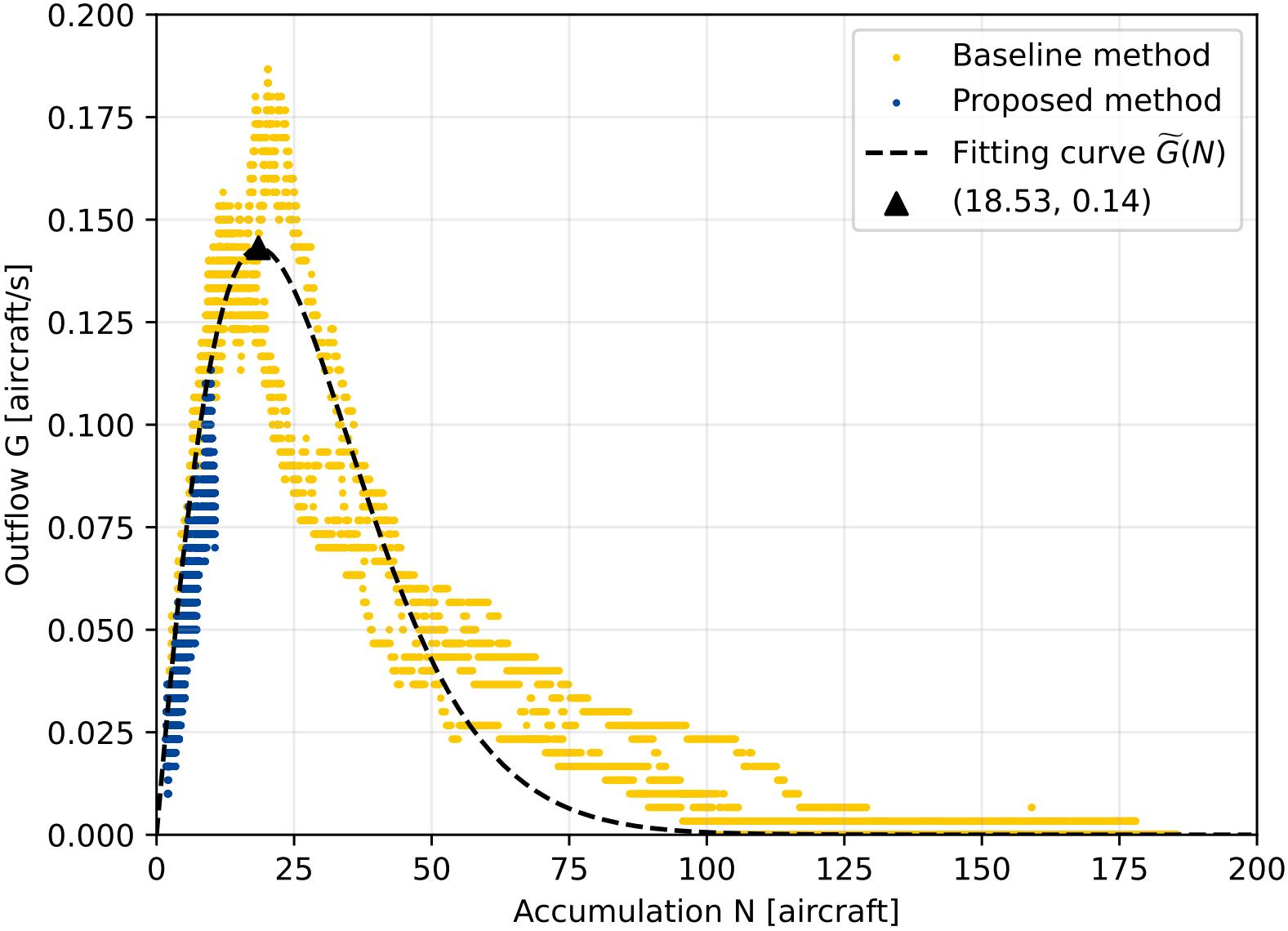}
    \captionsetup{font=scriptsize}
    \subcaption{2D examples with $z = 500\,\mathrm{m}$}
    \label{}
\end{subfigure}
\begin{subfigure}[t]{0.329\linewidth}
    \centering
    \includegraphics[width=\linewidth]{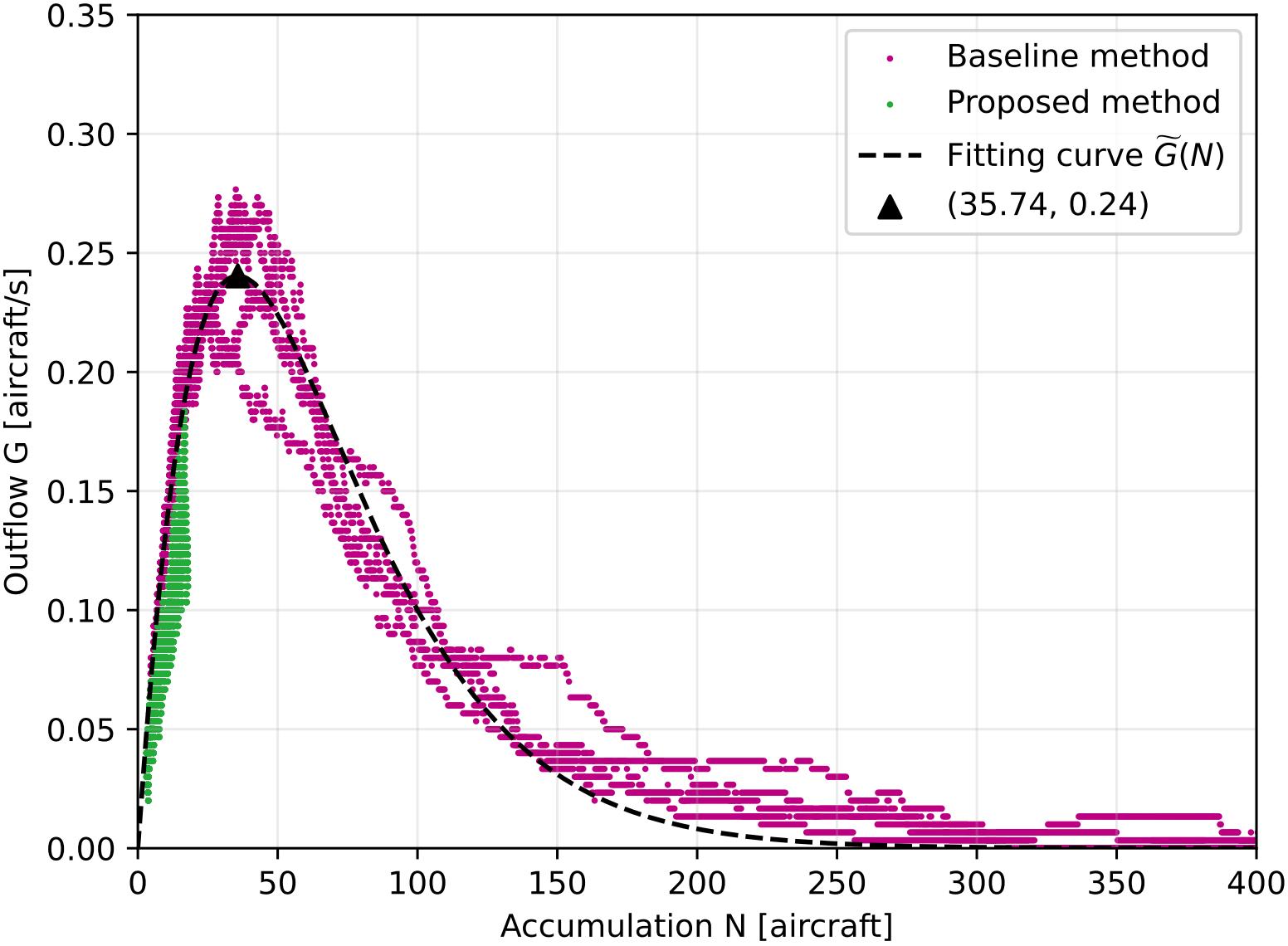}
    \captionsetup{font=scriptsize}
    \subcaption{3D examples with $z \in [450, 550]\,\mathrm{m}$}
    \label{}
\end{subfigure}
\begin{subfigure}[t]{0.329\linewidth}
    \centering
    \includegraphics[width=\linewidth]{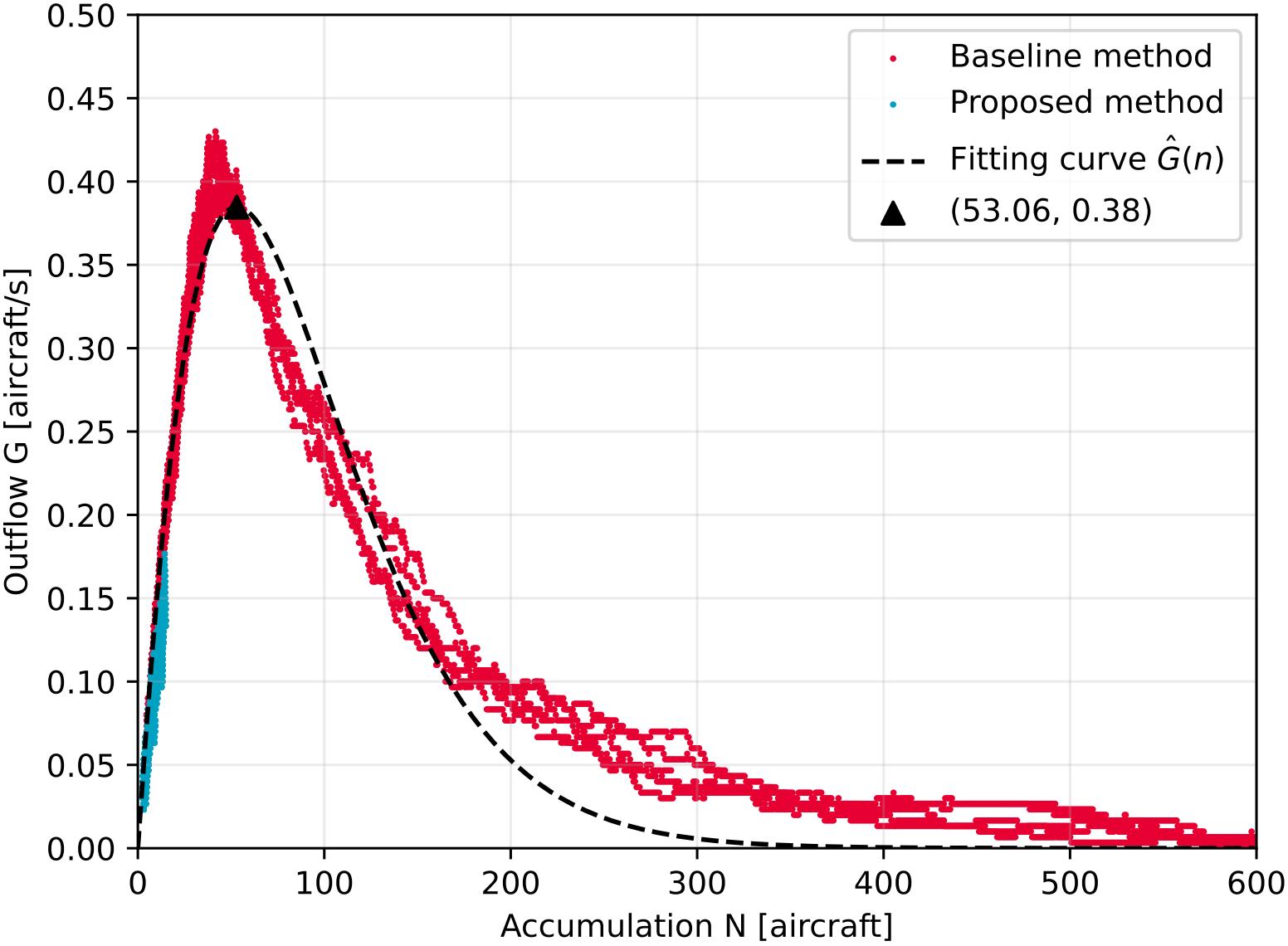}
    \captionsetup{font=scriptsize}
    \subcaption{3D examples with $z \in [400, 600]\,\mathrm{m}$}
    \label{}
\end{subfigure}
\vspace{-2mm}
\caption{The results of outflow-accumulation relationships in the ``$*$'' type scenario.}
\label{fig:mfd*}
\end{figure}
\begin{figure}[!h]
\vspace{-4mm}
\centering
\begin{subfigure}[t]{0.329\linewidth}
    \centering
    \includegraphics[width=\linewidth]{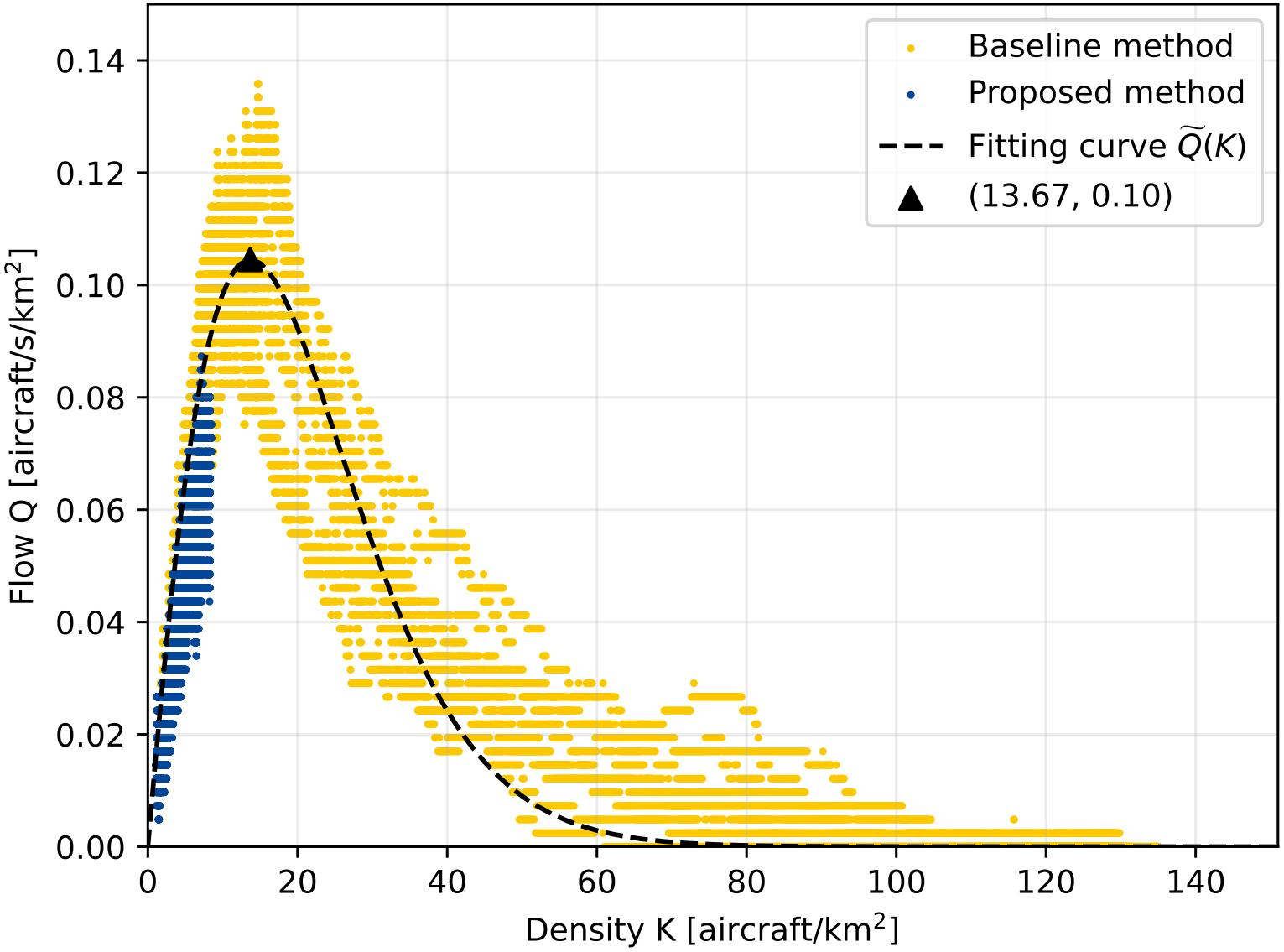}
    \captionsetup{font=scriptsize}
    \subcaption{2D cases with $z = 500\,\mathrm{m}$}
    \label{}
\end{subfigure}
\begin{subfigure}[t]{0.329\linewidth}
    \centering
    \includegraphics[width=\linewidth]{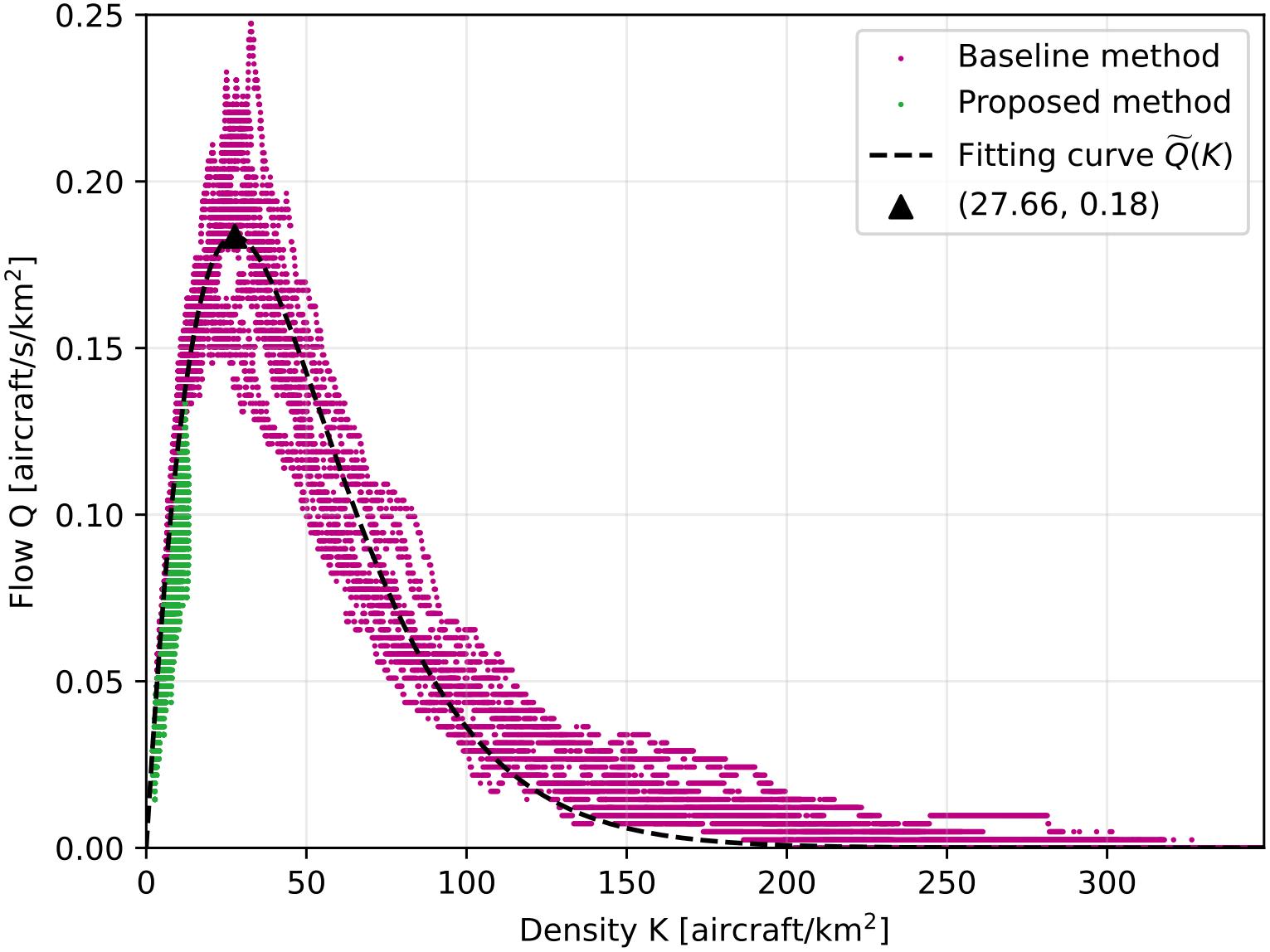}
    \captionsetup{font=scriptsize}
    \subcaption{3D cases with $z \in [450, 550]\,\mathrm{m}$}
    \label{}
\end{subfigure}
\begin{subfigure}[t]{0.329\linewidth}
    \centering
    \includegraphics[width=\linewidth]{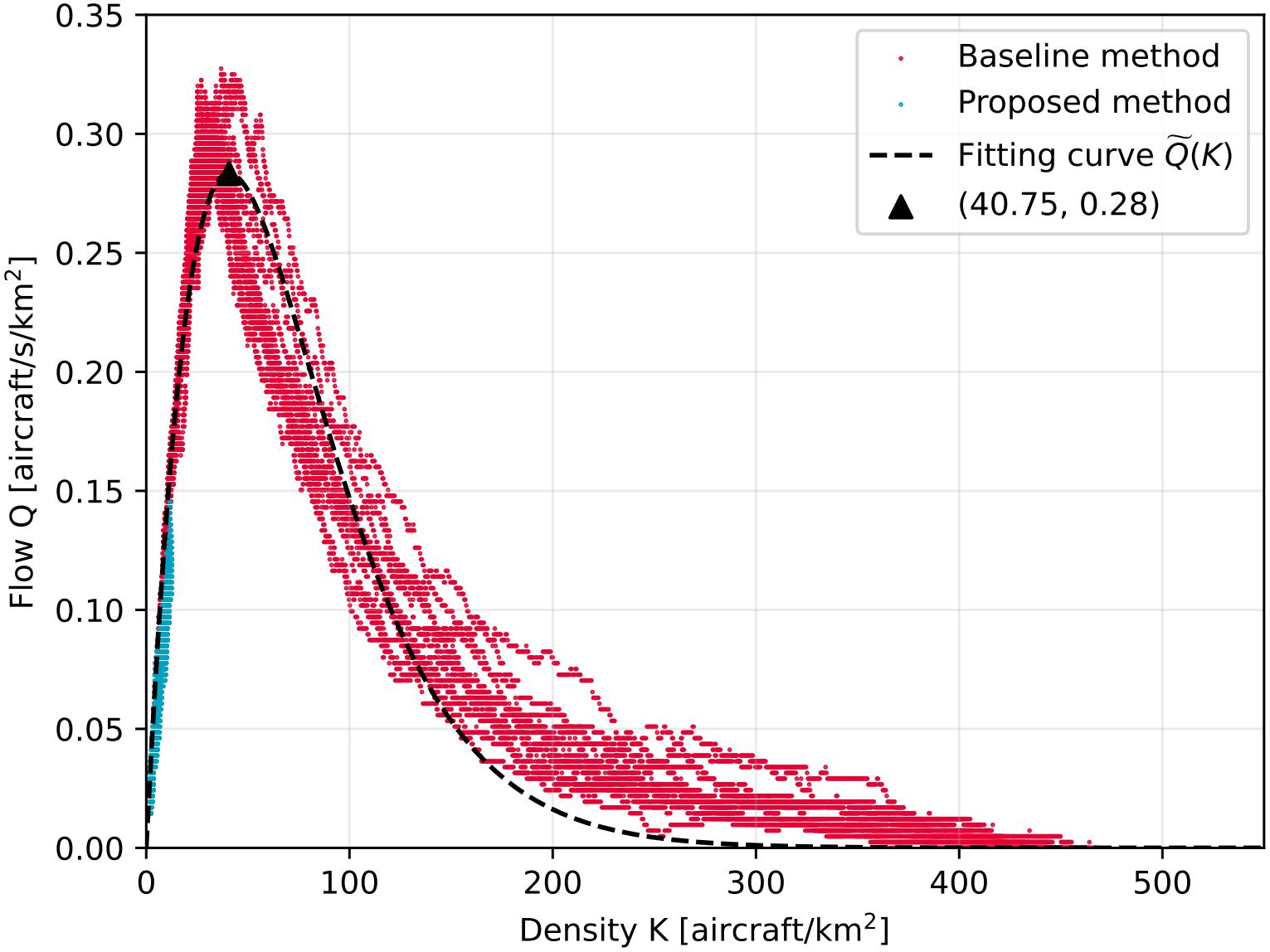}
    \captionsetup{font=scriptsize}
    \subcaption{3D cases with $z \in [400, 600]\,\mathrm{m}$}
    \label{}
\end{subfigure}
\vspace{-2mm}
\captionsetup{font=footnotesize}
\caption{The results of flow-density relationships.}
\label{fig:airsapce_capacity}
\end{figure}
}

\newpage
Under the demand setting illustrated in \autoref{fig:Aircraft_inflow}, the proposed framework demonstrates advantages over the baseline, primarily due to the adverse impact of traffic congestion on network performance for the baseline case.  
To this end, we adjust the demand to a fixed level close to the critical flow and repeat the previous tests.
Five performance metrics are selected, i.e., the average minimum separation, the average travel speed, the trip completion rate, the energy consumption, and the computational efficiency, as summarized in \autoref{tab:metrics}.
The energy consumption is computed based on aircraft velocities, following the approach described in \cite{safadi2024holistic,safadi2025optimal}.
Compared with the baseline, the proposed framework achieves substantial improvements in traffic safety and efficiency, with increases in the average minimum separation (+98.2\%), the average travel speed (+70.2\%), and the trip completion rate (+130\%), along with a reduction in the energy consumption (-23.0\%).
Interestingly, the proposed framework requires additional computation for path planning but achieves higher computational efficiency (+433\%).  
This can be explained from two aspects.  
First, the FAMs proposed in \autoref{sec:FAM} are computationally efficient for path planning.
Second, the optimal path planning reduces the occurrence of aircraft conflicts (as evidenced by the increased average minimum separation), thereby reducing the computational effort required for collision avoidance.

\begin{table}[!h]
    \centering
    \begin{threeparttable}[h]
    \captionsetup{font=footnotesize}
    \caption{Performance comparison between the proposed framework and the baseline.$^\dagger$}
    \label{tab:metrics}
    \renewcommand\arraystretch{1.25}
    \footnotesize
    \begin{tabular}{l p{2cm}p{0.5cm}p{1.2cm}}
    % \hline
    \toprule
    & Baseline method & \multicolumn{2}{l}{Proposed method} \\
    % \hline
    \midrule
    Average Minimum Separation [m] & 153.3 & 303.9 &(+98.2\%)\\
    % \hline
    Average Travel Speed [m/s] & 11.39 & 19.39 &(+70.2\%)\\
    % \hline
    Trip Completion Rate [trip/s] & 0.083 & 0.191 &(+130\%)\\
    % \hline
    Energy Consumption  [kWh/trip] & 0.0148 & 0.0114 &(-23.0\%)\\
    % \hline
    Computational efficiency [iteration/s] & 0.91 & 4.85 &(+433\%)\\
    % \hline
    \bottomrule
    \end{tabular}
    \begin{tablenotes}
    \item[$\dagger$] Higher values indicate better performance, except for energy consumption.
    \end{tablenotes}
    \end{threeparttable}
\end{table}

\subsubsection{Results of the two-layer airspace network} \label{sec:two-layer}

We further evaluate the effectiveness of the proposed framework within a two-layer airspace network.  
The two airspace layers are connected via two vertical air tubes, linking region $R_{10}$ to $R_{29}$ and region $R_{16}$ to $R_{35}$.  
The demand OD pairs are randomly generated between regions $R_{10}$ and $R_{16}$, with altitudes concentrated in the lower layer.
\autoref{fig:comparison_layer} illustrates the air traffic evolution of the proposed framework across different airspace network configurations.
The single-layer case is used here for comparison.

Although the proposed framework shows great potential in mitigating local traffic congestion, its performance is ultimately constrained by the airspace capacity. 
For example, when the demand level exceeds the capacity of a single layer, even uniformly distributed air traffic becomes susceptible to congestion, as illustrated in \autoref{fig:single-layer}.  
In particular, the route guidance mechanism is unable to alleviate traffic accumulation within regions $R_{10}$ and $R_{16}$, where oversaturated demand is directly imposed.
A two-layer network configuration is expected to alleviate this limitation by increasing the overall network capacity.
Hence, we activate an additional airspace layer and conduct simulations under the same settings.  
The results presented in \autoref{fig:two-layer} confirm the conjecture.
With the aid of vertical tubes, the route guidance strategy upgrades from two-dimensional (horizontal) to three-dimensional (horizontal and vertical) path planning.
The air traffic, initially concentrated in the lower layer, is redistributed across the entire network by the proposed framework.
This significantly alleviates traffic congestion within the lower layer and enhances overall network performance, as evidenced by increases in the average minimum separation (+56.3\%), the average travel speed (+38.9\%) and the trip completion rate (+40.1\%).

A brief summary of the key results from \autoref{sec:single-layer} and \autoref{sec:two-layer} is provided below.
The proposed air traffic management framework is validated to be capable of achieving safe and efficient UAM operations, particularly in reducing the occurrence of aircraft conflicts and mitigating local traffic congestion.
The framework’s effectiveness in promoting air traffic homogeneity has important implications for advancing research on the airspace MFD \citep{haddad2021traffic,safadi2023macroscopic,cummings2024airspace,cummings2024comparing}.
The results of both single-layer and two-layer airspace networks highlight the potential of the proposed framework to accommodate more complex network configurations.
However, further investigation and analysis of complex airspace network configurations is left for future work.

{
\newpage
\begin{figure}[!h]
\centering
\begin{subfigure}[t]{\linewidth}
    \centering
    \includegraphics[height=4.5cm]{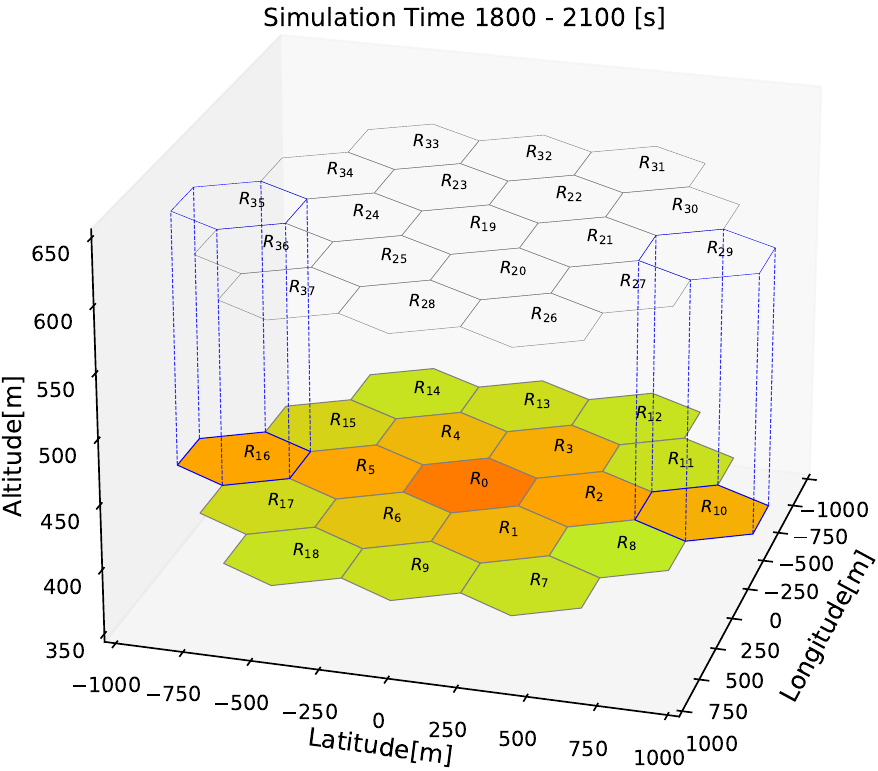}
    \hspace{-1.7mm}
    \includegraphics[height=4.5cm]{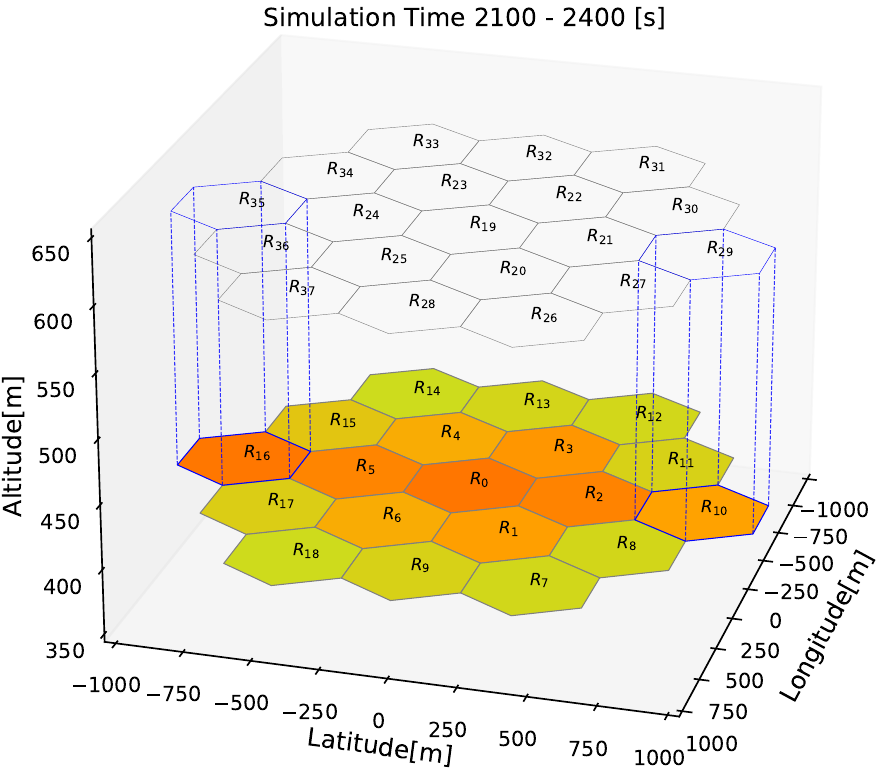}
    \hspace{-1.7mm}
    \includegraphics[height=4.5cm]{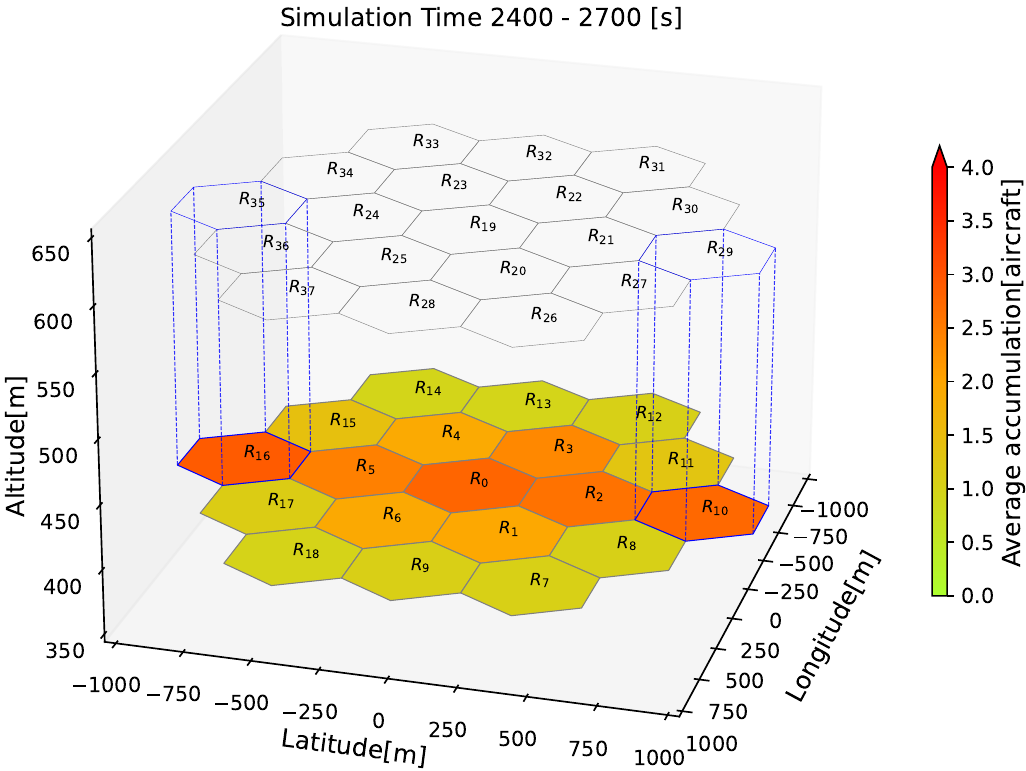}
    \\
    \vspace{3mm}
    \includegraphics[height=4.5cm]{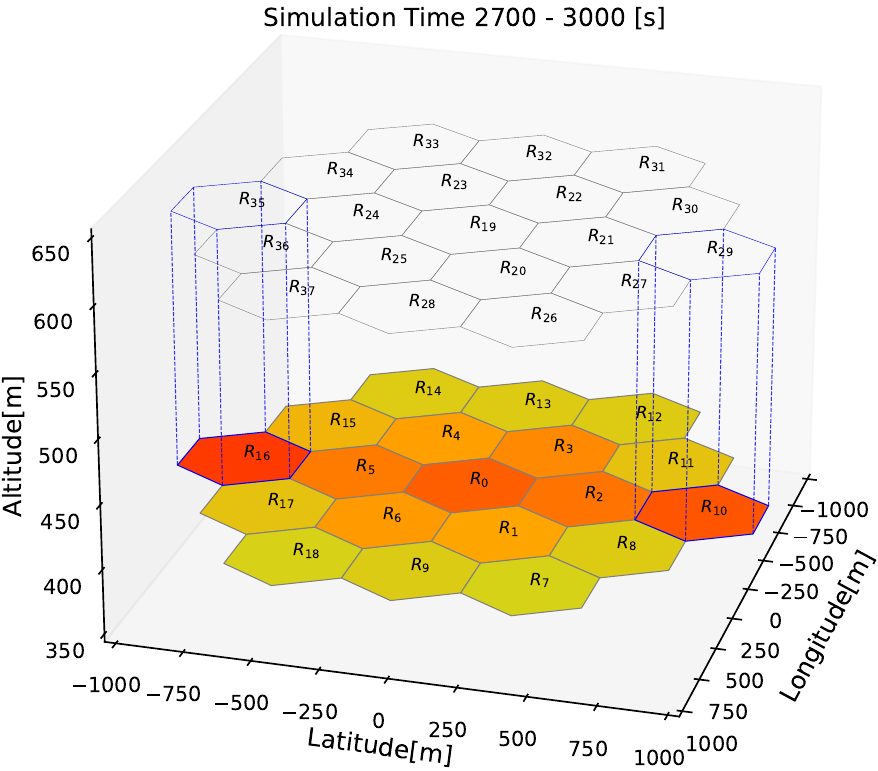}
    \hspace{-1.7mm}
    \includegraphics[height=4.5cm]{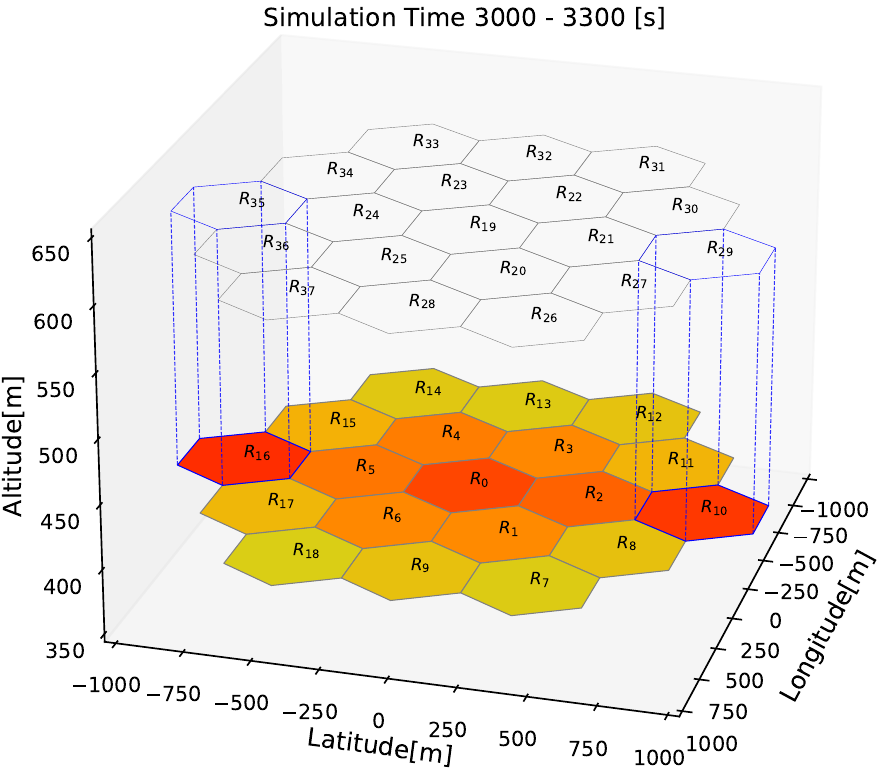}
    \hspace{-1.7mm}
    \includegraphics[height=4.5cm]{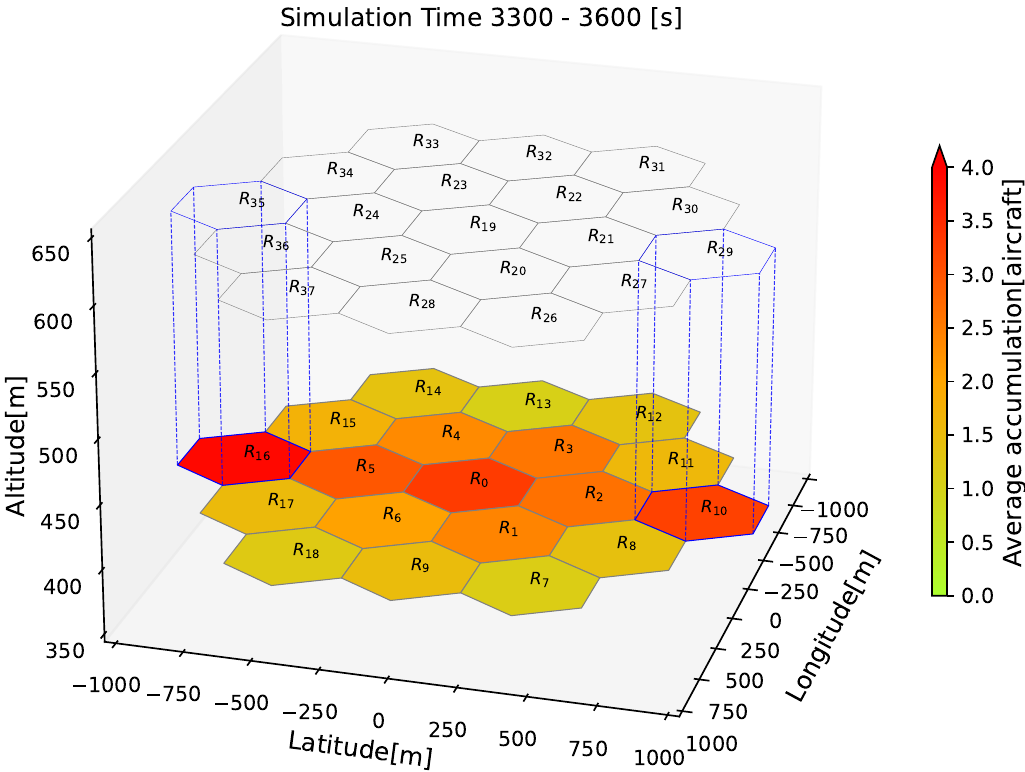}
    \captionsetup{font=footnotesize}
    \subcaption{Traffic evolution in a single-layer airspace network}
    \label{fig:single-layer}
\end{subfigure}
\\
\vspace{4mm}
\begin{subfigure}[t]{\linewidth}
    \centering
    \includegraphics[height=4.5cm]{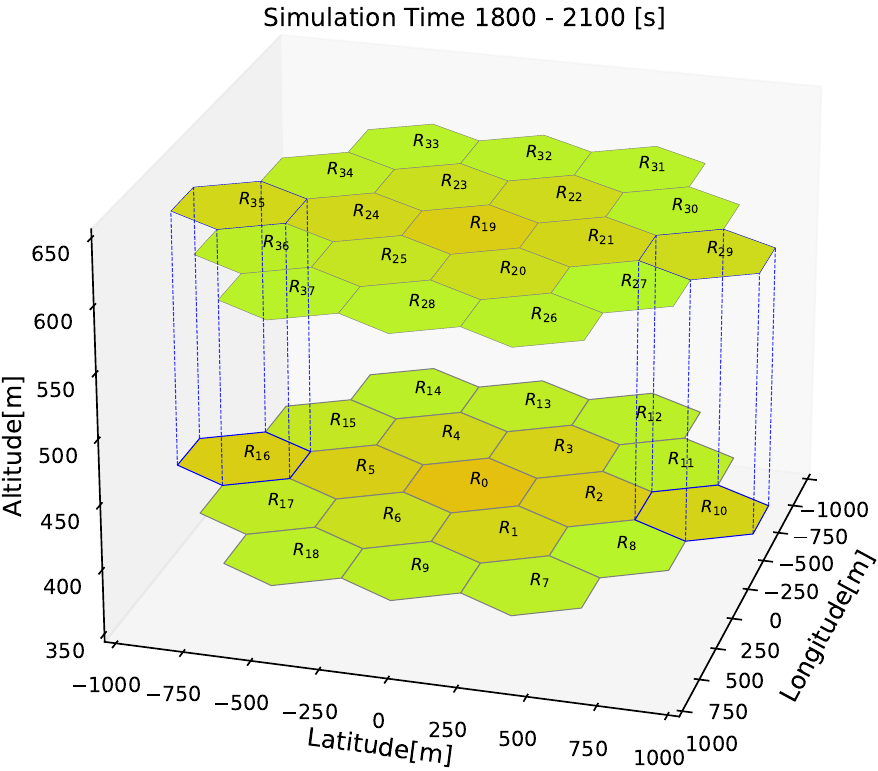}
    \hspace{-1.7mm}
    \includegraphics[height=4.5cm]{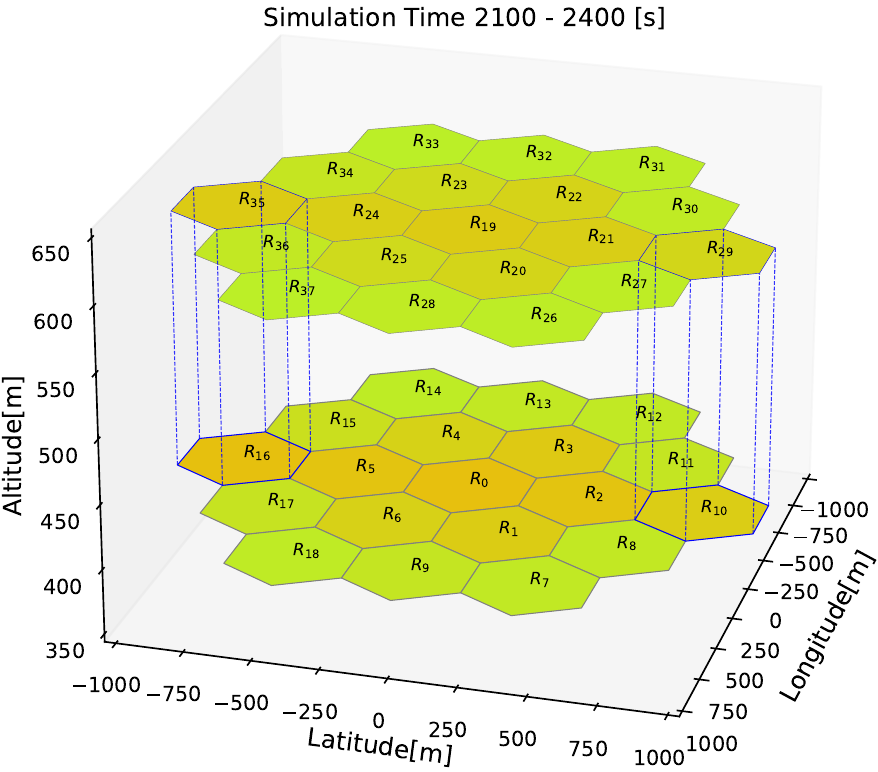}
    \hspace{-1.7mm}
    \includegraphics[height=4.5cm]{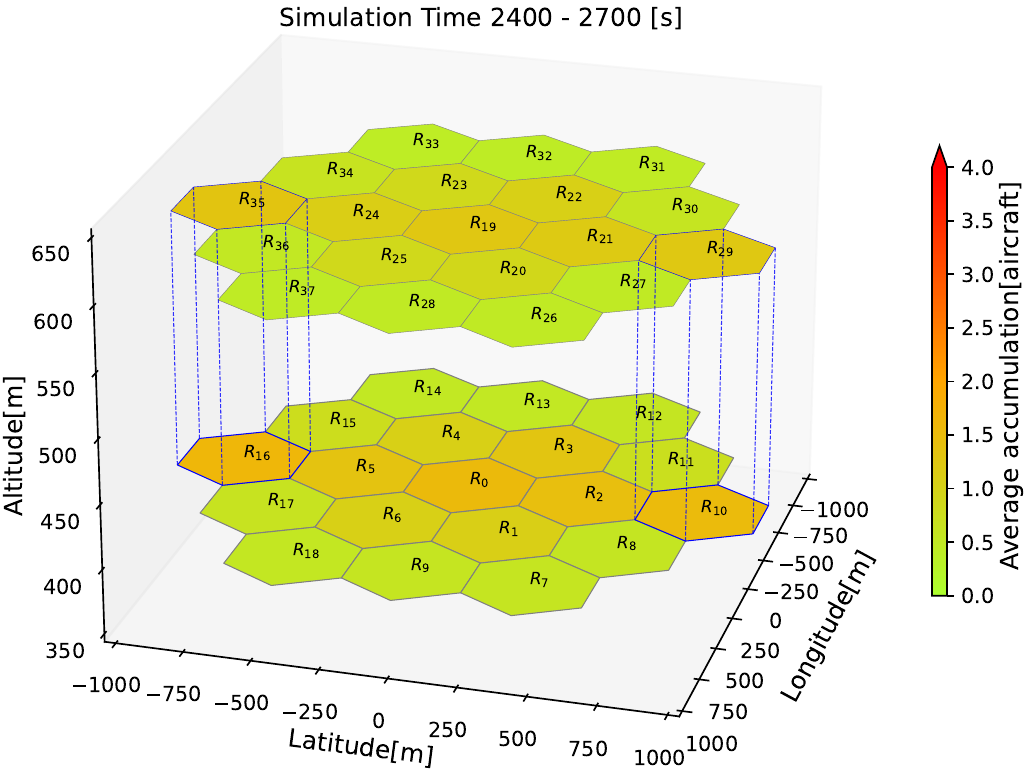}
    \\
    \vspace{3mm}
    \includegraphics[height=4.5cm]{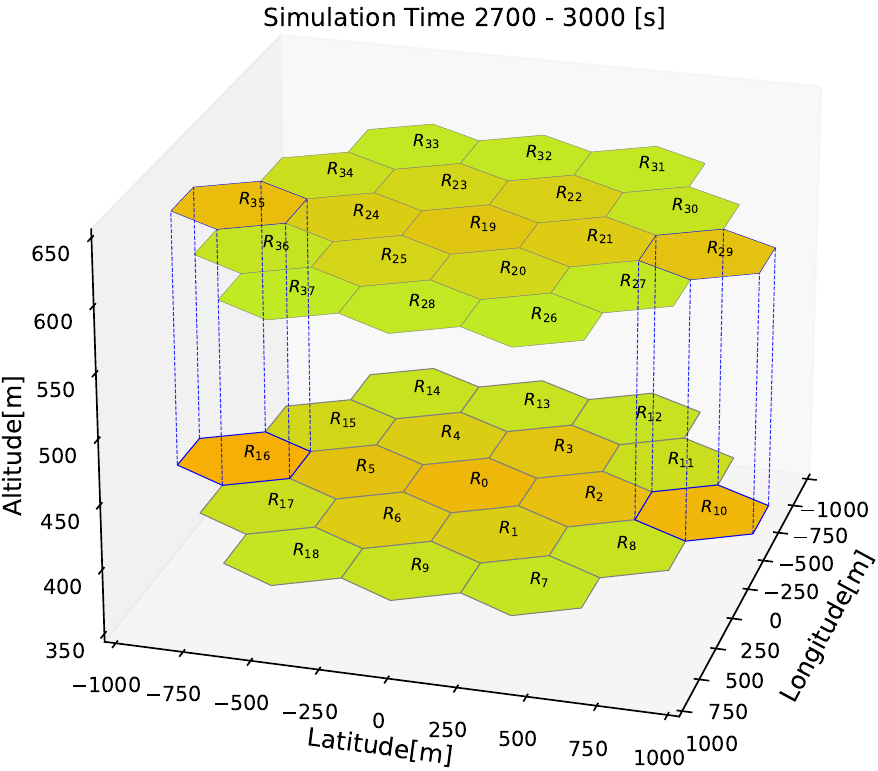}
    \hspace{-1.7mm}
    \includegraphics[height=4.5cm]{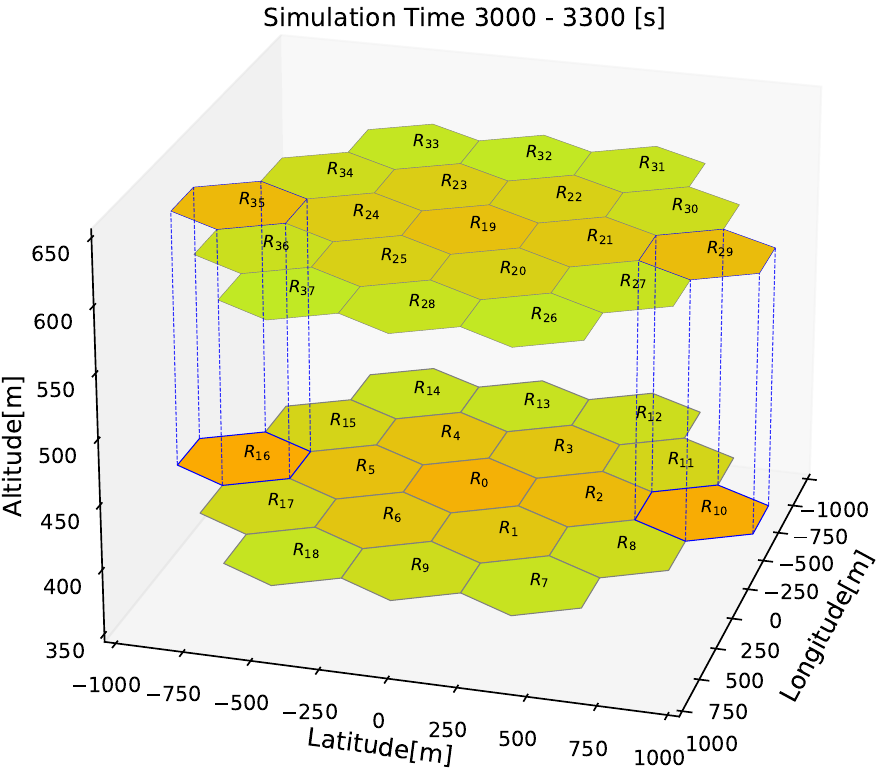}
    \hspace{-1.7mm}
    \includegraphics[height=4.5cm]{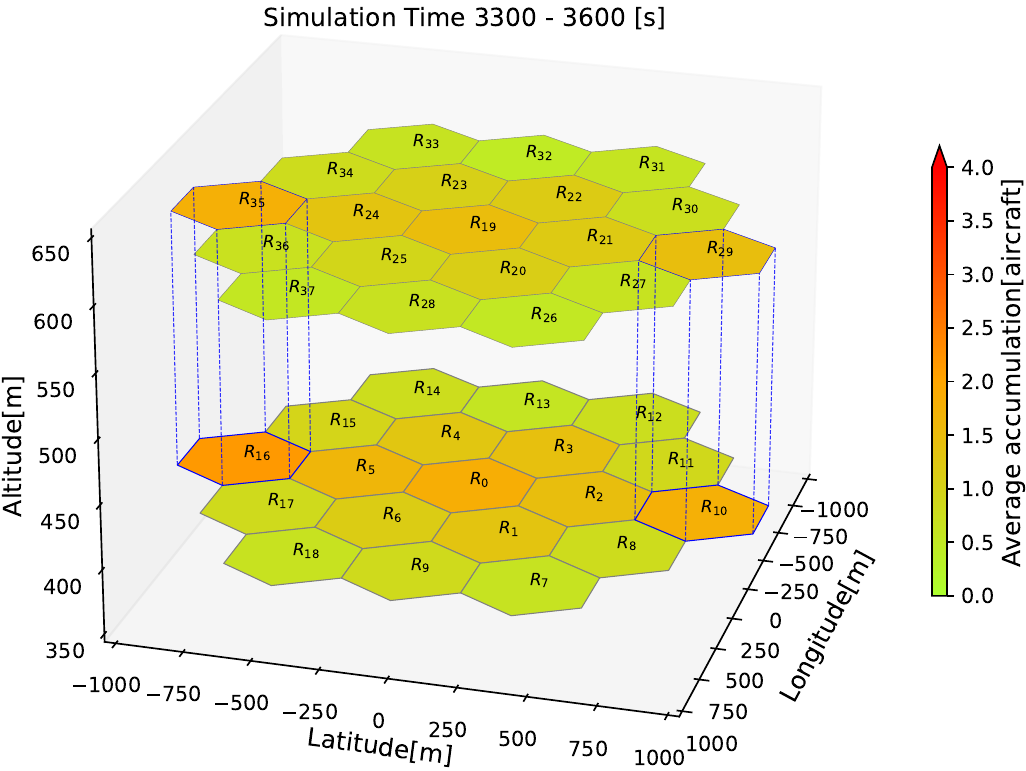}
    \captionsetup{font=footnotesize}
    \subcaption{Traffic evolution in a two-layer airspace network}
    \label{fig:two-layer}
\end{subfigure}
\captionsetup{font=footnotesize}
\caption{Performance comparison of the proposed framework across different airspace network configurations, with blue wireframes indicating vertical air tubes connecting the two layers.}
\label{fig:comparison_layer}
\end{figure}
}

\subsubsection{Enabling flexible UAM operations with the proposed framework} \label{sec:flexible UAM operations}

\begin{figure}[!h]
\centering
\includegraphics[height=4.7cm]{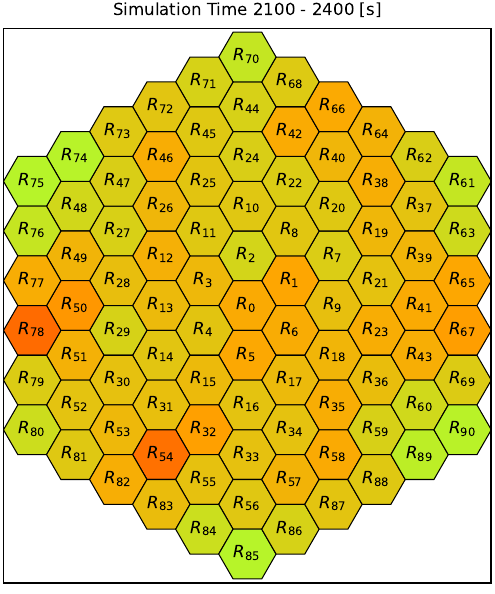}
\hspace{-1.8mm}
\includegraphics[height=4.7cm]{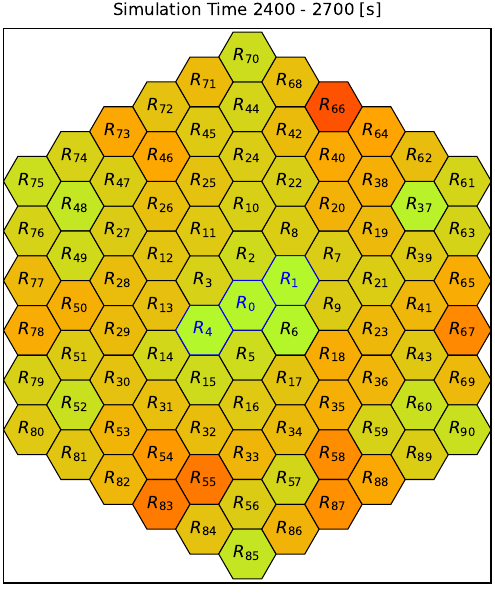}
\hspace{-1.8mm}
\includegraphics[height=4.7cm]{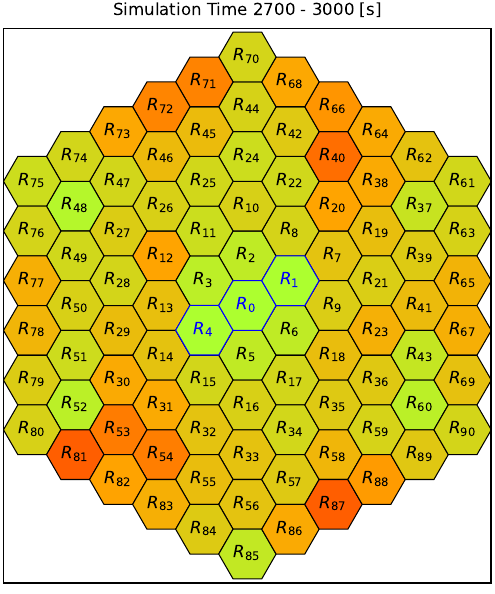}
\hspace{-1.8mm}
\includegraphics[height=4.7cm]{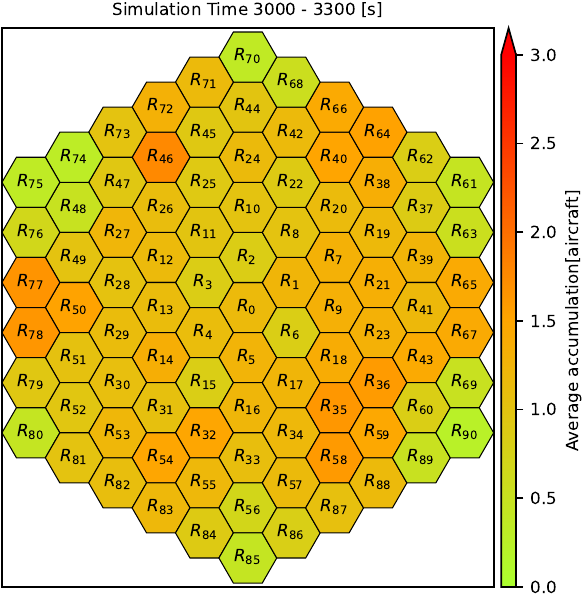}
\captionsetup{font=footnotesize}
\caption{Dynamic no-fly zones in regions $R_0$, $R_1$, and $R_4$ during $t \in [2400, 3000]\,\mathrm{s}$.}
\label{fig:airspace_clearance}
\end{figure}

In this section, we present examples to demonstrate the potential of the proposed framework in enabling flexible UAM operations.
Under the route guidance mechanism proposed in \autoref{sec:Route_guidance}, reducing the airspace capacities of specific regions would prevent aircraft from entering them.
This implies that it is possible to achieve any desired distribution of air traffic (not limited to the homogeneous distribution as discussed previously).
For example, no-fly zones can be dynamically established by adjusting the capacities of the corresponding airspace regions, as illustrated in \autoref{fig:airspace_clearance}.
During the simulation period of $t \in [2400, 3000]\,\mathrm{s}$, the route guidance strategy prevents aircraft from entering airspace regions $R_0$, $R_1$, and $R_4$, thereby creating dynamic no-fly zones. 
This helps the proposed framework address practical challenges such as no-fly zones induced by wind gusts in real-world environments.

\begin{figure}[!ht]
\centering
\begin{subfigure}[t]{0.44\linewidth}
    \centering
    \includegraphics[height=6.5cm]{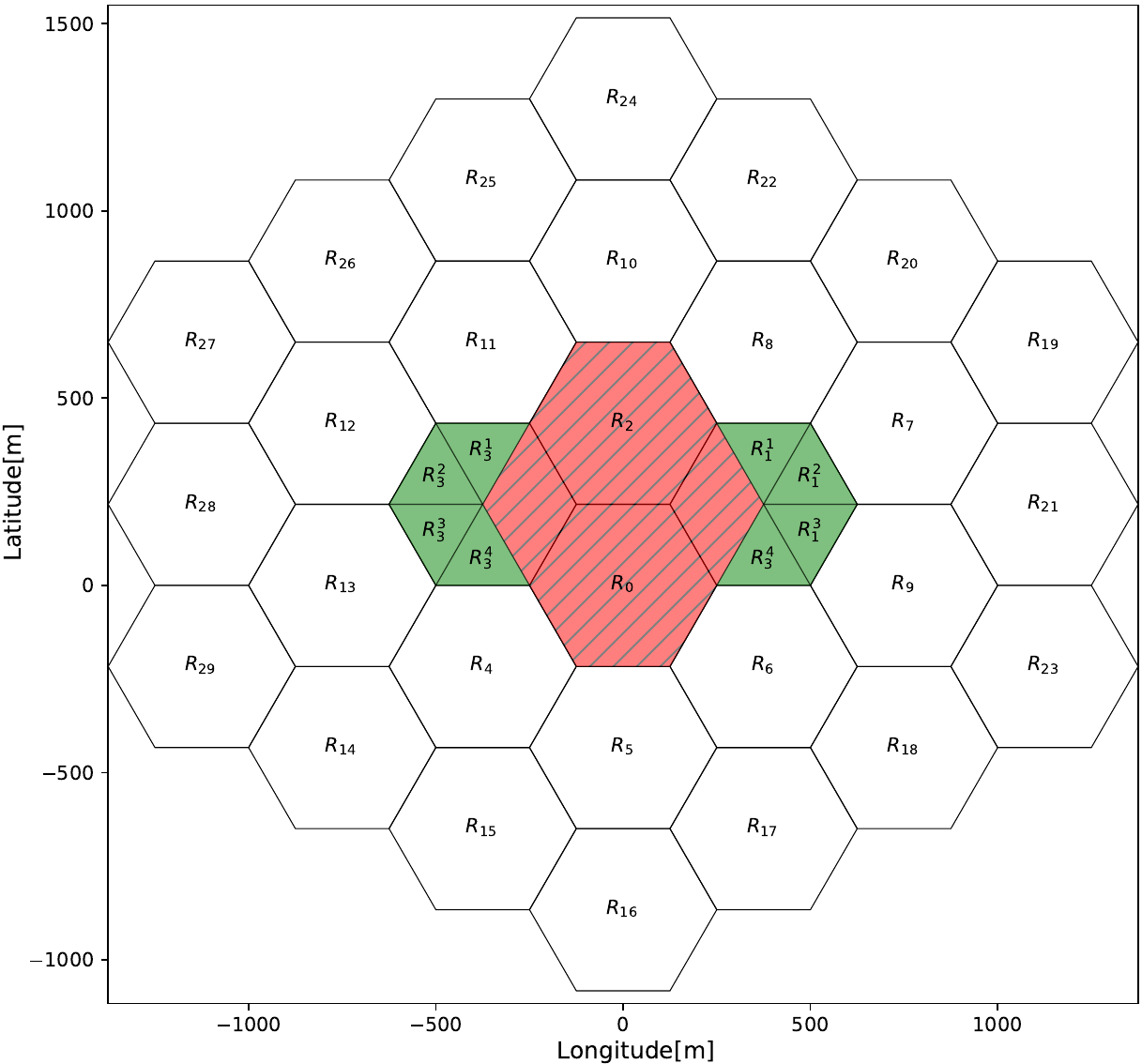}
    \captionsetup{font=footnotesize}
    \subcaption{Region division}
    \label{fig:}
\end{subfigure}
\hspace{1mm}
\begin{subfigure}[t]{0.44\linewidth}
    \centering
    \includegraphics[height=6.5cm]{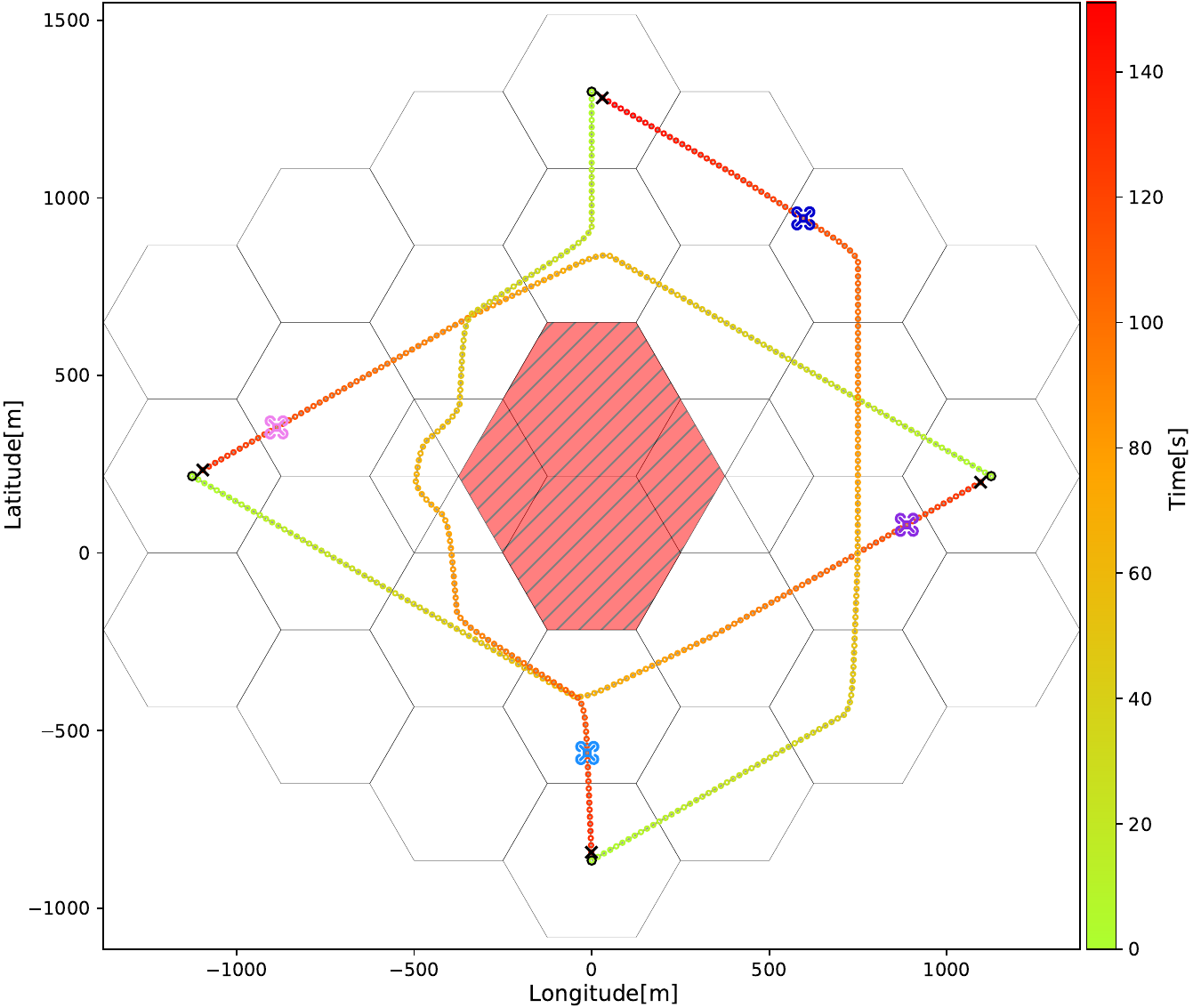}
    \captionsetup{font=footnotesize}
    \subcaption{Aircraft trajectories}
    \label{fig:no-fly2}
\end{subfigure}
\captionsetup{font=footnotesize}
\caption{An example of no-fly zones (red shaded areas) with hexagonal airspace regions divided into triangular sub-regions (green shaded areas).}
\label{fig:no-fly_zone}
\end{figure}

Furthermore, no-fly zones may require more precise shapes (not just the union of hexagons) to accommodate realistic obstacles.  
An example is presented in \autoref{fig:no-fly_zone} to demonstrate the potential of the proposed framework in achieving such no-fly zones.  
By subdividing hexagonal airspace regions into smaller triangular sub-regions, a spindle-shaped no-fly zone (marked by red shade) is created.  
Compared with the union of hexagons, this no-fly zone preserves an additional portion of usable airspace (marked by green shade).  
To enable aircraft to utilize the additional portion of usable airspace, the route guidance strategy is modified, with sampled aircraft trajectories illustrated in \autoref{fig:no-fly2}.  
In this example, the spindle-shaped no-fly zone helps reduce part of the aircraft's travel distance, particularly for flights between region $R_{16}$ and $R_{24}$.

The above results demonstrate the effectiveness of advanced routing algorithms for dynamic UAM management.
Similar findings have been reported in prior studies, particularly under dynamic or uncertain airspace conditions.
For example, \cite{pang2020concept} introduced AirMatrix network configurations and proposed risk cost maps to guide routing, allowing aircraft to avoid high-risk areas such as no-fly zones.
\cite{heidari2021collision,du2025unmanned} investigated optimal control approaches to aircraft path planning that account for varying wind loads and unknown geofences in uncertain environments.
In addition, \cite{li2024bi} developed a hybrid offline-online algorithm that allows aircraft to adapt paths in response to newly added no-fly zones and dynamic obstacles.
Together, these studies highlight that updating network configurations in response to traffic or environmental conditions facilitates dynamic airspace management through rerouting, consistent with the approach presented in this paper.
% \cite{heidari2021collision} proposed an optimal control approach for multi-rotor aircraft path planning under dynamic environments with varying wind loads.
% \cite{abdul2024congestion} developed a rule-based routing algorithm that enables aircraft to select alternative local paths in surrounding unoccupied airspace, thereby alleviating congestion along nominal paths.

\subsubsection{Evaluation of the fast approximation methods} \label{sec:Evaluation_FAM}

\begin{figure}[!ht]
    \centering
    \includegraphics[width=0.85\linewidth]{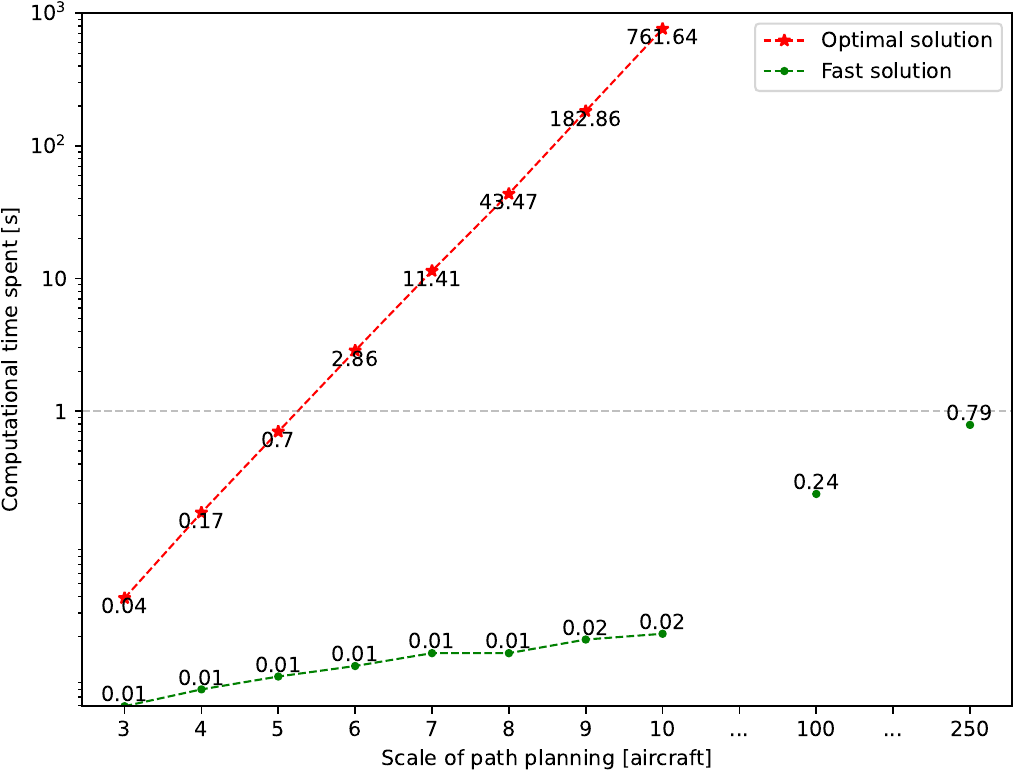}
    \captionsetup{font=footnotesize}
    \caption{Comparison of computational time between the exhaustive search and the fast approximation method.}
    \label{fig:computational_comparison}
\end{figure}

\begin{figure}[!h]
\begin{subfigure}[t]{0.4\linewidth}
    \centering
    \includegraphics[width=\linewidth]{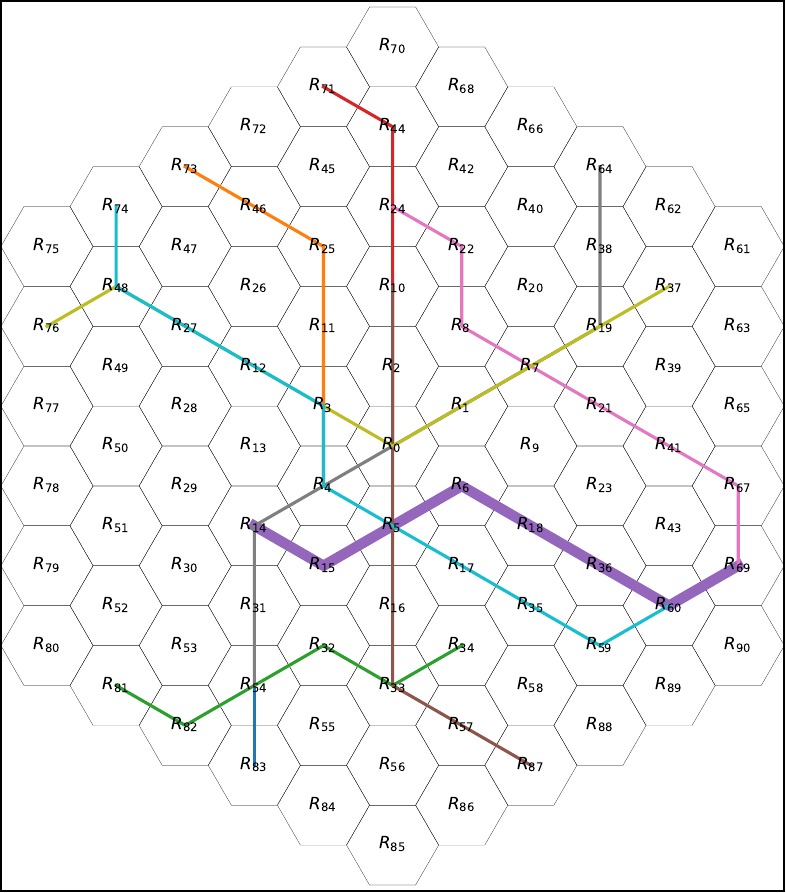}
    \captionsetup{font=footnotesize}
    \subcaption{Optimal solution}
    \label{fig:optimal}
\end{subfigure}
\hspace{2mm}
\begin{subfigure}[t]{0.4\linewidth}
    \centering
    \includegraphics[width=\linewidth]{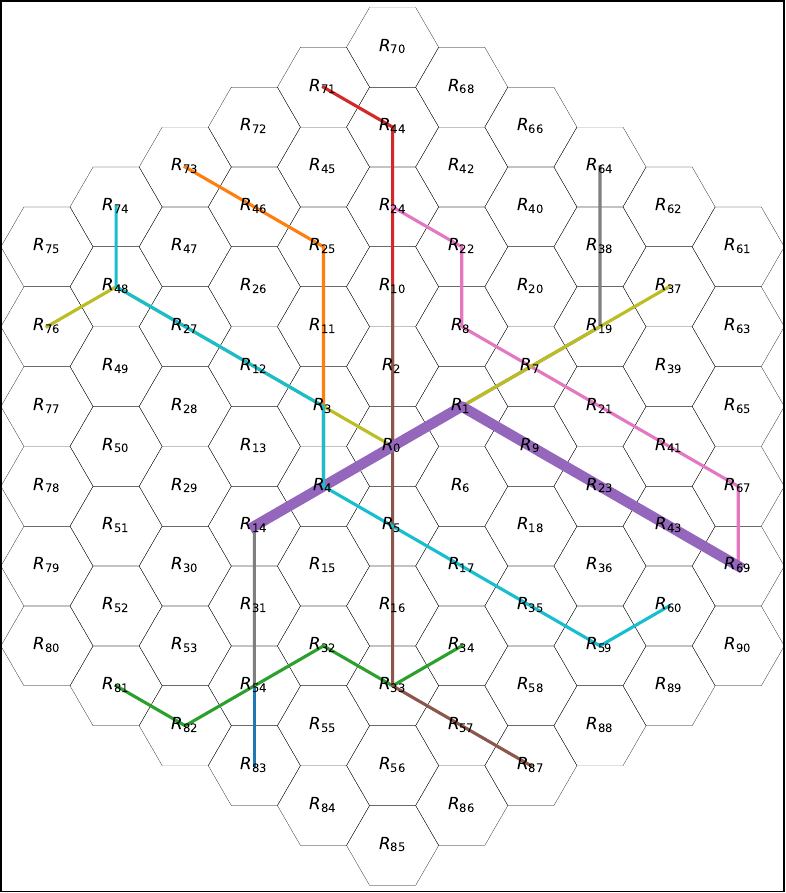}
    \captionsetup{font=footnotesize}
    \subcaption{Approximation solution}
    \label{fig:fast}
\end{subfigure}
\captionsetup{font=footnotesize}
\caption{
Comparison of the approximation solution and the optimal solution: the difference between the two solutions involves only one aircraft (with its path marked by a bold purple line). 
}
\label{fig:comparison_path}
\end{figure}

Lastly, we evaluate the performance of the FAMs proposed in \autoref{sec:FAM}, in terms of computational efficiency and accuracy.
\autoref{fig:computational_comparison} compares the computational time of the exhaustive search and the FAMs.  
As the problem scale increases, the computational time required for the FAMs grows very slowly.
In contrast, the computational time required for the optimal solution (obtained through exhaustive search) increases exponentially.  
With a high-performance computing cluster, the proposed algorithms can be implemented to solve large-scale UAM problems involving far more than 250 aircraft.  
Furthermore, the approximate solution provided by the FAMs performs well, as it is very close to the optimal solution.
As illustrated in \autoref{fig:comparison_path}, the difference between the two solutions involves only one aircraft (with its path marked by a bold purple line) at a problem scale of ten aircraft.  
Additionally, the discrepancy in travel cost between the two solutions is less than $1.5\%$.
These results demonstrate the great potential of the fast approximation methods in addressing large-scale path planning problems.

\newpage
\section{Conclusion}\label{sec:Conclusion}
This paper focused on real-time traffic simulation and management for urban air mobility (UAM).
Given that UAM is likely to be deployed between specific urban areas, dense point-to-point operations could increase the risk of aircraft collisions and air traffic congestion, particularly at large conflict points similar to roadway junctions.
To address this, we proposed an air traffic management framework that integrates route guidance and collision avoidance for large-scale UAM operations.
The proposed route guidance strategy provides aircraft with time-efficient paths composed of waypoints, while the collision avoidance algorithm generates safe trajectories between given waypoints.
The proposed framework achieves great efficiency in large-scale UAM operations and retains an elegant property of the Macroscopic Fundamental Diagram (MFD) for air traffic while guaranteeing air traffic safety.

To the best of our knowledge, this work is one of the first to combine route guidance and collision avoidance. 
By introducing the route guidance mechanism, aircraft are appropriately assigned across both spatial and temporal dimensions, which ensures air traffic homogeneity even for spatially heterogeneous demand.
The simulation results demonstrate our framework's capability to guarantee the critical assumption of traffic homogeneity in the MFD model for air traffic \citep{geroliminis2011properties,safadi2023macroscopic}.
Moreover, the results indicate that the proposed framework could enable efficient and flexible UAM operations, including air traffic assignment, local congestion mitigation, and dynamic no-fly zone management.
Compared with a collision-free baseline strategy, the proposed framework achieves considerable improvements in traffic safety and efficiency, with increases in the average minimum separation (+98.2\%), the average travel speed (+70.2\%), and the trip completion rate (+130\%), along with a reduction in the energy consumption (-23.0\%).
These findings highlight the framework's potential for supporting real-time traffic management in large-scale UAM systems.

Since the current approaches are developed based on a simplified and abstract urban environment, several promising directions can be pursued in future work to address these limitations.
One potential direction is to enhance the collision avoidance algorithm to account for environmental uncertainties, such as those caused by weather conditions, communication disturbances, or actuator limitations.
In addition, further investigation into structured airspace design that incorporates realistic urban features, such as tall buildings and dynamic obstacles, would be valuable for supporting the early deployment of UAM systems.

% {\color{red} The proposed framework can be easily transferred to multi-agent systems.
% Noting that this work has addressed route guidance in a compromised way, further studies in multi-agent path planning with the consideration of traffic congestion (or local deadlocks) would be a promising direction \citep{dergachev2021distributed}.
% Back to the field of UAM, interesting directions for future work would include (\rmnum{1}) investigating how to achieve macroscopic traffic flow control through microscopic aircraft control, e.g., by route guidance or speed control; and (\rmnum{2}) developing adaptive and dynamic region division methods for route guidance. }

\section*{Acknowledgement}
The work in this paper was jointly supported by research grants from the National Natural Science Foundation of China (No. 72071214 \& 62203239) and the Science and 
Technology Planning Project of Guangdong Province (No. 2023B1212060029).

% \begin{appendices}
% \input{Sections/Appendix/Part-A}
% \input{Sections/Appendix/Part-B}
% \end{appendices}

\bibliographystyle{elsarticle-harv} 
\bibliography{reference}
\end{document}